\documentclass[11pt]{cernrep}
\usepackage{graphicx}
\usepackage{here}
\begin{document} 
\def    \nn             {\nonumber} 
\def    \=              {\;=\;} 
\def    \frac           #1#2{{#1 \over #2}} 
\def    \ret            {\\[\eqskip]} 
\def    \ie             {{i.e.} } 
\def    \etal             {{et al.} } 
\def    \eg             {{e.g.} } 
\def\MS{\hbox{$\overline{\rm MS}$}}
\def    \lsim {\raisebox{-3pt}{$\>\stackrel{<}{\scriptstyle\sim}\>$}}  
\def    \gsim {\raisebox{-3pt}{$\>\stackrel{>}{\scriptstyle\sim}\>$}}  
\def    \gtrsim {\raisebox{-3pt}{$\>\stackrel{>}{\scriptstyle\sim}\>$}}  
\def    \esim {\raisebox{-3pt}{$\>\stackrel{-}{\scriptstyle\sim}\>$}}  
\def\wup{{W^+}}
\def\wum{{W^-}}
\def\wupm{{W^\pm}}
\def\wump{{W^\mp}}
\newcommand     \be     {\begin{equation}} 
\newcommand     \ee     {\end{equation}} 
\newcommand     \ba     {\begin{eqnarray}} 
\newcommand     \ea     {\end{eqnarray}} 
\newcommand     \sst            {\scriptstyle} 
\newcommand     \sss            {\scriptscriptstyle} 
\newcommand     \auno{a^{(1)}} 
\newcommand     \avg[1]         {\left\langle #1 \right\rangle} 
\newcommand     \ptmin     {\ifmmode p_{\scriptscriptstyle T}^{\sss min} \else 
                           $p_{\scriptscriptstyle T}^{\sss min}$ \fi} 
\def     \muf           {\mbox{$\mu_{\sss F}$}} 
\def     \mur            {\mbox{$\mu_{\sss R}$}} 
\def    \muo            {\mbox{$\mu_0$}} 
\newcommand\as{\alpha_{\sss \mathrm S}} 
\newcommand\astwo{\alpha_{\sss \mathrm S}^2} 
\newcommand\asthree{\alpha_{\sss \mathrm S}^3} 
\newcommand\asfour{\alpha_{\sss \mathrm S}^4} 
\newcommand\aem{\alpha_{\mathrm em}} 
\newcommand\cb{\overline{c}} 
\newcommand\bb{\overline{b}} 
\newcommand\tb{\overline{t}} 
\newcommand\Qb{\overline{Q}} 
\newcommand\qq{{\scriptscriptstyle Q\overline{Q}}} 
\def \asopi{\mbox{$\frac{\as}{\pi}$}} 
\def \oafour {\mbox{${\cal O}(\asfour)$}} 
\def \oacube {\mbox{${\cal O}(\asthree)$}} 
\def \oatwo {\mbox{${\cal O}(\astwo)$}} 
\def \oas   {\mbox{${\cal O}(\as)$}} 
\def \bbbar {\mbox{$b \bar b$}} 
\def \ccbar {\mbox{$c \bar c$}} 
\def \pt   {\mbox{$p_{\scriptscriptstyle T}$}} 
\def \et   {\mbox{$E_{\scriptscriptstyle T}$}} 
\def \Emu  {\ifmmode{E_{\mu}
    }\else{$E_{\mu}$}\fi} 
\def \Enu  {\ifmmode{E_{\nu}
    }\else{$E_{\nu}$}\fi} 
\def \nudis {$\nu$DIS}
\def \nubar {\bar{\nu}}
\def \nue  {\ifmmode{\nu_e}\else{$\nu_e$}\fi} 
\def \numu  {\ifmmode{\nu_{\mu}}\else{$\nu_{\mu}$}\fi} 
\def \nufact {$\nu$-Factory}
\def \rap   {\mbox{$\eta$}} 
\def \deltar {\mbox{$\Delta R$}} 
\def \dphi {\mbox{$\Delta \phi$}} 
\def \to   {\mbox{$\rightarrow$}} 
\def    \mb             {\mbox{$m_b$}} 
\def    \mc             {\mbox{$m_c$}} 
\def    \mt             {\mbox{$m_t$}} 
\newcommand \jpsi{\ifmmode{J/\psi 
    }\else{$J/\psi$}\fi} 
\def\calF{{\cal F}} 
\def\calP{{\cal P}} 
\def\calM{{\cal M}} 
\def\calO{{\cal O}} 
\def        \mW         {\mbox{$m_W$}} 
%%%%%%%%% MELE MACROS %%%%%%%%%%%%%%% 
\def\lta{\;\raisebox{-.5ex}{\rlap{$\sim$}} \raisebox{.5ex}{$<$}\;} 
\def\gta{\;\raisebox{-.5ex}{\rlap{$\sim$}} \raisebox{.5ex}{$>$}\;} 
\newcommand{\permille}{$^0 \!\!\!\: / \! _{00}\;$} 
\newcommand{\GeV}{GeV} 
\newcommand{\mw}{M_{W}} 
\newcommand{\md}{m_{d}} 
\newcommand{\ms}{m_{s}} 
\newcommand{\mz}{M_{Z}} 
 
\newcommand{\ibidem}{{\it ibidem\/},} 
\newcommand{\into}{\;\;\to\;\;} 
\newcommand{\pT}{\mbox{$p_{T}$}} 
\newcommand{\mZ}{\mbox{$m_{Z}$}} 
\newcommand{\ET}{\mbox{$E_{T}$}} 
\newcommand{\ETmiss}{\etmiss} %\mbox{$E_{T}^{miss}$}} 
\def\half{\frac{1}{2}} 
%%% GAMBINO MACROS
\newenvironment{comment}[1]{}{}
\newcommand{\equ}[1]{Eq.~(\ref{#1})}
\newcommand{\eqs}[1]{Eqs.~(\ref{#1})}
\newcommand{\Eqs}[2]{Eqs.~(\ref{#1}) and~(\ref{#2})}
\newcommand{\ccur}{\mbox{$\hat{c}^2$}}
\newcommand{\ew}{electroweak~}
\newcommand{\non}{\nonumber}
\newcommand{\msbar}{\overline{\mathrm MS}}
\newcommand{\smallms}{{\scriptscriptstyle \mathrm MS}}
\newcommand{\smallmsbar}{\overline{\smallms}}
\newcommand{\smallz}{{\scriptscriptstyle Z}} %  a smaller Z
\newcommand{\smallw}{{\scriptscriptstyle W}} %
\newcommand{\smallh}{{\scriptscriptstyle H}} %
\newcommand{\smallv}{{\scriptscriptstyle V}}
\newcommand{\acur}{ {\hat \alpha} }
\newcommand{\mh}{M_\smallh}
%\newcommand{\mt}{M_t}
%\newcommand{\mb}{m_b}
% BENEDIKT MACROS
\newcommand{\feynmandag}{\slash\hspace{-0.55em}}
\newcommand{\rmd}{\mathrm{d}}
\newcommand{\rmi}{\mathrm{i}}
\newcommand{\alphas}{\alpha_{\mathrm{s}}}
\newcommand{\xBj}{x_{\mathrm{Bj}}}

% START FRONT PAGE FOR HEP-PH (comment away for the CERN Report)
\pagestyle{plain}
\begin{titlepage}
\nopagebreak
{\flushright{
        \begin{minipage}{5cm}
        CERN-TH/2001-131\\
        {\tt hep-ph/0105155}\\
        \end{minipage}        }
        
}
\vfill
\begin{center}
{\LARGE { \bf \sc 
PHYSICS AT THE FRONT-END OF A NEUTRINO FACTORY: A QUANTITATIVE
  APPRAISAL}

}
\vfill                                                       
  M.L. Mangano~$^{a}$ (convener), 
  S.I.~Alekhin~$^{b}$, M.~Anselmino~$^{c}$,
  R.D.~Ball~$^{a,d}$, M.~Boglione~$^{e}$,
  U.~D'Alesio~$^{f}$, S.~Davidson~$^{g}$,
  G.~De~Lellis~$^{h}$, J.~Ellis~$^{a}$,
  S.~Forte~$^{i}$, P.~Gambino~$^{a}$,
  T.~Gehrmann~$^{a}$, A.L.~Kataev~$^{a,j}$, A.~Kotzinian~$^{a,k}$,
  S.A.~Kulagin~$^{j}$, B.~Lehmann-Dronke~$^{l}$,
  P.~Migliozzi~$^{h}$, F.~Murgia~$^{f}$, G.~Ridolfi~$^{a,m}$
 \\
{\small
  $^{a}$~Theoretical Physics Division, CERN, Geneva, Switzerland\\
  $^{b}$~Institute for High Energy Physics, Protvino, Russia\\
  $^{c}$~Dipartimento di Fisica Teorica dell'Universit\`a e Sezione INFN
  di Torino, Turin, Italy\\
  $^{d}$~Dept. of Physics and Astronomy, University of Edinburgh, Scotland\\
  $^{e}$~Dept. of Physics and IPPP, University of Durham, U.K.\\
  $^{f}$~Dipartimento di Fisica dell'Universit\`a e Sezione INFN di
  Cagliari, Cagliari, Italy\\
  $^{g}$~Theoretical Physics, Oxford University, U.K.\\
  $^{h}$~INFN, Sezione di Napoli, Naples, Italy\\
  $^{i}$~INFN, Sezione di Roma III, Rome, Italy\\
  $^{j}$~Institute for Nuclear Research, Academy of Sciences, Moscow, Russia\\
  $^{k}$~JINR, Dubna, Russia\\
  $^{l}$~Institut f\"ur Theoretische Physik, Univerist\"at Regensburg,
  Germany\\
  $^{m}$~INFN, Sezione di Genova, Genoa, Italy }

\end{center}                                   
\nopagebreak
\vfill
%\vskip 3cm
\begin{abstract} 
We present a quantitative appraisal of the physics potential for
neutrino experiments at the front-end of a muon storage ring. We
estimate the forseeable accuracy in the determination of several
interesting observables, and explore the consequences of these
measurements.  We discuss the extraction of individual quark and
antiquark densities from polarized and unpolarized deep-inelastic
scattering. In particular we study the implications for the
undertanding of the nucleon spin structure.  We assess the
determination of $\as$ from scaling violation of structure functions,
and from sum rules, and the determination of $\sin^2\theta_W$ from
elastic $\nu e$ and deep-inelastic $\nu p$ scattering. We then
consider the production of charmed hadrons, and the measurement of
their absolute branching ratios. We study the polarization of
$\Lambda$ baryons produced in the current and target fragmentation
regions.  Finally, we discuss the sensitivity to physics beyond the
Standard Model.
\end{abstract}                                                
\vskip 1cm
CERN-TH/2001-131\hfill \\
\today \hfill  
\vfill       
\end{titlepage}
% END FRONT PAGE FOR HEP-PH

\tableofcontents

\title{PHYSICS AT THE FRONT-END OF A NEUTRINO FACTORY: A QUANTITATIVE
  APPRAISAL} 
%PROSPECTS FOR NEUTRINO DEEP-INELASTIC SCATTERING MEASUREMENTS AT
%   THE NEUTRINO FACTORY}
%
% 

\author{M.L. Mangano~$^{a}$ (convener), 
  S.I.~Alekhin~$^{b}$, M.~Anselmino~$^{c}$,
  R.D.~Ball~$^{a,d}$, M.~Boglione~$^{e}$,
  U.~D'Alesio~$^{f}$, S.~Davidson~$^{g}$,
  G.~De~Lellis~$^{h}$, J.~Ellis~$^{a}$,
  S.~Forte~$^{i}$, P.~Gambino~$^{a}$,
  T.~Gehrmann~$^{a}$, A.L.~Kataev~$^{a,j}$, A.~Kotzinian~$^{a,k}$,
  S.A.~Kulagin~$^{j}$, B.~Lehmann-Dronke~$^{l}$,
  P.~Migliozzi~$^{h}$, F.~Murgia~$^{f}$, G.~Ridolfi~$^{a,m}$ } 

\institute{
  $^{a}$~Theoretical Physics Division, CERN, Geneva, Switzerland\\
  $^{b}$~Institute for High Energy Physics, Protvino, Russia\\
  $^{c}$~Dipartimento di Fisica Teorica dell'Universit\`a e Sezione INFN
  di Torino, Turin, Italy\\
  $^{d}$~Dept. of Physics and Astronomy, University of Edinburgh, Scotland\\
  $^{e}$~Dept. of Physics and IPPP, University of Durham, U.K.\\
  $^{f}$~Dipartimento di Fisica dell'Universit\`a e Sezione INFN di
  Cagliari, Cagliari, Italy\\
  $^{g}$~Theoretical Physics, Oxford University, U.K.\\
  $^{h}$~INFN, Sezione di Napoli, Naples, Italy\\
  $^{i}$~INFN, Sezione di Roma III, Rome, Italy\\
  $^{j}$~Institute for Nuclear Research, Academy of Sciences, Moscow, Russia\\
  $^{k}$~JINR, Dubna, Russia\\
  $^{l}$~Institut f\"ur Theoretische Physik, Univerist\"at Regensburg,
  Germany\\
  $^{m}$~INFN, Sezione di Genova, Genoa, Italy
}
\maketitle
\begin{abstract}
We present a quantitative appraisal of the physics potential for
neutrino experiments at the front-end of a muon storage ring. We
estimate the forseeable accuracy in the determination of several
interesting observables, and explore the consequences of these
measurements.  We discuss the extraction of individual quark and
antiquark densities from polarized and unpolarized deep-inelastic
scattering. In particular we study the implications for the
undertanding of the nucleon spin structure.  We assess the
determination of $\as$ from scaling violation of structure functions,
and from sum rules, and the determination of $\sin^2\theta_W$ from
elastic $\nu e$ and deep-inelastic $\nu p$ scattering. We then
consider the production of charmed hadrons, and the measurement of
their absolute branching ratios. We study the polarization of
$\Lambda$ baryons produced in the current and target fragmentation
regions.  Finally, we discuss the sensitivity to physics beyond the
Standard Model.
\end{abstract}

\section{INTRODUCTION}
The use of  intense neutrino beams as a way of exploring the deep
structure of hadrons has long been recognized~\cite{Geer:1998iz} as a
big added value of a muon-collider~\cite{mucoll} and neutrino-factory
(\nufact) complex. Recent documents \cite{bkbook,Albright:2000xi} have
outlined with great care the areas where deep-inelastic-scattering
(DIS) experiments operating closely downstream of the muon ring could
provide significant contributions to our understanding of the nucleon
structure.

These studies pointed out the potential for measurements of
unparalleled precision of both unpolarized and polarized neutrino
structure functions (SFs), leading to an accurate
decomposition of the partonic content of the nucleon in terms of
individual (possibly spin-dependent) flavour densities.
In addition to the measurements of SFs, the large
rate of charm production, allowed even with muon beam energies as low
as  50~GeV, gives an opportunity for accurate studies of 
the spectrum and decay properties of charmed systems (mesonic and
baryonic), as well as for an improved determination of the CKM matrix
element $V_{cd}$. Operation at muon beam energies in excess of 500~GeV
would allow similar studies using $b$-flavoured hadrons.
The large neutrino fluxes will also make large-statistics $\nu_{\mu}
e$ and  $\nu_{e} e$ scattering experiments possible. These
measurements may provide very accurate determinations of the weak
interaction parameter $\sin^2\theta_W$, complementing in terms of
accuracy and systematics the current determinations from higher-energy
measurements in $Z^0$ decays and from DIS. 

The goal of the work performed within our Working Group was to address
some of the topics proposed in~\cite{bkbook,Albright:2000xi} in a
quantitative way, and carry out a concrete appraisal of the impact
that measurements done at the \nufact\  could have on relevant
observables. A first study in this direction, limited to the case of
SFs, has recently appeared in~\cite{Ball:2000qd}.

In Section~2 of this document we  review our notation and
describe the benchmark beam and detector parameters used in this
study. In Section~3 we discuss the determination of unpolarized SFs, 
and of their flavour decomposition, using a
next-to-leading-order (NLO) global fit
analysis. The importance of the NLO analysis is not related to the
cross-section changes induced by NLO corrections, which are marginal
when evaluating at this stage the expected event rates, 
but to the mixing between quark and gluon contributions
which arise at NLO. This mixing leads to a potential loss of accuracy
in the extraction of individual flavours. A similar NLO study is
documented in Section~4 for the polarized case; there we study both
the accuracy in the determination of the individual shapes of
polarized parton distributions, and the accuracy in the extraction of
the proton axial charges. We shall put these results in the framework
of the ability to distinguish between different scenarios for the
description of the proton spin. The relevance of the NLO effects is 
even more significant  in
this case than in the unpolarized case, because
of the larger uncertainties on the polarized gluon contribution.
In that section we also analyse the use
of tagged charm final states to study the strange quark polarized
distribution. In Section~5 we discuss the prospects
for extractions of $\as$ from global SF fits, as well as from 
the GLS and unpolarized Bjorken sum
rules, and in Section~6 we analyse the nuclear effects involved in the
extraction of charged-current (CC) neutrino SFs from heavy
targets. New prospects in this area are opened by the availability of
new SFs, whose nuclear corrections have sizes
different from those studied with data available today.
In Section~7 we discuss the extraction of $\sin^2\theta_W$ from
$\nu e$  scattering and DIS. The large statistics will enable
measurements of an accuracy similar to that available today from 
LEP, and will provide important and complementary tests of the
Standard Model (SM).
In Section~8 we study measurements
involving charm quarks. In Section~9 we consider the application
of polarization measurements in semi-exclusive final states
to the study of polarized nucleon densities. In Section~10
we finally consider the potential of the \nufact\ for the detection of
indirect evidence of new physics from precise measurements of
SM observables.

\section{GENERALITIES}
\begin{figure} 
\begin{center} 
\centerline{
\includegraphics[width=0.8\textwidth,clip]{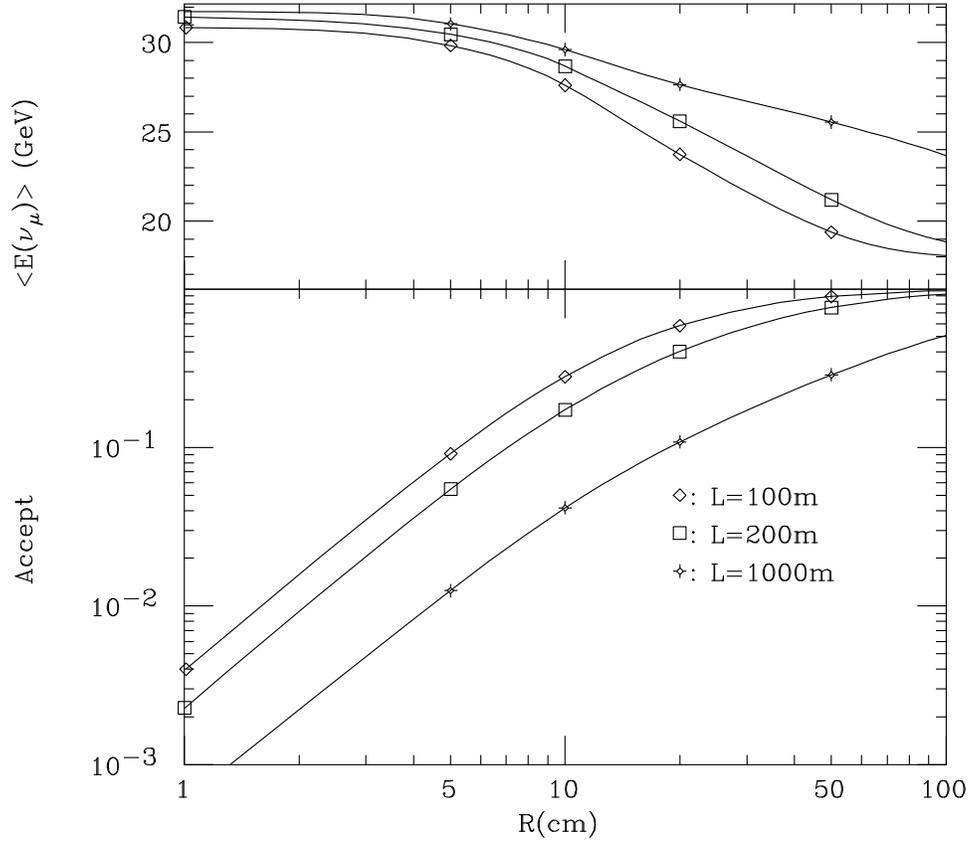} }
\vskip -0.4cm 
\caption{Bottom: for 50~GeV muons decaying along a
  straight section of length $L=100,200$ and 1000m, 
  we plot the fraction of the muon-neutrino
  flux contained within a circle of radius $R$, at a distance of 30m
  from the end of the straight section. Top: average \numu\ energy
  for the three beam configurations, as a function of $R$.
\label{fig:acceptR}}
\end{center} 
\end{figure} 
For our studies (and unless otherwise indicated) we shall assume the
following default specifications.  Muon beam energy, $E_{\mu}=50$~GeV;
length of the straight section, $L=100$~m; distance of the
detector from the end of the straight section, $d=30$~m; number
of muon decays per year along the straight section, $N_{\mu}=10^{20}$;
muon beam angular divergence, $0.1\times m_{\mu}/E_{\mu}$, $m_\mu$
being the muon mass; muon beam transverse size
$\sigma_x=\sigma_y=1.2$~mm.  We also assume a cylindrical detector,
with azimuthal symmetry around the beam axis, with a target of
radius $R=50$~cm and a density of $100~{\mathrm g}/{\rm cm}^2$ ($10~{\mathrm 
  g}/{\mathrm cm}^2$ in the case of polarized targets).  The statistics,
then, 
scale linearly with the detector length, while the dependence of other
parameters, such as the radius or the length of the straight section, is
clearly more complex. Some examples are given in
Fig.~\ref{fig:acceptR}.

\begin{figure} 
\begin{center} 
\includegraphics[width=0.8\textwidth,clip]{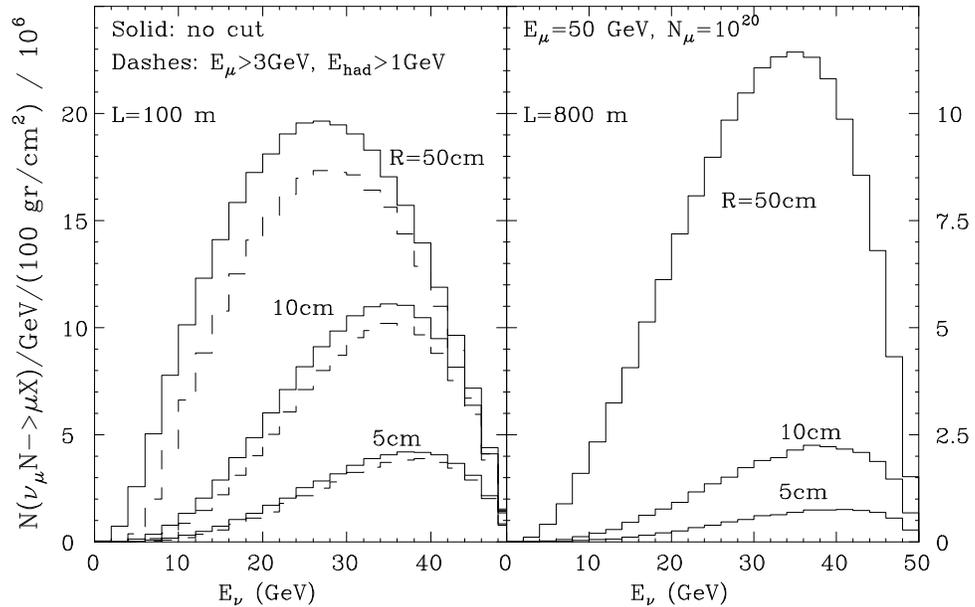} 
\end{center} 
\vskip -1cm 
\caption{CC event rates, in units of $10^6$, as function of Lab-frame 
  neutrino spectra, for several detector and beam configurations. The
  dashed lines on the left include cuts on the final-state muon
  ($\Emu>3$~GeV) and on the final-state hadronic energy
  ($E_{had}>1$~GeV). The solid lines have no energy-threshold cuts
  applied. The three set of curves correpsond to different detector
  radiuses (50, 10 and 5~cm, from top to bottom).
\label{fig:50gev}} 
\end{figure}

\begin{figure} 
\begin{center} 
\includegraphics[width=0.8\textwidth,clip]{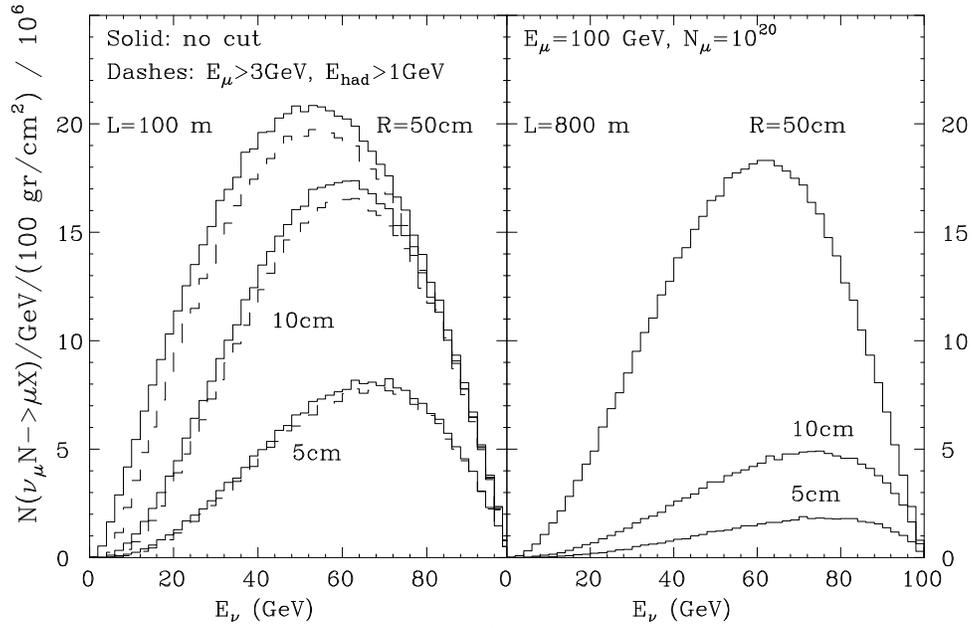} 
\end{center} 
\vskip -1cm 
\caption{Same as Fig.~\ref{fig:50gev}, but for 100~GeV muon beams.
\label{fig:100gev}} 
\end{figure}

\begin{figure} 
\begin{center} 
\includegraphics[width=0.8\textwidth,clip]{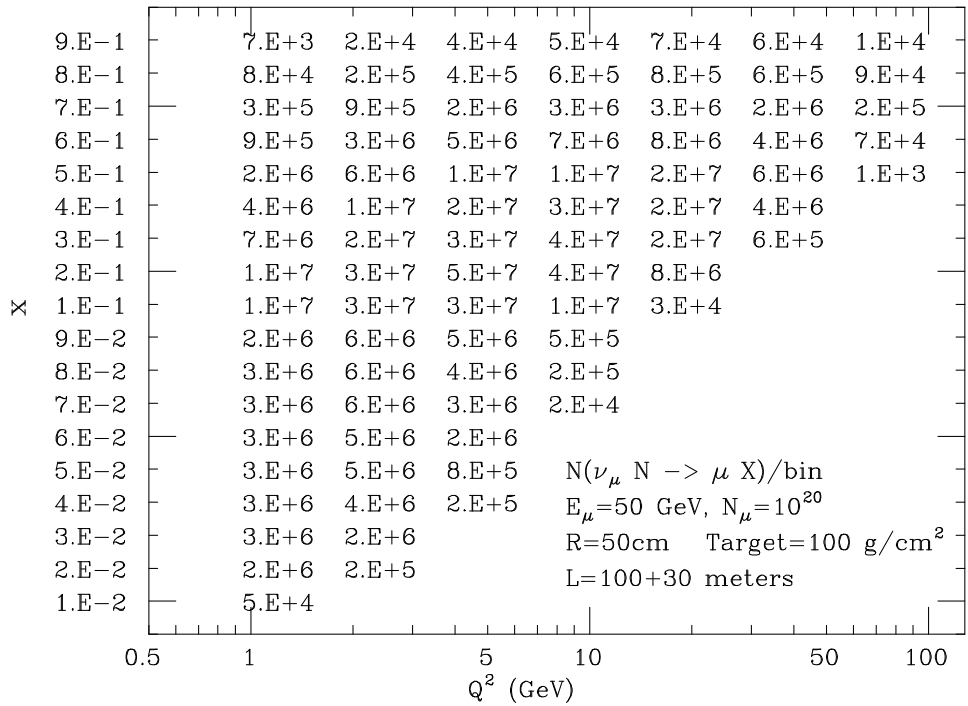} 
\end{center} 
\caption{Event rates, in different
bins of $(x,Q^2)$, for the default beam and detector configuration
($\Emu=50$~GeV, $L=100$~m, $R=50$~cm).
\label{fig:xqrates}} 
\end{figure}

The neutrino spectra are calculated using standard expressions for the
muon decays (see e.g.~\cite{bkbook}). 
For simplicity (with the exception of the $\nu e$ scattering studies), 
we shall confine ourselves to the case of $\nu_{\mu}$ and
$\bar{\nu}_{\mu}$ CC DIS. 
The laboratory-frame neutrino spectra, convoluted with the CC
interaction cross-sections, are shown for several detector and beam
configurations in Fig.~\ref{fig:50gev} ($\Emu=50$~GeV) 
and~Fig.~\ref{fig:100gev} ($\Emu=100$~GeV). The number of events, in different
bins of $(x,Q^2)$, are shown in Fig.~\ref{fig:xqrates}. 

\section{UNPOLARIZED STRUCTURE FUNCTIONS}
\label{sec:F2F3}
\subsection{Formalism}
Unpolarized CC SFs are defined 
through the decomposition of unpolarized differential CC 
cross-sections into invariant functions of 
the momentum of the struck quark ($x$) and the momentum transfer
squared of the $W$ boson ($Q^2$): the standard definitions give
\be \label{xsectot}
\frac{d^2\sigma}{dx dy} = 
\frac{G_F^2 S}{2\pi(1+Q^2/M_W^2)^2}
\left[(1-y)F_2 
+ y^2 xF_1
\pm y \left(1-\frac{y}{2} \right) xF_3\right],\label{eq:sfn}
\ee
where $S=2m\Enu$ is the nucleon--neutrino centre-of-mass energy, 
$m$ is the nucleon mass, $\Enu$ is the neutrino 
beam energy, assumed to be $\gg m$, $y$ is
the fractional lepton energy loss, or $(E_\nu-E_\ell)/E_\nu$,
and the $\pm$ signs 
refer to the sign of the CC: $W^+$ 
exchange for $\nu$ scattering and $W^-$ for $\bar\nu$. 
In neutrino scattering, $x,y$, and $Q^2$ can all be determined 
simply by measuring the outgoing lepton energy and direction, 
and the hadronic energy in the event. If both $\nu$ and $\bar \nu$
beams are available, there 
 are then, in principle, six independent SFs for each target. 

We now wish to determine the expected statistical
accuracy with which the individual SFs, 
and their flavour components,
can be determined. To do this, we shall exploit the different $y$
dependences of the cross section on the various $F_{i}$.
The advantage of the neutrino beams from muon decays is their
wide-band nature. This allows us to modulate the $y$ dependence for fixed
values of $x$ and $Q^2$ using the neutrino energy:
\be
y=\frac{Q^2}{2xm \Enu} \; .
\ee
We produced $y$ distributions by generating events within
different bins of $x$ and $Q^2$, and performed minimum-$\chi^2$ fits
of the generated data using the cross-section \equ{xsectot}. For each
bin, the values of $x$ and $Q^2$ at which we quote the results are
obtained from the weighted average of the event rate. As an input, we
used the CTEQ4D set of parton distributions~\cite{Lai:1997mg}. 
The dependence on
the parameterization of the parton distributions is very small, and
will be neglected here. We verified that other recent sets of parton
distributions give similar results. 
The absolute number of events expected in each bin is scaled by the
total number of muon decays; this number of events determines the
statistical error on the individual SFs obtained
through the fit.  

\begin{figure} 
\begin{center} 
\includegraphics[width=0.45\textwidth,clip]{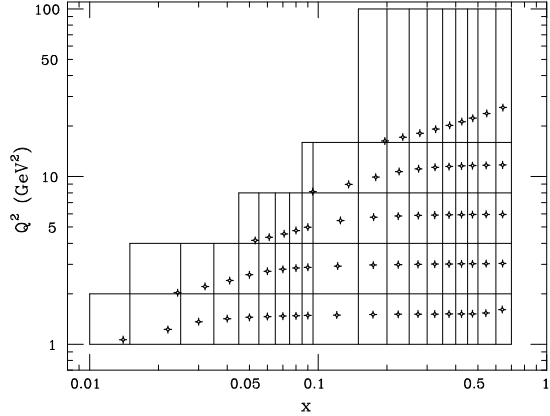}
\caption{$x$ and $Q^2$ binning for the generation of CC
events. The crosses correspond to the weighted bin centers.
\label{fig:xqrange}}
\end{center} 
\end{figure}

We generate events in the ($x,Q^2$) bins shown in
Fig.~\ref{fig:xqrange}.  Twenty
equally-spaced bins in the range $0\leq y\leq 1$ are used for the $y$
fit. The total number of $x$ bins varies in different $Q^2$ bins
because of kinematic acceptance and minimum energy cuts.
The statistical errors returned by the fits are used as estimates of
the statistical errors in the extraction of the $F_{i}$. 
In the parton model, four of the SFs 
are related through the Callan--Gross relations $F_2=2xF_1$:
the longitudinal SF $F_L=F_2-2xF_1$ begins at 
$O(\alpha_s)$ in perturbation theory. We
considered both the cases of three-component fits (leaving $F_{1}$ and
$F_{2}$ uncorrelated, and fitting all three SFs), and
two-component fits (assuming the Callan--Gross relation, and fitting
for $F_{2}$ and $F_{3}$). 

\begin{figure} 
\begin{center} 
\includegraphics[width=0.8\textwidth,clip]{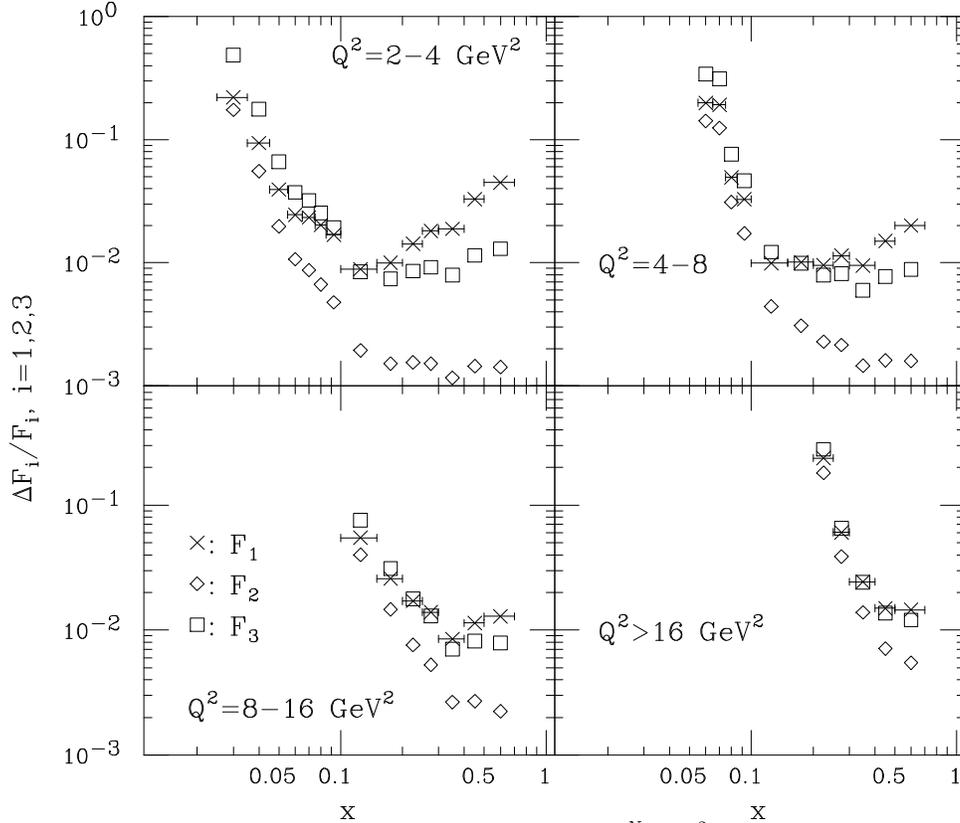} 
\end{center} 
\vskip -1cm 
\caption{Expected errors on the determination of the individual
  nucleon SFs $F^{\nu N}_i(x,Q^2)$ ($i=1,2,3$), for various ranges of
  $Q^2$. The horizontal bars indicate the range in $x$ defining the
  bins within which the statistical errors are determined.
\label{fig:F123err}} 
\end{figure}

\begin{figure}
\begin{center} 
\vglue 5cm
\includegraphics[width=0.8\textwidth,clip]{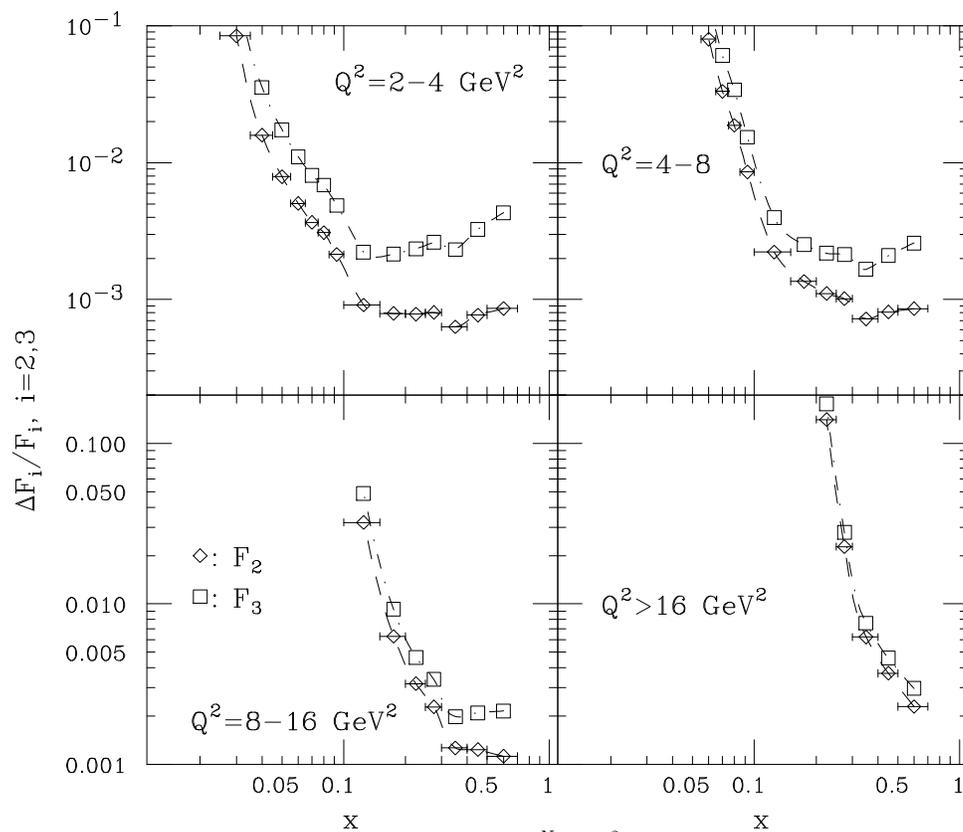} 
\end{center} 
\vskip -1cm 
\caption{Expected errors on the determination of the
nucleon SFs $F^{\nu N}_i(x,Q^2)$ ($i=2,3$), assuming the Callan--Gross
relation, for various ranges of
  $Q^2$. The horizontal bars indicate the range in $x$ defining the
  bins within which the stastical errors are determined.
(to be included)
\label{fig:F23err}} 
\end{figure}

The three-component fit for the nucleon SFs, 
obtained assuming one year
exposure of a deuterium target to a muon--neutrino beam, is shown in
Fig.~\ref{fig:F123err}. Notice that the errors of $F_1$ are always
better than 10\%. $F_2$ is determined very well at large $x$ since in
this region one is more sensitive to the $y\to 0$ limit, where the
contribution from $F_1$ and $F_3$ is suppressed (see Eq.~\ref{eq:sfn}).
Figure~\ref{fig:F23err} shows the result obtained assuming the
Callan--Gross relation. Now the relative errors are better than
1\% for most regions of $x$ and $Q^2$. The significant improvement in
the determination of $F_3$ at large $x$ results from the Callan--Gross
constraint on $F_1$, and the fact that, as pointed out above,
$F_2$ is very well determined at large $x$.

\subsection{Leading-order results}
\label{sec:unpolsfres}
The parton content of the 
SFs $F_1$, 
$F_2$ and $F_3$ depends on the charge of the exchanged gauge boson.
At leading order, we have
\ba
\label{part1}
 F_1^\wup &=&\bar u + d + s + \bar c,
\quad\quad\quad\qquad F_1^\wum = u + \bar d + \bar s + c,\nn \\
\label{part2}
 F_2^\wup &=&2x(\bar u + d + s + \bar c),
\quad\qquad F_2^\wum = 2x(u + \bar d + \bar s + c),\\
xF_3^\wup &=& 2x(-\bar u + d + s - \bar c),
\qquad xF_3^\wum =2x(u -\bar d - \bar s + c) \nn.
\label{part3}
\ea
\begin{figure}
\begin{center} 
\includegraphics[width=0.8\textwidth,clip]{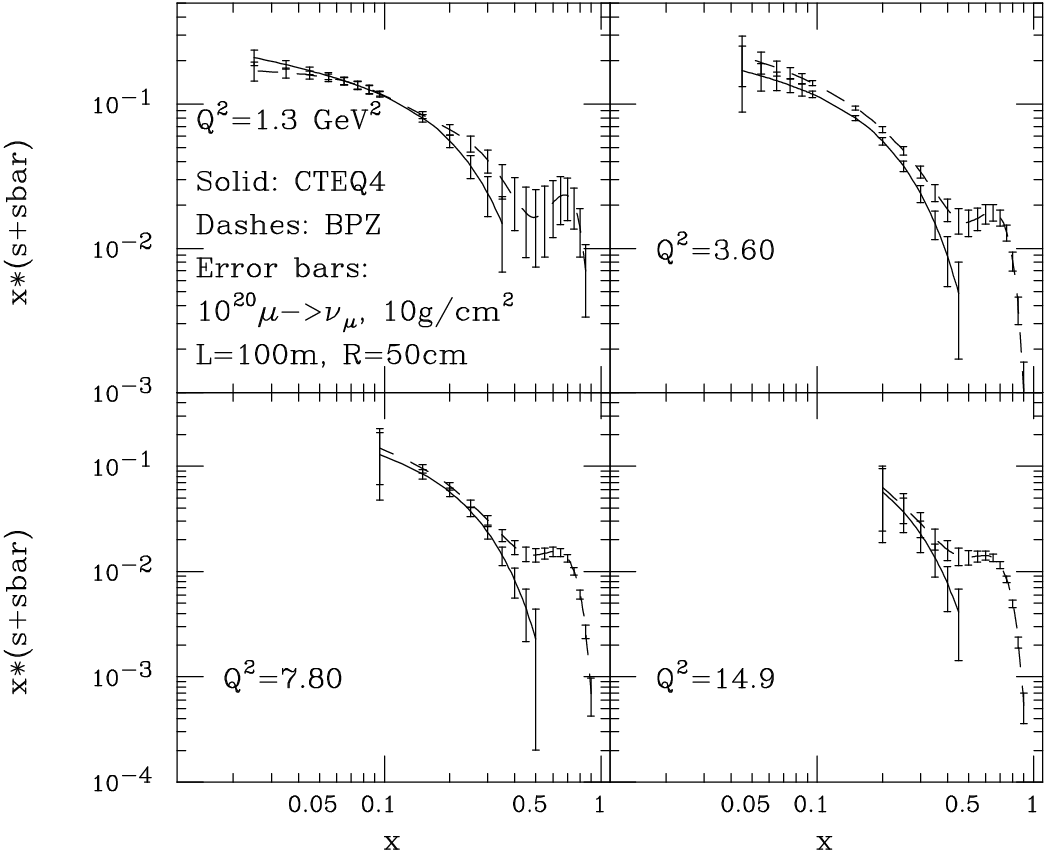} 
\end{center} 
\vskip -1cm 
\caption{Expected errors on the determination of $s(x)+\bar
  s(x)$. Error bars from the \nufact\  are superimposed on the
  current $s(x)+\bar
  s(x)$ fits from CTEQ and from Barone, Pascaud and Zomer (BPZ).
\label{fig:strange}} 
\end{figure} 
The corresponding expressions for a neutron target are obtained
(assuming isospin invariance) by replacing $u\leftrightarrow d$. It is
thus not difficult to see that, constructing appropriate 
linear combinations of the eight independent SFs
$(F_2^\wupm)_{p,n}$ and $(xF_3^\wupm)_{p,n}$,
it is possible to
disentangle $u\pm \bar u$, $d\pm \bar d$ and 
$s\pm\bar s$, provided only that $c\pm\bar c$ can be determined
independently, either theoretically or empirically.
More explicitly, we have
\ba
\label{decomunpol1}
(F_2^{\wup+\wum})_p=(F_2^{\wup+\wum})_n
&=& x(u+\bar u + d+\bar d + s+\bar s + c+\bar c), \\ 
(xF_3^{\wup-\wum})_p-(xF_3^{\wup-\wum})_n 
&=& -2x( u+\bar u-( d + \bar d)), \\
\label{eq:F3ssbsum}
(xF_3^{\wup-\wum})_p+(xF_3^{\wup-\wum})_n 
&=& 2x( s+\bar s-( c + \bar c)), \\
(xF_3^{\wup+\wum})_p=(xF_3^{\wup+\wum})_n
&=& x(u-\bar u + d-\bar d + s-\bar s + c-\bar c),\\ 
(F_2^{\wup-\wum})_p-(F_2^{\wup-\wum})_n 
&=& -2x( u-\bar u-( d - \bar d)), \\
\label{eq:F2ssbdif}
(F_2^{\wup-\wum})_p+(F_2^{\wup-\wum})_n 
&=& 2x( s-\bar s-( c - \bar c)), 
\label{decomunpol2}
\ea
where $F_i^{\wup\pm\wum}\equiv \half (F_i^\wup \pm F_i^{\wum})$.
The first and fourth of these equations are the SFs 
$F_2^{\wup+\wum}$ and $F_3^{\wup+\wum}$ 
normally measured in neutrino scattering (though on heavy targets). 
The second and third equations allow flavour decomposition of the total 
$q+\bar q$ distributions, and the fifth and sixth equations
allow a similar decomposition for the valence distributions. 

\begin{figure}
\begin{center} 
\includegraphics[width=0.8\textwidth,clip]{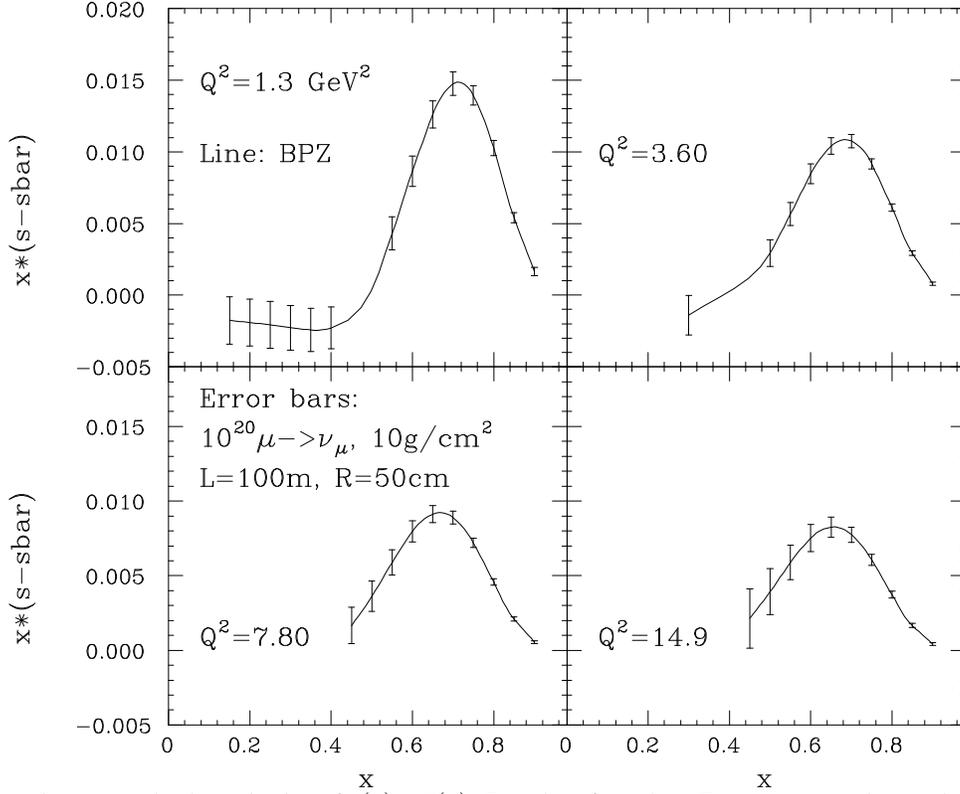} 
\end{center} 
\vskip -1cm 
\caption{Expected errors on the determination of $s(x)-\bar
  s(x)$. Error bars from the \nufact\  are superimposed on the
  current $s(x)-\bar
  s(x)$ fits from Barone, Pascaud and Zomer (BPZ).
\label{fig:ssbdif}} 
\end{figure}

A detailed study of the strange sea is especially important,
since this distribution is very poorly known at present.
In a leading-order analysis one can extract the
individual $s(x)$ and $\bar s(x)$ at the \nufact\ 
using \Eqs{eq:F3ssbsum}{eq:F2ssbdif}.
In order to disentangle the
strange and charm contributions, it is necessary
either to tag charm in the final state, or to assume that the charm
contribution is generated dynamically by perturbative evolution.
Assuming that the charm contribution can be either neglected or
independently determined,
a 2-year running on a nucleon target (1 year each for $\mu^+$ and
$\mu^-$ beams) would result in errors in the
determination of $s(x)+\bar s(x)$ as shown in
Fig.~\ref{fig:strange}. There we compare the size of the predicted
uncertainties with the values of the $s(x)+\bar s(x)$ densities
currently estimated in the analysis of the CTEQ group 
(set CTEQ4,~\cite{Lai:1997mg}), and
in the recent study by Barone, Pascaud and Zomer (BPZ, \cite{Barone:2000yv}).
This last
study, based on a fit to CDHS neutrino data, finds evidence for an
intrinsic strange component of the proton, which results in an enhancement
at large $x$ relative to the CTEQ fits. As shown in the figure,
this evidence could be firmly
established, and its size very accurately determined, using neutrino
factory data. Equation~(\ref{eq:F2ssbdif}) can then be used to estimate the
errors on the determination of the difference $s(x)-\bar s(x)$. These
are shown in Fig.~\ref{fig:ssbdif}, superimposed to the recent fits
by BPZ.

\subsection{NLO extraction of parton densities}
\begin{figure}[ht]
\begin{center}
\includegraphics[width=14cm,height=10cm,clip]{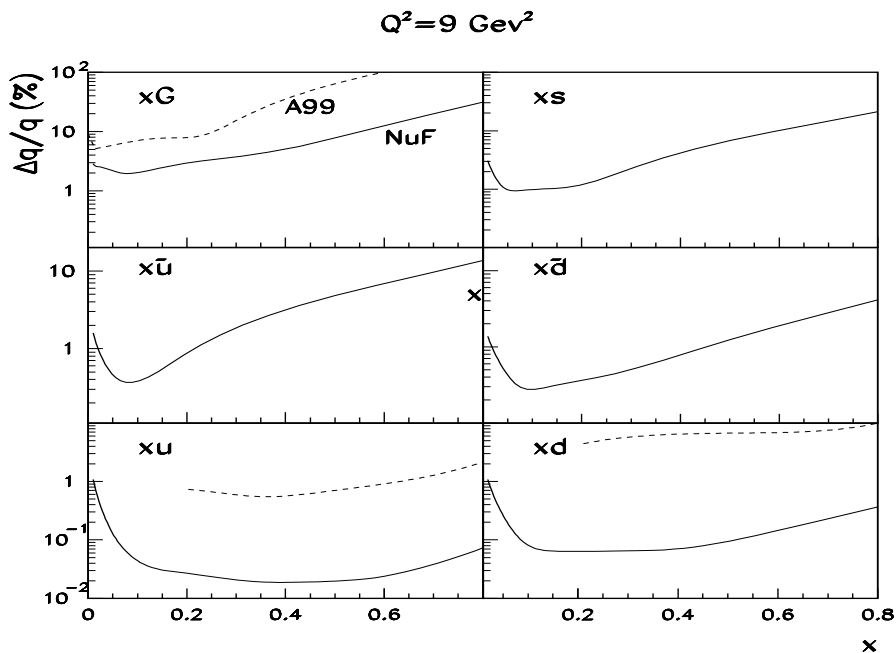} 
\end{center}
\vskip -1cm 
\caption{The relative statistical errors on PDFs (\%) accessible at the 
\nufact~
(full lines). The dashed lines give statistical 
errors on the PDF set A99 obtained from the fit to the existing 
charged leptons DIS data \protect\cite{Alekhin:2001ch}.}
\label{fig:pdfserrors} 
\end{figure}  
\begin{figure}[ht]
\begin{center}
\includegraphics[width=14cm,height=14cm,clip]{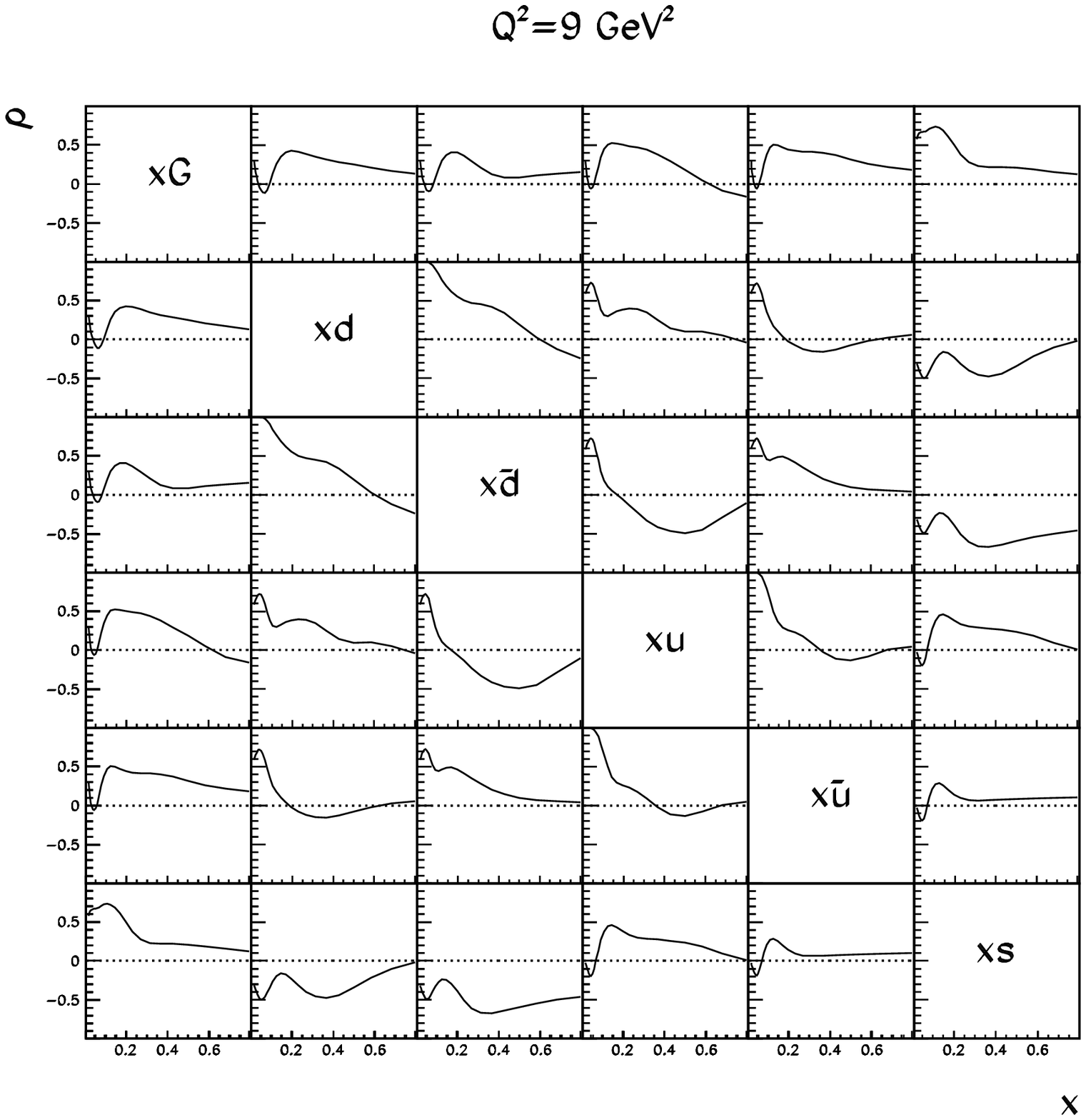} 
\end{center}
\vskip -1cm 
\caption{The $x$-dependence of the correlation coefficients $\rho$
for the PDFs extracted from the analysis of the generated \nufact~ data.
The labels in the diagonal mark the rows and columns of the matrix.}
\label{fig:pdfcorr} 
\end{figure}  
The leading-order study presented in the previous subsection
already provides a useful set of benchmark accuracies that can be
achieved at the \nufact. However, an estimate of the
precision in the extraction of the individual parton distribution functions
(PDFs) requires a full NLO analysis.
In order to estimate the statistical uncertainties in the extraction
of PDFs from \nufact~ data,
we used the errors calculated in the previous section to generate  
8 sets of `fake' data for the SFs $F_{2,3}$ 
for neutrino/antineutrino beams and for hydrogen/deuterium targets.
The central values of the fake data were obtained using 
the PDFs extracted from 
the existing charged leptons DIS data~\cite{Alekhin:2001ch}. 
The SFs $F_{2,3}$ calculated from the 
PDFs, parametrized as 
\ba
&& xd_V(x,Q_0)={A^{V}_d}x^{a_d}(1-x)^{b_d},\quad
xd_S(x,Q_0)={A_d^S}x^{a_{sd}}(1-x)^{b_{sd}} \; ,
\\
&&
xu_{V}(x,Q_0)={A^{V}_u}x^{a_u}(1-x)^{b_u}(1+\gamma_2^{u}x),~
xu_S(x,Q_0)={A_u^S}x^{a_{su}}(1-x)^{b_{su}} \; ,
\\
&&
xG(x,Q_0)=A_Gx^{a_G}(1-x)^{b_G}(1+\gamma_1^{G}\sqrt{x}+\gamma_2^{G}x),\quad
xs(x,Q_0)={A_s}x^{a_{ss}}(1-x)^{b_{ss}} \; ,
\label{eqn:pdfs}
\ea
at $Q_0^2=9~{\rm GeV}^2$ and evolved within the NLO QCD approximation,
were then fitted to these fake data, varying the PDFs parameters.
The form of Eq.~(\ref{eqn:pdfs})
was motivated in Ref.~\cite{Alekhin:2001ch}, 
but, contrary to that analysis, the PDFs parameters 
$a_{su},~A_s,~a_{ss},~b_{ss}$ were also released in the fit, while
the parameters $A^{V}_d,~A^{V}_u,~A_G $ were constrained
by conservation of momentum and fermionic numbers, as 
in Ref.\cite{Alekhin:2001ch}. 
The indices $V$ and $S$ are used to denote
valence  and sea distributions, respectively; the functions 
$u,d,s,G$ give $u$-, $d$-, $s$-quarks, and gluon distributions;
and $s=\overline s$ is assumed at this stage of the analysis.

The statistical errors on PDFs obtained in the fit to fake data are 
given in Fig.~\ref{fig:pdfserrors}. For comparison, we present
in the same figure
the errors on the PDFs obtained from the analysis of 
Ref.~\cite{Alekhin:2001ch}.
One can see that the precision in the determination of 
the gluon distribution has improved by 
about an order of magnitude, and even more
for the valence $u$- and $d$-quark distributions. 
This improvement may be extremely helpful, for example in
expanding the capabilities of the LHC in searches for 
new phenomena characterized by particles of large mass.
The precision of the determination of the sea quark distributions
attainable at the \nufact~ cannot be directly compared 
with the corresponding errors on the PDFs given in 
Ref.~\cite{Alekhin:2001ch}; there, several 
parameters describing the sea quark distributions were 
fixed at reasonable values,
since available data do not allow their independent 
extraction. In particular, the strange sea was assumed 
equal to about a  
half of the non-strange sea, while the behavior of the non-strange sea-quark
distribution at small $x$ was assumed to be universal. 
Thanks to the avaliability of
eight independent combinations of parton distributions,
a complete separation 
of individual flavor distributions is now possible 
without additional constraints. 
This is crucial for precise studies of the flavour
content of the nucleon. 
The correlation matrix for the PDFs extracted from the QCD 
fit to the fake \nufact~ data 
is given in Fig.~\ref{fig:pdfcorr}. Indeed, one can see that  
in general the absolute values of the 
correlation coefficients do not exceed 0.3 in the whole range of $x$.

\begin{figure}[ht]
\begin{center}
\includegraphics[width=14cm,height=10cm,clip]{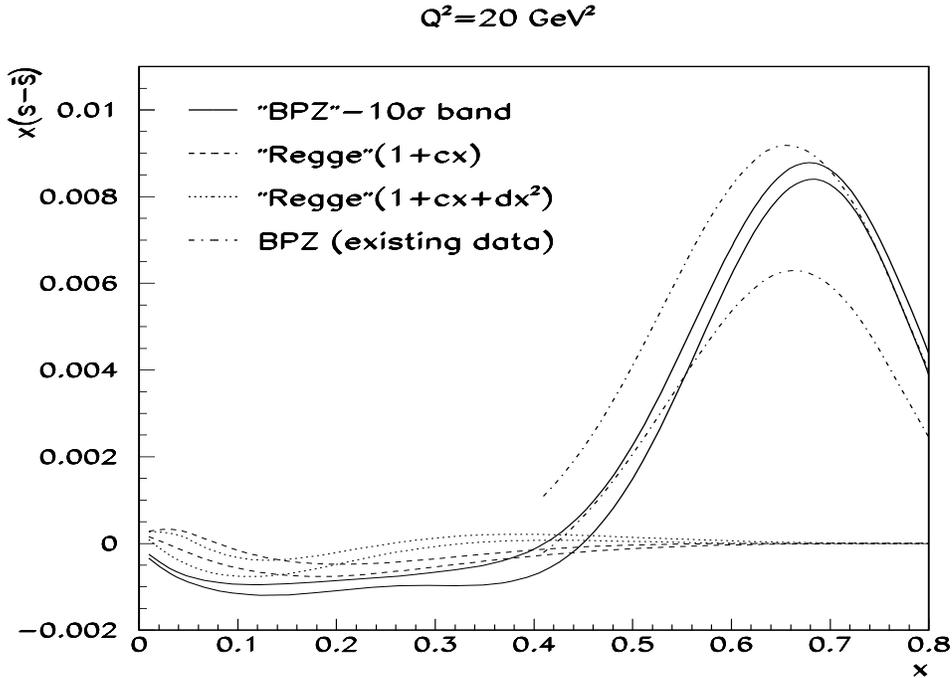} 
\end{center}
\vskip -1cm 
\caption{The error bands for $x(s-\overline s)$ difference accessible 
at the \nufact~ for the various forms used for the 
parameterization of this difference (full line: the BPZ-like form, 
dashed lines: the Regge-like form with linear polynomial factor, 
dotted line: the Regge-like form with quadratic polynomial factor).
The $1\sigma$ bands are given for the the Regge-like parameterizations
and $10\sigma$ bands for the BPZ-like one.}
\label{fig:saserrors} 
\end{figure}

We now turn to a study
of the potential of the \nufact~ data to determine the asymmetry of
the strange sea, which is not accessible in neutral-current (NC)
DIS experiments. In this analysis,
the strange sea was chosen to be of the form 
\begin{equation}
xs(x,Q_0)={A_s}x^{a_{ss}}(1-x)^{b_{ss}}(1+c_{ss}x^{d_{ss}}),~~
x\overline s(x,Q_0)={A_s}x^{a_{ss}}(1-x)^{b_{sas}}(1+c_{sas}x^{d_{sas}}),
\label{eqn:BPZpdf}
\end{equation}
which is motivated by the results of Ref.~\cite{Barone:2000yv}.
We now assume that the $s-\bar{s}$ difference is as given in 
Ref.~\cite{Barone:2000yv}, by choosing similar values of the parameters,
namely
\begin{equation}
A_s=0.06,\quad b_{ss}=5.6,\quad b_{sas}=5.4,\quad c_{ss}=11000,\quad d_{ss}=12,
\quad d_{sas}=7.4.
\label{eqn:BPZpar}
\end{equation}
The parameter $a_{ss}$ was set equal to $a_{sd}$, and the parameter 
$c_{sas}$ was calculated from the constraint
\begin{equation}
\int_0^1 dx\, [s(x)-\overline s(x)]=0.
\label{eqn:sasconst}
\end{equation}
We cannot simply combine these $s$ and $\bar s$ distributions with those of 
Ref.~\cite{Alekhin:2001ch}, because 
the $xs$ distribution of Eq.~(\ref{eqn:BPZpdf})
at large $x$ is comparable to the contribution from valence quarks,  
and correlated to it.
Therefore, we repeated the fit of Ref.~\cite{Alekhin:2001ch}
with the $s$ and $\overline s$ distributions
of Eqs.~(\ref{eqn:BPZpdf},\ref{eqn:BPZpar}).
We then used the results of this fit to generate a new set of 
fake data, and repeated a QCD fit to these fake data using
the PDFs of Eq.~(\ref{eqn:pdfs}),
but with the $s$ and $\bar s$ distributions of
Eq.~(\ref{eqn:BPZpdf}).
The errors on $x(s-\overline s)$ obtained in this way are given in 
Fig.~\ref{fig:saserrors}. For comparison, we recall that
the error on $x(s-\overline s)$
obtained in the BPZ analysis are ${\cal O}(0.002)$ at $x\sim 0.7$
(see Fig.~10); this means that an improvement of more than an
order of magnitude in the precision may be achieved at the \nufact~.

Alternatively, one can choose different forms for the $x(s-\overline s)$ 
difference; for example, it may be constructed using Regge phenomenology 
considerations. According to this approach, the behavior of non-singlet 
parton distributions at small $x$ is governed by meson trajectories,
and is generally $\sim \sqrt{x}$. We constructed two variants of the 
parameterization of $x(s-\overline s)$
based on these arguments, namely
\begin{equation}
  x(s-\overline s)(x,Q_0) = 
   A_{\Delta s}x^{a_{\Delta s}}(1-x)^{b_{\Delta s}}(1+c_{\Delta s}x)
\end{equation}
and 
\begin{equation}
\label{eq:dsSA}
x(s-\overline s)(x,Q_0)=A_{\Delta s}x^{a_{\Delta s}}(1-x)^{b_{\Delta s}}
(1+c_{\Delta s}x+d_{\Delta s}x^2).
\end{equation}
The starting PDF set of Eq.~\ref{eqn:pdfs}
was then modified by substituting
$xs \rightarrow xs + x(s-\overline s)/2$ and 
$x\overline s \rightarrow x\overline s - x(s-\overline s)/2$.
The initial values of the parameters used to generate fake data
were chosen as $a_{\Delta s}=0.5$, $A_{\Delta s}=0.02$ and $b_{\Delta s}=7.5$.
The initial value of the parameter $d_{\Delta s}$ of \equ{eq:dsSA} was set
equal to 100 and, in both cases, the parameter $c_{\Delta s}$ was defined 
from the constraint of Eq.~(\ref{eqn:sasconst}).

The errors on $x(s-\overline s)$ obtained 
using a Regge-like form of $x(s-\overline s)$
are given in Fig.~\ref{fig:saserrors}.
One can see that they are quite different from those obtained by
assuming the BPZ form for $x(s-\bar{s})$. 
The main difference between Regge-like and 
BPZ forms, is that the strange sea for the latter is very 
large at high $x$, even larger than the $d$-quark sea.
As a result, the precision of the strange-sea determination 
for the BPZ set is comparable to the precision in the determination
of the valence quark, namely ${\cal O}(0.1\%)$. Since
$s\gg\bar{s}$ in the BPZ fit, 
$\Delta(s-\bar s)/(s-\bar s) \approx \Delta s/s$.
Thus, the high precision obtained using the BPZ form
simply originates from the fact that $s$ is very large at high $x$.

\begin{figure}
\begin{center}
\includegraphics[width=14cm,height=10cm,clip]{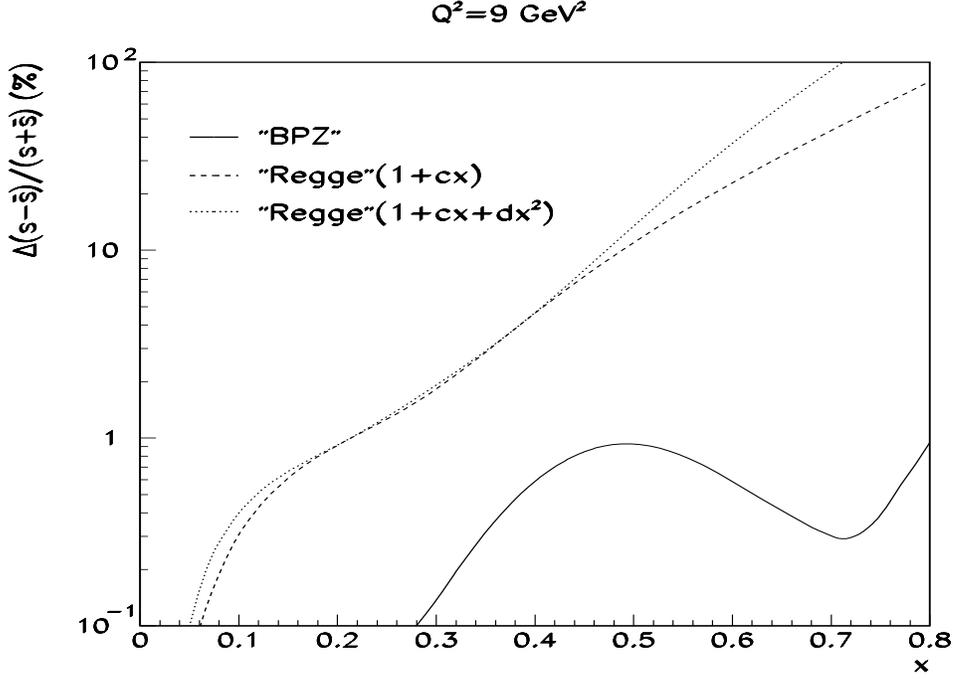} 
\end{center}
\vskip -1cm 
\caption{Accuracy in the determination of $(s-\overline s)$ relative to
the total amount of strange sea, for the different strange
parameterizations.}
\label{fig:sasr} 
\end{figure}

The precision of the determination of $x(s-\overline s)$ accessible at
the \nufact\ will however always be at the level of 1\%, or
better, in the
region of its maximum, regardless of the functional form used to
parameterize this difference.
The accuracy relative to
the total amount of strange sea is given, for the different strange
parameterizations, in fig.~\ref{fig:sasr}.


\section{POLARIZED STRUCTURE FUNCTIONS}
\subsection{Formalism}
\label{sec:psfform}
Polarized SFs may be defined in analogy to the 
unpolarized ones through asymmetries in the polarized 
cross sections (see \cite{bt} for a recent review).
The polarized cross-section difference (with a proton helicity
$\lambda_p=\pm 1$),
\be\label{deltadef}
\Delta\sigma\equiv
\sigma(\lambda_p=-1)-\sigma(\lambda_p=+1),
\ee
 is given by
\ba
{d^2\Delta\sigma^{\lambda_\ell}(x,y,Q^2)\over dx dy} &=&
{G^2_F  \over \pi (1+Q^2/m_W^2)^2}{Q^2\over xy}
\Biggl\{
  \left[
     -\lambda_\ell\, y (2-y)  x g_1- (1-y) g_4- y^2 x g_5
  \right]  \nonumber \\
\label{xsecas}
&\quad& 
+2xy{m^2\over Q^2}
  \left[\lambda_\ell x^2y^2 g_1 +\lambda_\ell  2 x^2 y g_2
       +\left(1-y-x^2y^2\frac{m^2}{Q^2}\right) xg_3
  \right. \\
 &&
\left. \qquad\quad
   -x\left(1-\frac{3}{2}y-x^2 y^2 \frac{m^2}{Q^2}\right)
g_4-x^2y^2 g_5 \right] \Biggr\} \; , \nonumber
\ea
where $\lambda_{\ell}$ is the lepton helicity.
With these definitions~\cite{Forte:2001ph}
\footnote{There are many variants 
in the literature: see \cite{bk} for a compilation. In particular,
the conventions used here
are the same as those used in ref.~\cite{Ball:2000qd}, 
except for the signs of $g_4$ and $g_5$.},
$g_2$ and $g_3$ drop out in the high energy limit $E\gg m$,
and we are left with an expression of the same form as the 
unpolarized decomposition (\ref{eq:sfn}), but with $F_1\to -g_5$, 
$F_2\to -g_4$ and $F_3\to \; 2g_1$. Thus we again have only three 
partonic SFs for each $\nu$ and $\bar{\nu}$:
\be \label{eq:polasy}
{d^2\Delta\sigma^{\lambda_\ell}(x,y,Q^2)\over dx dy}
= {G^2_F  \over \pi (1+Q^2/m_W^2)^2}{Q^2\over xy}
  \left[
     -\lambda_\ell\, y (2-y)  x g_1- (1-y) g_4- y^2 x g_5
  \right]
\ee
The two remaining SFs $g_2$ and 
$g_3$ have no simple partonic interpretation
and are contaminated by twist-3 contributions: their twist-2
components are fixed by the Wandzura--Wilczek relation
(giving $g_2$ in terms of $g_1$) and a similar relation
gives $g_3$ in terms of $g_4$. They are
determined by measuring 
asymmetries with a transversely polarized target.
In the parton model, $g_4$ and $g_5$ are related by an 
analogue~\cite{Dicus} of the Callan--Gross relation: 
$g_4 = 2x g_5(1+O(\alpha_s))$.  Even though this relation is violated
beyond leading order, the SFs $g_4$ and $g_5$ 
still measure the same combination of parton distributions, albeit
with different coefficient functions. Therefore at leading twist
there are only two
independent polarized SFs, conventionally taken to be
$g_1$ and $g_5$.

The flavour decomposition of the SFs $g_1$   and 
$g_5$ may be expressed in terms of 
parton densities~\cite{flavcomp} as
\ba
g_1^\wup &=&  \Delta\bar u + \Delta d + \Delta s 
+ \Delta\bar c,
\qquad\qquad\quad g_1^{\wum} = \Delta u +\Delta\bar d + 
\Delta\bar s + \Delta c, \\
g_5^\wup &=& \Delta\bar u - \Delta d -\Delta s +\Delta \bar c,
\qquad\qquad g_5^{\wum} = -\Delta u + \Delta \bar d 
+ \Delta \bar s - \Delta c \; ,
\label{polpart}
\ea
in precise analogy with the unpolarized case.
Again, by constructing appropriate 
linear combinations of all eight independent SFs
(conventionally taken as $(g_1^\wupm)_{p,n}$ and 
$(g_5^\wump)_{p,n}$), obtained by longitudinally polarized $\nu$ 
and $\bar\nu$ scattering on proton and neutron 
(or deuteron) targets, it is possible 
to separately disentangle $\Delta u\pm \Delta\bar u$,
$\Delta d\pm \Delta\bar d$ and $(\Delta s\pm\Delta\bar s)$ just 
as in Eqs.(\ref{decomunpol1})--(\ref{decomunpol2}),
provided only that $\Delta c\pm \Delta \bar{c}$ can be determined.

Some combinations of polarized SFs are of particular 
interest. For example, writing 
$g_i^{\wup\pm\wum}\equiv g_i^\wup\pm g_i^{\wum}$, the 
first moment of
\be
\Delta\Sigma=(g_1^{\wup+\wum})_p=(g_1^{\wup+\wum})_n
=\Delta u+\Delta\bar u +\Delta d+\Delta\bar d 
+\Delta s+\Delta\bar s +\Delta c+\Delta\bar c
\label{eq:a0} 
\ee
is the singlet axial charge $a_0$. This is a much more direct measurement 
than the traditional one through electron--proton or deuteron DIS,
since in the latter case one must first subtract the octet charge 
$a_8$ which is then only determined indirectly through hyperon decays.
Thus in $\nu$-DIS one would 
have a direct check on the anomalous suppression of $a_0$. Similarly,
first moments of
\ba
\label{eq:a3}
&& 6\left[(g_1^{\gamma^*})_p-(g_1^{\gamma^*})_n\right]
=(g_5^{\wup-\wum})_p-(g_5^{\wup-\wum})_n 
= \Delta u+\Delta\bar u -(\Delta d + \Delta\bar d) = \Delta\,q_3\\
&& 6\left[(g_1^{\gamma^*})_p+(g_1^{\gamma^*})_n\right]-
\frac{5}{3}\left(g_1^{W^++W^-}\right)_p
=(g_5^{\wup-\wum})_p+(g_5^{\wup-\wum})_n 
= -(\Delta s+\Delta\bar s)+(\Delta c + \Delta\bar c) \; ,
\nonumber\\
\label{decompolb} 
\ea
give direct measurements of the axial charge $a_3$
and of the contribution of strange 
quarks to the nucleon spin, as would the 
tagging of charm in the final state. 
The first moment of the combination
\be
\Delta\,q_8=\Delta u+\Delta\bar u +\Delta d+\Delta\bar d 
-2(\Delta s+\Delta\bar s)
\label{eq:a8}
\ee
is the octet axial charge $a_8$, currently determined by hyperon decays
using SU(3) symmetry, thereby
allowing a test of SU(3) symmetry violation, and specifically 
a distinction  between different models for it~\cite{sutr}.

Flipping the signs, we can also determine the contribution of 
valence quarks to the spin, since
\ba
(g_5^{\wup+\wum})_p=(g_5^{\wup+\wum})_n
&=& -\Delta u+\Delta\bar u -\Delta d + \Delta\bar d 
-\Delta s+\Delta\bar s -\Delta c+\Delta\bar c,\nonumber\\ 
(g_1^{\wup-\wum})_p-(g_1^{\wup-\wum})_n 
&=& -2(\Delta u-\Delta\bar u-(\Delta d - \Delta\bar d)),\\
(g_1^{\wup-\wum})_p+(g_1^{\wup-\wum})_n 
&=&2(\Delta s-\Delta\bar s -(\Delta c - \Delta\bar c)),\nonumber
\label{decompolc} 
\ea  
so one could even check for intrinsic strange polarization 
$\Delta s-\Delta\bar s$. None of these valence polarizations can be 
cleanly measured in current polarization experiments.

It should be pointed out that the flavour separations shown above receive
important contributions from NLO corrections, most notably from
contributions from the polarized gluon density. In particular, $g_1$
is given at NLO by
\be \label{gnlo}
g^{{\rm NLO}}_1(x,Q^2) =\Delta C_q\otimes
g^{{\rm LO}}_1+2[n_f/2]\,\Delta C_g\otimes\Delta g \; ,
\ee
where the $\Delta C_i$ are appropriate coefficient functions.
Also, beyond leading order the relation between parton distributions  and
SFs is ambiguous because of the factorization scheme ambiguity.
This problem is particularly relevant in the case of Eq.~(\ref{gnlo}),
because, as is by now well
known~\cite{bt}, the scheme dependence of the first moment of the singlet polarized quark
distributions is unsuppressed as $\as\to0$, due to the fact that at
leading order the first moment $\Delta g(1)$ of the gluon distribution
evolves as $1/\as$.  A 
consequence of this is that it is possible to choose the factorization
scheme in such a way that the first moment of the singlet quark
distribution is scale independent (to all perturbative orders), even
though this is not the case in the \MS\ scheme. This is the case
e.g. in the so-called Adler-Bardeen (AB)
scheme~\cite{bfr}. The first moment of any quark distribution in the
\MS\ and AB schemes are related by
\be\label{abtoms}
{\Delta q_i^{\rm\overline{MS}}
(1,Q^2)=\Delta q_i^{\rm AB}(1)- {\as\over 4\pi} \Delta g(1,Q^2),}
\ee
where $\Delta q(1,Q^2)=\int_0^1 dx \Delta q(x,Q^2)$.
Theoretical motivations for this choice will be
discussed in Sect.~\ref{sec:pspin} below. 

It follows that care should be taken when comparing AB and \MS\ scheme
polarized parton distributions, since the quark distributions will
differ by an O(1) gluon contribution. Furthermore, in a generic
scheme, a meaningful determination of polarized parton distributions
requires at least the inclusion of NLO corrections.  The full NLO
analysis of parton distributions in neutrino DIS, using the known
anomalous dimensions \cite{NLO} and coefficient functions \cite{dfs},
is presented in Ref.~\cite{Forte:2001ph}.

\subsection{Positivity bounds on polarized densities}
\begin{figure}
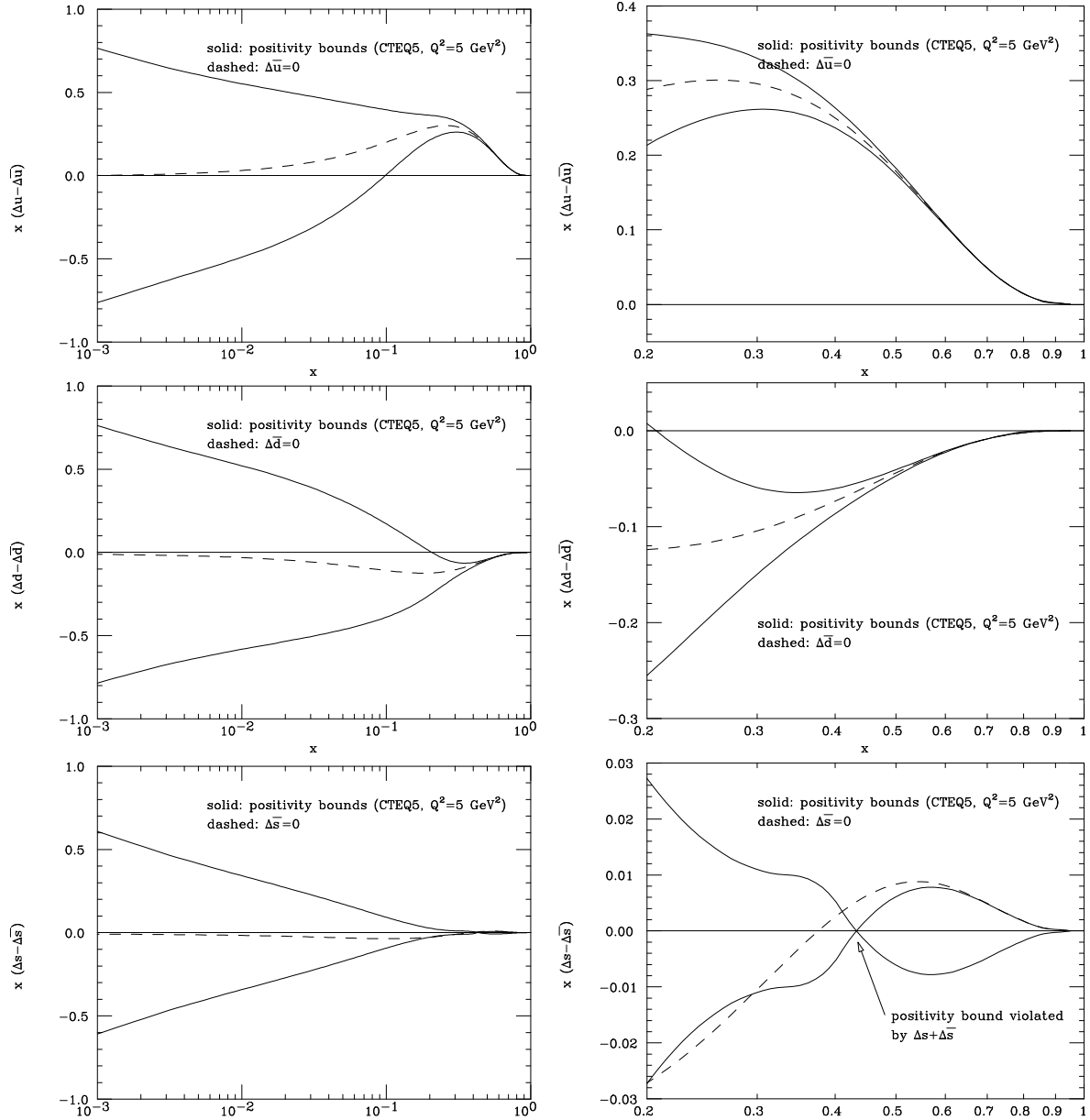
 
\begin{center} 
\includegraphics[width=0.48\textwidth,clip]{dupos1.eps} \hfil
\includegraphics[width=0.48\textwidth,clip]{dupos2.eps} \\
\includegraphics[width=0.48\textwidth,clip]{ddpos1.eps} \hfil
\includegraphics[width=0.48\textwidth,clip]{ddpos2.eps} \\
\includegraphics[width=0.48\textwidth,clip]{dspos1.eps} \hfil
\includegraphics[width=0.48\textwidth,clip]{dspos2.eps} 
\end{center} 
\vskip -1cm 
\caption{Positivity bounds on $x(\Delta q-\Delta\bar q)$ ($q=u,d,s$), compared
  with the $\Delta\bar q = 0$ and  $\Delta q=\Delta\bar q$ hypotheses.
  Unpolarized densities taken from the CTEQ5
  parametrization. 
\label{fig:dqpos}} 
\end{figure}

Cuurently available NC 
polarized DIS data can be used, in conjunction with 
specific hypotheses on the form of $\Delta q-\Delta\bar q$, to
explore potential scenarios to be probed with polarized neutrino
scattering at the \nufact.
To start with, it is interesting
to study the constraints set by positivity~\cite{pos}
 on the values of $\Delta
q-\Delta\bar q$ for individual flavours. At leading order in QCD, the
positivity of both left- and right-handed quark densities implies the
following obvious relations: 
\be \label{eq:dqpos}
\vert \Delta q(x) \vert < q(x) \; , 
\quad \vert \Delta \bar q(x) \vert < \bar q(x)  \; .
\ee 
These constraints are shown in Fig.~\ref{fig:dqpos}
for $x(\Delta q-\Delta\bar q)$, with $q=u,d$. The
allowed region is confined between the two continuous lines.  Use was
made of the most recent polarized fits from
ABFR~\cite{Altarelli:1998nb} and of the  CTEQ5~\cite{Lai:2000wy} 
unpolarized densities. The figures show that, while the assumption
$\Delta\bar q = 0$ is consistent with the positivity bounds, the
hypothesis $\Delta q=\Delta\bar q$ badly violates them as soon as
$x\gsim 0.1$--0.2. As a result we conclude that 
$\vert \Delta q \vert \gg
\vert \Delta\bar q \vert$ for $x \gsim 0.1$, which is reasonable in
view of \equ{eq:dqpos} and of the fact that $q \gg \bar q $
in this region of $x$.

\begin{figure} 
\begin{center} 
\includegraphics[width=0.8\textwidth,clip]{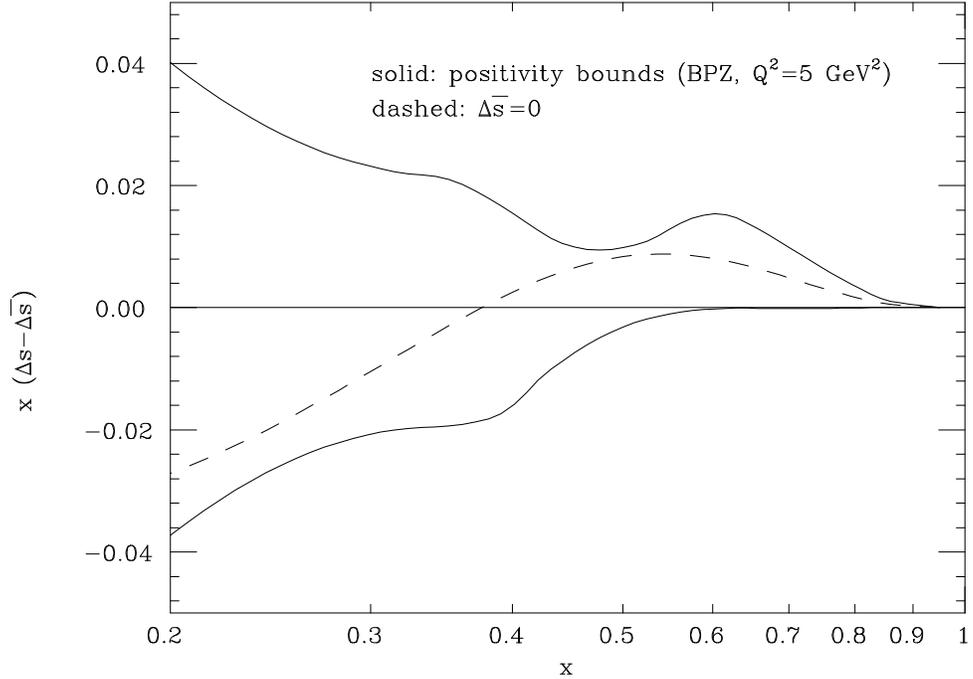} 
\end{center} 
\vskip -1cm 
\caption{Positivity bounds on $x(\Delta s-\Delta\bar s)$, compared
  with the $\Delta\bar s = 0$ and  $\Delta s=\Delta\bar s$
  hypotheses. $s(x)$ unpolarized densities taken from the BPZ
  parametrization. 
\label{fig:dspos3}} 
\end{figure}

In the case of the strange quark, it turns out that the combination
$\Delta s+\Delta \bar s$ from the ABFR fit already violates the
positivity bound obtained using the unpolarized strange distributions
from CTEQ5. This bound is instead satisfied if the unpolarized
strange distribution from the BPZ fit is used. In such a case, the bound
turns out to be satisfied for both $\Delta\bar q = 0$ and $\Delta
q=\Delta\bar q$ (see Fig.~\ref{fig:dspos3}). The NLO
corrections to the positivity bounds turn out to be negligible
(see Ref.~\cite{Forte:2001ph} for a detailed discussion).

\begin{figure}
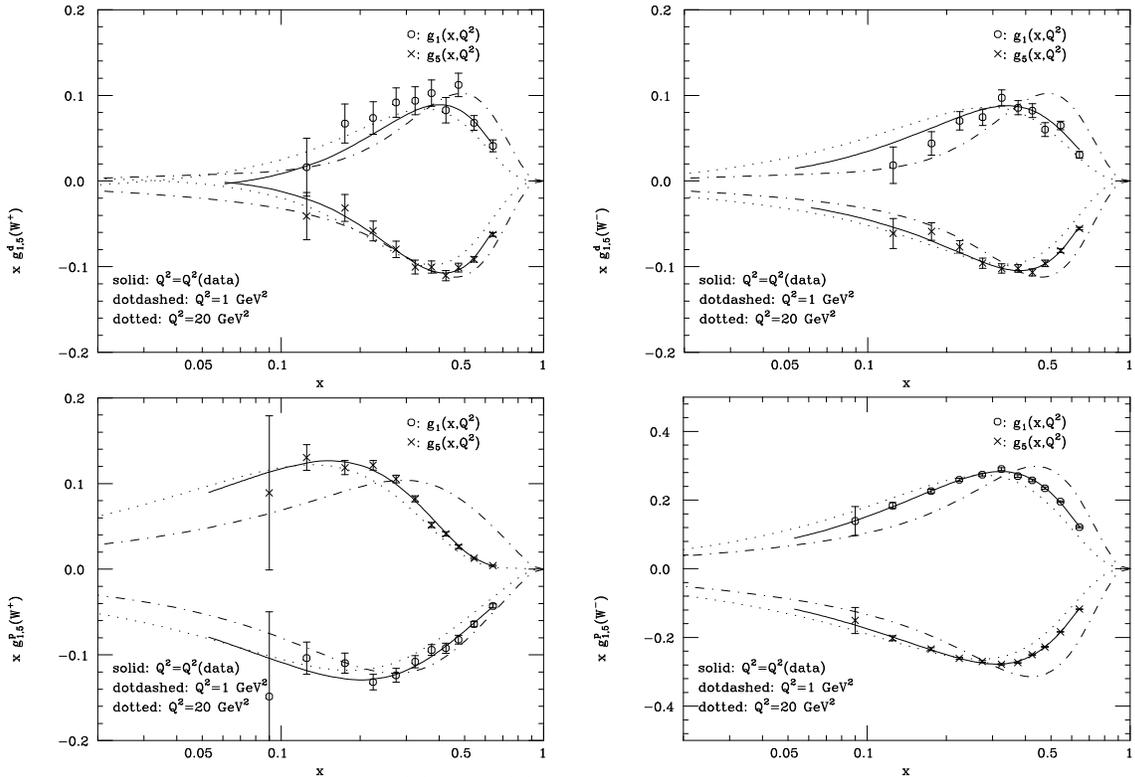
 
\begin{center} 
\includegraphics[width=0.45\textwidth,clip]{dwp.eps} \hfil
\includegraphics[width=0.45\textwidth,clip]{dwm.eps} \\
\includegraphics[width=0.45\textwidth,clip]{pwp.eps} \hfil
\includegraphics[width=0.45\textwidth,clip]{pwm.eps} \\
\end{center} 
\vskip -1cm 
\caption{Accuracy in the determination of $g_1(x)$ and $g_5(x)$ for
  deutoren (above) and proton (below) targets, in $\nu$ (left) and
  $\bar\nu$ (right) scattering
\label{fig:nucleon-wminus}} 
\end{figure} 

\subsection{Theoretical scenarios and the spin of the proton}
\label{sec:pspin}
One of the main reasons of interest in polarized quark distributions
is the unexpected smallness of the nucleon axial charge, which has been
determined in the first generation of polarized DIS experiments. A
clarification of the physics behind this requires a determination of
the detailed polarized parton content of the nucleon. It is useful to sketch
here the various scenarios to describe the polarized content of the
nucleon, which are
representative of possible theoretical alternatives and could
be tested in future experiments.

Firstly, it should be noticed that even though current data give a
value of the axial charge which is compatible with zero, they cannot
exclude a value as large as $a_0(10$~GeV$^2)=0.3$~\cite{cracovia}. 
Also, the
current value is obtained by using information from hyperon
$\beta$ decays  and
SU(3) symmetry. Clearly, the theoretical implications of an exact zero
are quite different from those of a value that is just smaller than
expected in quark models. It is thus important to have a direct
determination of the axial charge. If a small value 
is confirmed, it could be understood as the consequence of a
cancellation~\cite{anom} between a large value of the scale-independent
first moment of the quark 
(discussed at the end of Sect.~\ref{sec:psfform}) 
and a large first moment of the gluon.
In this (`anomaly')
scenario the up, down and strange polarized distributions in the
AB-scheme are close to their expected quark-model values, so in
particular the strange distribution is much smaller than the up and
down distributions. In Ref.~\cite{topsup}, this cancellation of quark and
gluon components has been derived from the topological properties of
the QCD vacuum (and thus further predicted to be a universal
property of all hadrons).

If instead the polarized gluon distribution is small, the smallness of
the singlet axial charge can only be explained with a large and
negative strange distribution. In this case, the scale-independent
first moment of the singlet quark distribution is also small. This
scale-independent suppression of the axial charge might be explained
by invoking non-perturbative mechanisms based on an instanton-like
vacuum configuration~\cite{inst}. In this `instanton' scenario the strange
polarized distribution is large and equal to the antistrange
distribution, since gluon-induced contributions must come in
quark-antiquark pairs.

Another scenario is possible, where the smallness of the singlet axial
charge is due to intrinsic strangeness, \ie\ the C-even strange
combination is large, but the sizes of $\Delta s$ and $\Delta \bar s$
differ significantly from each other. Specifically, it has been
suggested that while the strange distribution (and specifically its
first moment) is large, the antistrange distribution is much smaller,
and does not significantly contribute to the nucleon axial
charge~\cite{ints}. This way of understanding the nucleon spin structure is
compatible with Skyrme models of the nucleon, and thus we will refer
to this as a `skyrmion' scenario~\cite{strange}.

Therefore, the main qualitative issues that are relevant to the
nucleon spin structure are to assess how small the axial charge is, to
determine whether the polarized gluon distribution is large, and then
whether the strange polarized distribution is large, and whether the
strange polarized quark and antiquark distributions are equal to each
other or not.  More detailed scenarios might then be considered, once
the individual quark and antiquark distributions have all been
accurately determined. For instance, while the up and down antiquark
distributions are small, they need not be zero, and in fact they could
be different from one another~\cite{glumod}, just like their unpolarized
counterparts appear to be. Investigating these issues could shed
further light on the detailed structure of polarized nucleons.

\subsection{Statistical errors on polarized densities at the \nufact}
The fit to the $y$ distributions
at fixed $x$ and $Q^2$ for a fully polarized target gives the
value of the combinations $F_2\pm 2xg_5$ and $F_3\pm 2g_1$.
Polarization asymmetries are extracted by combining
data sets obtained using targets with different orientations of the
polarization. The statistical accuracies with which the combinations
can be performed depend on the statistical content of each individual
data set.  Since the polarization asymmetries are small with respect to
the unpolarized cross sections, the {\it absolute} statistical
uncertainties on the extraction of polarized SFs will
have a very mild dependence on the value of the polarized SFs
 themselves; they will be mostly determined by the value of
the unpolarized SFs (which to first approximation fix
the overall event rate), and by the polarization properties of the
target.

Therefore, for simplicity, we directly use the expected
statistical errors $\sigma_{F_2,F_3}$ 
obtained in Section~\ref{sec:unpolsfres} for
the extraction of $F_2$ and
$F_3$ form unpolarized targets, assuming in this case a target
thickness of 10g/cm$^2$. We then  relate these to the errors on the
polarized cross sections by using the following relation given
in~\cite{Ball:2000qd}: 
\be
\sigma_{g_i} = F^{tgt}_{\nu ,\bar\nu}
\sqrt{2} \; \alpha_{ij} \; \frac{\sigma_{F_j}}{2} \; ,
\ee
where
$\alpha_{ij}=1$ for $(i,j)=(1,3)$ and $\alpha_{ij}=1/x$ for
$(i,j)=(5,2)$, and where $F^{tgt}_{\nu ,\bar\nu}$ is a correction
factor (always larger than 1) that accounts for the ratio of the
target densities to $H_2$ or $D_2$, for the incomplete target
polarization, and for the dilution factor of the target, namely the
$\nu$ (or $\bar\nu$) cross-section weighted ratio of the polarized
nucleon to total nucleon content of the target.  The factor of
$\sqrt{2}$ in the numerator reflects the need to subtract the
measurements with opposite target polarization.

The values
of the uncertainties in the determination of the eight
CC SFs ($g_1$ and $g_5$ with the two
available beams and targets) are assigned at the cross-section
weighted bin centres. To obtain the absolute errors on the
SFs for proton and deuterium
targets we use the p-butanol and D-butanol
target~\cite{Adams:1999qy}
correction factors given in Ref.~\cite{Ball:2000qd}, namely
$F^{p}_{\nu}=2.6$, $F^{p}_{\bar\nu}=1.6$ and $F^{D}_{\nu,\bar\nu}=4.4$
(for a more complete discussion of polarized targets and their
complementary properties, see~\cite{Ball:2000qd} and references therein).  We
have assumed a luminosity of $10^{20}$ muons decaying in the straight
section of the muon ring for each charge, for each target, and for
each polarization. Assuming that only one polarization and one target
can run at the same time, this means eight years of run. While the
number of muons may not be dramatically increased, the integration
time can be reduced by a large factor if the target thickness can be
increased over the conservative 10~g/cm$^2$ assumed here, or if
different targets can be run simultaneously. 

\subsection{Structure-function fits}
We can now study how CC DIS data may be used to
determine the polarized parton content of the nucleon.
This study has been performed in Ref.~\cite{Forte:2001ph}, 
which we summarize here.
First,  we
need to make assumptions on the expected flavour content of the
nucleon, implementing the theoretical scenarios summarized in
Sect.~\ref{sec:pspin}.  

 We define two sets of C-even polarized densities consistent with
existing NC data (one set corresponding to a polarized
gluon consistent in size with the `anomaly' scenario, the second set
corresponding to a vanishing gluon polarization at $Q_0=1$~GeV), and
then to define three possible inputs for the C-odd densities,
consistent with the expectations of the three scenarios. We shall then
generate data according to these three alternatives, with the errors
defined as in the previous section, and study the
accuracy with which their parameters can be measured at the \nufact.

We start by describing the parametrization of the C-even densities,
for which we adopt  
the type-A fit of Ref.~\cite{cracovia}, 
defined as follows. The  quark distributions
$\Delta\Sigma^+$ (\ref{eq:a0}), $\Delta q_3$ (\ref{eq:a3}) and $\Delta
q_8$ (\ref{eq:a8}), 
and the polarized gluon distribution $\Delta g$ at the initial scale
$Q_0^2=1$~GeV$^2$ are all taken to be of the form
\be
\label{fitashape}
\Delta f(x,Q_0^2)= {\cal N}_f\eta_f
x^{\alpha_f}(1-x)^{\beta_f}(1+\gamma_f x^{\delta_f}) \; , 
\ee 
where the factor ${\cal N}_f$ is such that the parameter $\eta_f$ is
the first moment of $\Delta f$ at the initial scale. The non-singlet
quark distributions $\Delta q_3$ and $\Delta q_8$ are assumed to have
the same $x$ dependence, while the parameter $\eta_8$, corresponding
to the first moment of $\Delta q_8$, is fixed to the value
$\eta_8=0.579$ from octet baryon decay rates using SU(3) symmetry.
Furthermore, $\gamma_\Sigma=\gamma_g=10$, $\delta_3=\delta_8=0.75$,
$\delta_\Sigma=\delta_g=1$. All other parameters in
\equ{fitashape} are determined by the fitting procedure.  Recent
data~\cite{newdata} from the E155 collaboration, and the final data
set from the SMC collaboration have been added to those used in
Ref.~\cite{cracovia}, leading to a total of 176 NC data
points.

The best-fit values of the first moments of the C-even parton
distributions at the initial low scale of $Q_0^2=1$~GeV$^2$ thus obtained
are listed in the first column of Table~\ref{tab:fmr}, together with the errors
from the fitting procedure. The subsequent three rows of the table give the
values of the first moments of $\Delta q(x,Q^2)+\Delta{\bar q}(x,Q^2)$
at $Q^2=1$~GeV$^2$ for up, down, and strange, obtained by combining the
singlet and non-singlet quark first moments above.  Finally, we give in
the last row the value of the singlet axial charge $a_0$ at the scale
$Q^2=10$~GeV$^2$.

Because the first moment of the gluon distribution in this fit is
quite large, we can take this global fit as representative of the
`anomaly' scenario, even though the strange
distribution is not quite zero.  In order to construct parton
distributions corresponding to the other scenarios
we have also repeated this fit with the gluon distribution
 forced to vanish at the initial scale. This possibility is in fact
disfavoured by several standard deviations; however, once theoretical
uncertainties are taken into account a vanishing gluon distribution
can only be excluded at about two standard deviations~\cite{cracovia}, and
thus this possibility cannot be ruled out on the basis of present
data. The results of this fit for the various first moments are
displayed in the second column of Table~\ref{tab:fmr}.

We can now use these parton distributions to construct the unknown
C-odd parton distributions. We construct three sets of parton
distributions, corresponding to the three scenarios of 
Section~\ref{sec:pspin}. In
all cases, we assume $\Delta\bar u(x)=\Delta \bar
d(x)=0$. Furthermore, as the `anomaly' set we take the `generic' fit
of Table~\ref{tab:fmr} with the assumption $\Delta \bar s(x)=0$, the strange
distribution for this set being relatively small anyway. As
`instanton' and `skyrmion' parton sets we take the $\Delta g=0$ fit of
Table~\ref{tab:fmr}, 
with $\Delta s=\Delta\bar s$ in the former case, and $\Delta
\bar s=0$ in the latter case. 
The charm distribution is assumed to vanish below
threshold, and to be generated dynamically by perturbative evolution
above threshold.
With these choices all quark and
antiquark distributions are fixed, and thus all SFs
can be computed.

We generate for each of these three scenarios a set of pseudo data, by
assuming the availability of neutrino and antineutrino beams, and
proton and deuteron targets, in the ($x,Q^2$) bins of
Fig.~\ref{fig:xqrange}.
The data are gaussianly distributed about the values of the SFs
 at each data point in the three scenarios.  We obtain in
this way approximately 70 data points for each of the eight
CC SFs.

\begin{table}
\caption{\label{tab:fmr}
Best-fit values of the first moments for the data and
pseudodata fits discussed in the text.}
\hbox{\vbox{\tabskip=0pt \offinterlineskip
      \def\tablerule{\noalign{\hrule}}
      \halign to 450pt{\strut#&\vrule#\tabskip=.5em plus2em
                   &#\hfil &\vrule#
                   &\hfil#&\vrule#
                   &\hfil#&\vrule#
                   &\hfil#&\vrule#
                   &\hfil#&\vrule#
                  &\hfil#&\vrule#\tabskip=0pt\cr\tablerule
             &&\omit Par. \hidewidth
             &&\omit\hidewidth Generic fit\hidewidth
             &&\omit\hidewidth $\Delta g=0$ fit\hidewidth
             &&\omit\hidewidth `Anomaly' refit\hidewidth
             &&\omit\hidewidth `Instanton' refit \hidewidth
             &&\omit\hidewidth `Skyrmion' refit \hidewidth&\cr\tablerule
&& $\eta_\Sigma$&& $0.38\pm 0.03$ &&$0.31\pm 0.01$  &&$0.39\pm 0.01$   &&$0.321\pm 0.006$ && $0.324\pm 0.008$&\cr
&& $\eta_g$&& $0.79\pm 0.19$ &&$0$             &&$0.86\pm 0.10$   && $0.20\pm 0.06$  && $0.24\pm 0.08$  &\cr
&& $\eta_3$&&$1.110\pm 0.043$&&$1.039\pm 0.029$&&$1.097\pm 0.006$ &&$1.052\pm 0.013$ && $1.066\pm 0.014$&\cr
&& $\eta_8$&& $0.579$        &&$0.579$         &&$0.557\pm 0.011$ &&$0.572\pm 0.013$ && $0.580\pm 0.012$&\cr
\tablerule
&& $\eta_u$&& $0.777$        &&$0.719$         && $0.764\pm 0.006$&& $0.722\pm 0.010$&& $0.728\pm 0.009$&\cr
&& $\eta_d$&& $-0.333$       &&$-0.321$        &&$-0.320\pm 0.008$&&$-0.320\pm 0.009$&&$-0.325\pm 0.009$&\cr
&& $\eta_s$&& $-0.067$       &&$-0.090$        &&$-0.075\pm 0.008$&&$-0.007\pm 0.007$&&$-0.106\pm 0.008$&\cr
\tablerule
&& $a_0$   &&$0.183\pm 0.030$&&$0.284\pm 0.012$&&$0.183\pm 0.013$ &&$0.255\pm0.006$  &&$0.250\pm0.007$  &\cr
\tablerule
\tablerule
}}}\hfil\medskip
%\vfill
\end{table}
We proceed to fit a global set of data, which includes the original
NC data as well as the generated CC
data. We assign to the generated data the estimated statistical
errors, and fit including statistical errors only. The errors
assigned to the NC data are instead obtained, as in our
original fits, by adding in quadrature the statistical and systematic
errors given by the various experimental groups.

The fits are performed by adopting the same functional form and
parameters as in the original fit for the C-even parton
distributions, except that the normalization of the octet C-even
distribution $\eta_8$ is now also fitted. For the C-odd parton
distributions, we add six new parameters, namely the normalizations of
the up, down and strange C-odd distributions, and three small-$x$
exponents $\alpha$ (corresponding to an $x^\alpha$ small-$x$
behaviour). The shape is otherwise taken to be the same as that of the
C-even quark distributions.

\subsubsection{Results for first moments}
The best-fit values of all the normalization parameters are shown in
the last three columns of Table~\ref{tab:fmr}, 
where the rows labelled $\eta_u$,
$\eta_d$ and $\eta_s$ now give the best-fit values and errors on the
first moments of $\Delta q^-\equiv \Delta q-\Delta \bar{q}$. A comparison of
these values with those of our original fits leads to an assessment of
the impact of CC data on our knowledge of the polarized
parton content of the nucleon.

First, we see that the improvement in the determination of
the polarized gluon distribution is small, though significant.
This is because the gluon distribution is determined by
scaling violations, and the available range of $Q^2$ at a
$\Emu=50$~GeV \nufact\ is limited.

Let us now consider the C-even quark distributions. 
The error on the first moment of the singlet quark $\Delta\Sigma$ is reduced
 by the CC data by a factor of 3--5 relative to
the available NC data. 
This improvement is especially significant since the
determination of $\eta_\Sigma$ no longer requires knowledge of the
SU(3) octet component, unlike that from NC DIS, and it
is thus not affected by the corresponding theoretical uncertainty.
With this accuracy, it is possible to
experimentally refute or confirm the anomaly scenario by testing the
size of the scale-independent singlet quark first moment.
Correspondingly, the improvement  in knowledge of the gluon first
moment,  although modest, is sufficient
to distinguish between the two scenarios.

The determination of the singlet axial charge is improved by an amount
comparable to the improvement in the determination of the singlet
quark first moment. Its vanishing could thus be established at the
level of a few per cent. The determination of the isotriplet axial
charge is also significantly improved: the improvement is comparable
to that on the singlet quark; it is  due to the availability of the
triplet combination of CC SFs
given in Eq.~(\ref{eq:a3}).
This would allow an extremely precise test of the Bjorken
sum rule, and accordingly a very precise determination of the strong
coupling.  Finally, the octet C-even component is now also determined,
with an uncertainty of a few per cent.  Therefore, the strange C-even
component can be determined with an accuracy better than
10\%. Comparing this direct determination of the octet axial charge to
the value obtained from baryon decays would allow a test of different
existing models of SU(3) violation~\cite{sutr}.

Coming now to the hitherto unknown C-odd quark distributions, we see
that the up and down C-odd components can be determined at the level
of few per cent. This accuracy is just sufficient to establish whether
the up and down antiquark distributions, which are constrained by
positivity to be quite small, differ from zero, and whether or not they are
equal to each other.  Furthermore, the strange C-odd component
can be determined at a level of about 10\%, sufficient to test for
intrinsic strangeness, \ie\ whether the C-odd component is closer in
size to zero or to the C-even component.  The `instanton' and
`skyrmion' scenarios can thus also be distinguished at the level of
several standard deviations.

Of course, only experimental errors have been considered so far. In
Ref.~\cite{cracovia}\ it has been shown that theoretical uncertainties on
first moments are dominated by the small-$x$ extrapolation and
higher-order corrections. The error due to the small-$x$
extrapolation is a consequence of the limited kinematic coverage.
This will only be reduced once beam energies higher
than envisaged in this study will be achieved; otherwise,  
this uncertainty could
become the dominant one and hamper an accurate
determination of first moments.
 On the other hand, the
error due to higher-order corrections could be reduced, since it is
essentially related to the fact that available NC data
must be evolved to a common scale, and also errors are
amplified~\cite{bt} when extracting the singlet component from
NC data because of the need to take linear combinations
of SFs. Neither of these procedures is necessary if
CC data with the kinematic coverage considered here are
available.

\subsubsection{Results for $x$ distributions}
The best-fit SFs corresponding to the `anomaly' refit
(third column of Table~\ref{tab:fmr}) are displayed as functions of $x$ at the
scale corresponding to the bin 4~GeV$^2\leq Q^2\leq$~8~GeV$^2$, and
compared with the data in Fig.~\ref{fig:nucleon-wminus}.
Note that the SFs $g_1$ and $g_5$ always
have opposite signs because the (dominant) quark component in $g_1$
and $g_5$ has the opposite sign, while the antiquark component has the
same sign. For comparison, we also display the SFs at
the initial scale of the fits, and at a high scale. The good quality
of the fits is apparent from these plots.

\begin{figure} 
\begin{center} 
\centerline{
\includegraphics[width=0.48\textwidth,clip]{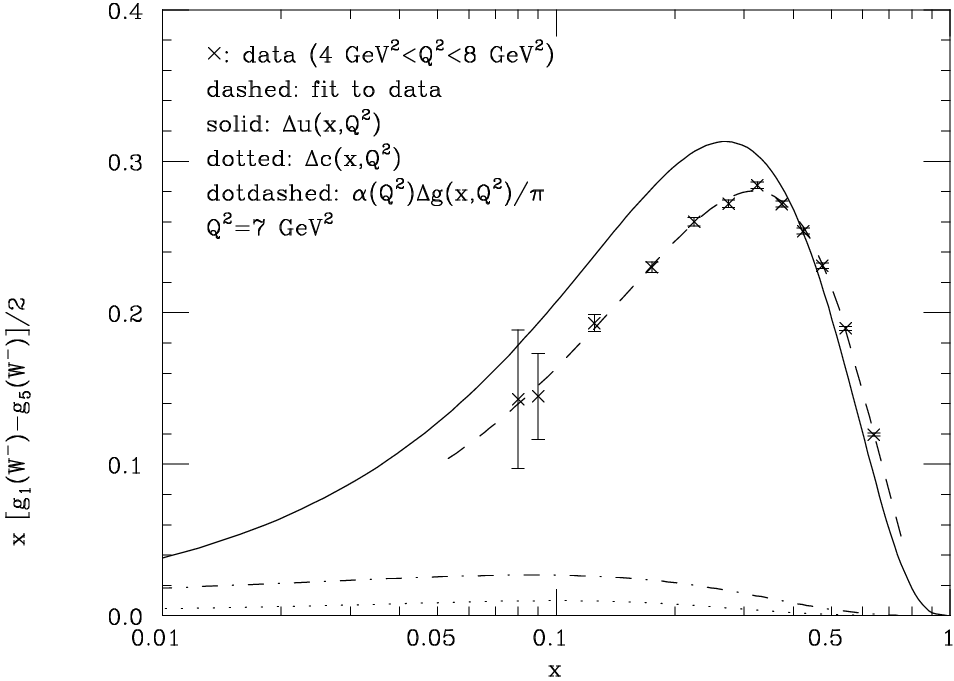} \hfil
\includegraphics[width=0.48\textwidth,clip]{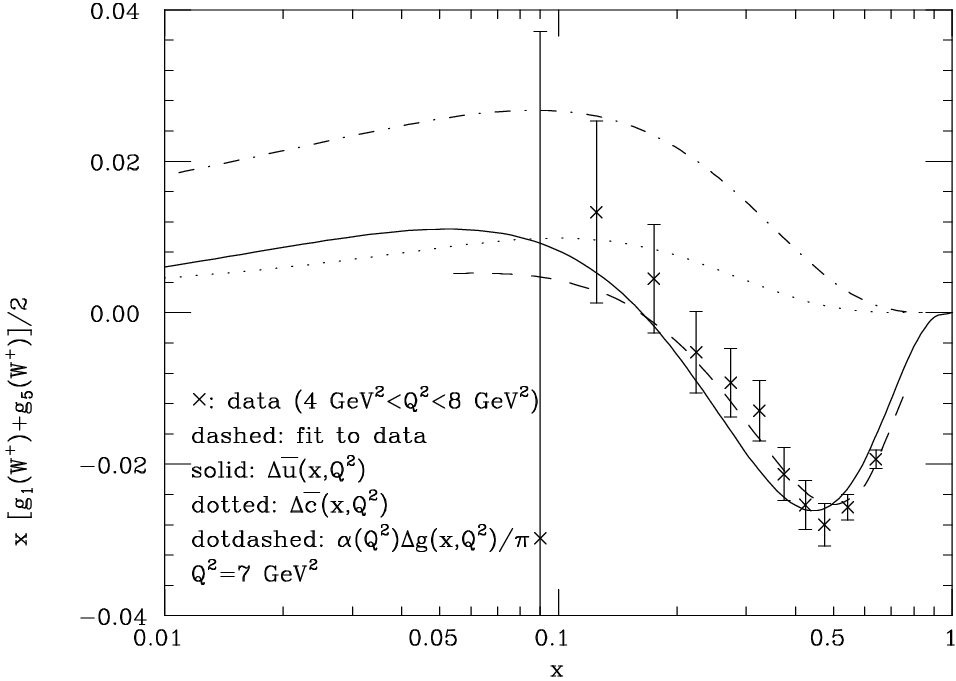} }
\vskip -0.4cm 
\caption{The combinations of SFs of
  \equ{flavcoma}, and the corresponding parton distributions.
\label{fig:deltau}} 
\end{center} 
\end{figure}
\begin{figure} 
\begin{center} 
\centerline{
\includegraphics[width=0.48\textwidth,clip]{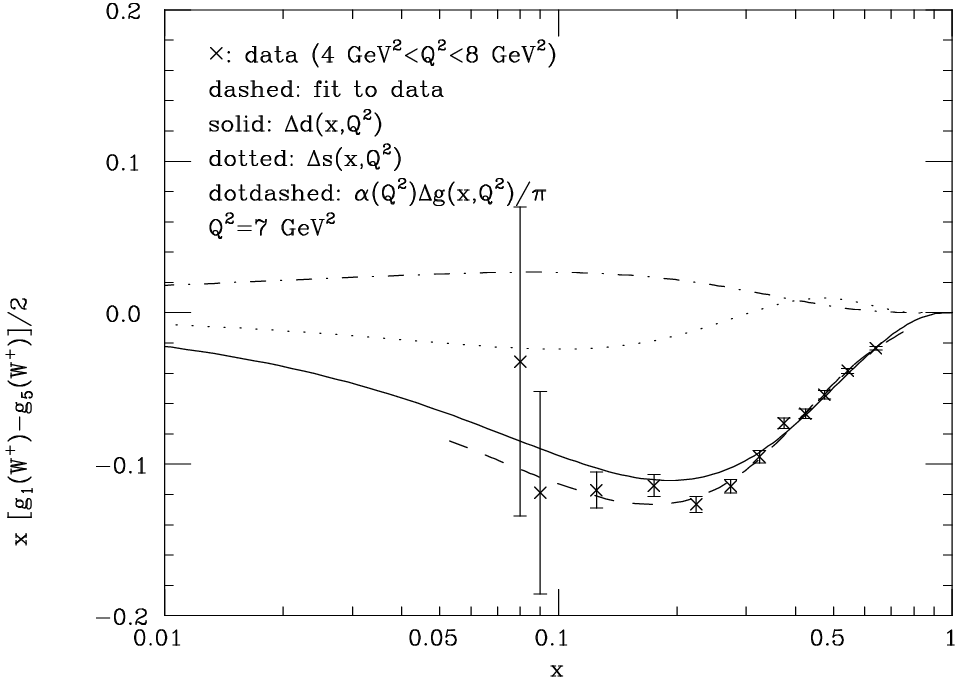} \hfil
\includegraphics[width=0.48\textwidth,clip]{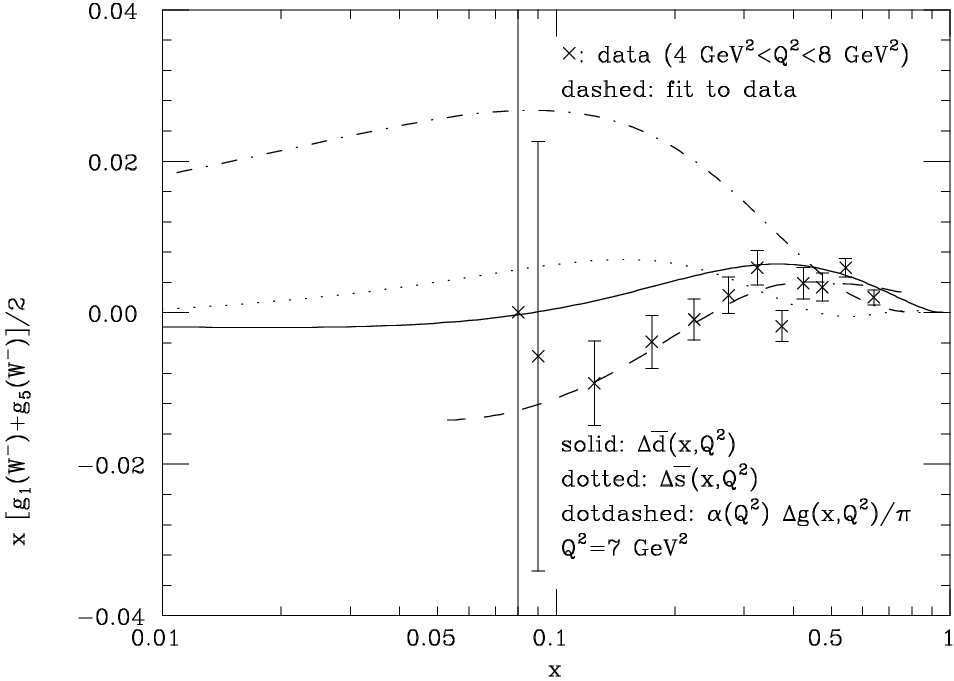} }
\vskip -0.4cm 
\caption{The combinations of SFs of
  \equ{flavcomb}, and the 
corresponding parton distributions.
\label{fig:deltad}} 
\end{center} 
\end{figure}

 
Given the poor quality of current knowledge of the shape of polarized
parton distributions, it is difficult to envisage detailed scenarios
and perform a quantitative analysis of the various shape parameters, as
we did for first moments. However, it is possible to get a rough
estimate of the impact of CC data on our knowledge of
the $x$ dependence of individual parton distributions by considering
the following combinations of SFs,
which, at leading order, are directly
related to individual parton distributions:
\ba
\label{flavcoma}
\frac{1}{2}\left(g_1^{\wum}-g_5^{\wum}\right)=\Delta u+\Delta c;&\quad&
\frac{1}{2}\left(g_1^{\wup}+g_5^{\wup}\right)=\Delta \bar u+\Delta \bar c;
\\
\label{flavcomb}
\frac{1}{2}\left(g_1^{\wup}-g_5^{\wup}\right)=\Delta d+\Delta s;&\quad&
\frac{1}{2}\left(g_1^{\wum}+g_5^{\wum}\right)=\Delta \bar d+\Delta \bar s.
\ea
In Figs.~\ref{fig:deltau}
and~\ref{fig:deltad} we show,
respectively, the combinations 
of \eqs{flavcoma} and (\ref{flavcomb}) for a
proton target, together with the pseudodata for the same combinations
of SFs, in the bin 4~GeV$^2\leq Q^2\leq$~8~GeV$^2$. In
each figure we also display the two parton distributions, which
contribute at leading order to the relevant combination of SFs
 at $Q^2=7$~GeV$^2$, as well as $(\as/\pi)\Delta g$ at the
same scale.

Let us consider the leftmost plot in Fig.~\ref{fig:deltau}.  It is
apparent that the expected statistical accuracy is very good for all
data with $x>0.1$.  This suggests that an accurate determination of
the shape of $\Delta u+\Delta c$ is possible. Furthermore, it is also
clear that $\Delta c$ (dotted curve) is extremely small with respect
to $\Delta u$ (solid curve). However, we observe that the difference
between the $\Delta u$ distribution (solid) and the data is of the
order of 15\% to 20\% for all $x$ below 0.4. This difference is
entirely due to NLO corrections. Specifically, the gluon
contribution (dot-dashed curve), which spoils the leading-order
identification of the quark parton distribution with the SF, as
discussed in Sect.~4.1, Eq.~(\ref{gnlo}), is small but non negligible.
Because the various contributions to NLO corrections (in
particular the gluon distribution) are affected by sizeable theoretical
uncertainties~\cite{cracovia}, this implies that $\Delta u$ can only
be determined with an error that is considerably larger than the
experimental one. At larger scales,  the subleading
corrections to coefficient functions are expected
to be smaller and smaller, while
a residual gluon contribution persists, because of the axial
anomaly~\cite{anom}.

A similar analysis of the left plot of Fig.~\ref{fig:deltau} tells us that a
determination of the shape of $\Delta \bar{u}$ is essentially
impossible.  This combination of SFs is the preferred
one for a determination of the charm distribution, since perturbatively
we expect $\Delta c=\Delta \bar{c}$, and $\Delta \bar u$ is much
smaller than $\Delta u$. Nevertheless, it is apparent from this figure
that even in this case a determination of the charm distribution
is out of reach.

A study of the down quark and antiquark distributions can be similarly
performed by looking at Fig.~\ref{fig:deltau}.  The conclusion for
$\Delta d$ is similar, although perhaps slightly less optimistic, to
that for $\Delta u$: a reasonable determination of its shape is
possible, but with sizeable theoretical uncertainties.
Likewise, the conclusion for $\Delta s$ is similar to that on $\Delta
c$, namely, a determination of its shape is out of reach.
The lower plot
shows that no significant information on the shape of $\Delta \bar d$
or $\Delta \bar s$ can be obtained from this analysis. An alternative
handle for the determination of $\Delta s$  and $\Delta \bar s$ 
is provided by the study of 
events with a tagged final-state charm, as discussed in the next
subsection.

\subsection{Extraction of $\Delta s(x)$ and $\Delta \bar{s}(x)$ from
  tagged charm}
At leading order, charm quarks are produced by CC DIS 
 off a strange or a down quark. The combination of strange 
and down quark distributions is determined by the CKM quark-mixing matrix and 
can be written as an effective distribution
\begin{equation}
s^c(x,Q^2) = |V_{cs}|^2 s(x,Q^2) + |V_{cd}|^2 d(x,Q^2) \;,
\end{equation}
where $|V_{cs}|^2\approx 0.95$ and $|V_{cd}|^2\approx 0.05$ are 
the squared CKM matrix elements. The same expression is valid for the 
antiquark distributions and for the polarized distributions.

For heavy-quark production, the $x$ and $Q^2$ determined from the 
momentum of the final-state lepton cannot be directly related to the 
momentum fraction and virtuality of the scattered parton. Taking kinematical 
corrections into account, one finds~\cite{Barnett:1976ak} that 
the parton distributions are probed at momentum fraction $\xi= 
x (1+ m_c^2/Q^2)$ and virtuality $\mu_c^2 = Q^2 + m_c^2$. 

The cross section for charm-quark production in CC DIS
 can be obtained from the inclusive cross
section~(\ref{xsectot}) by replacing the inclusive SFs with 
charm production ones, which read at 
leading order:
\begin{eqnarray}
F^{\wup}_{1,c} (x,Q^2) &=& s^c (\xi,\mu_c^2)
,\nonumber \\
F^{\wum}_{1,c} (x,Q^2) &=& \bar s^c (\xi,\mu_c^2)
,\nonumber \\
F^{\wup}_{2,c} (x,Q^2) &=& 2 \xi s^c (\xi,\mu_c^2)
,\nonumber \\
F^{\wum}_{2,c} (x,Q^2) &=& 2\xi\bar s^c (\xi,\mu_c^2)
,\nonumber \\
F^{\wup}_{3,c} (x,Q^2) &=& 2  s^c (\xi,\mu_c^2)
,\nonumber \\
F^{\wum}_{3,c} (x,Q^2) &=& -2 \bar s^c (\xi,\mu_c^2).
\end{eqnarray}
An alternative procedure to incorporate effects from the charm-quark mass
into the cross section has been proposed in~\cite{Aivazis:1994pi}. 
The predictions obtained in either prescription differ by less than 15\%. 

The CC
production of charmed hadrons in the final state off polarized 
targets allows a direct measurement of the polarized strange quark and 
antiquark distributions. The polarized cross-section difference 
can be obtained from the expression for inclusive CC DIS~(\ref{eq:polasy}),
 by replacing the SFs; 
the leading-order expressions read: 
\begin{eqnarray}
g^{\wup}_{1,c} (x,Q^2) &=& \Delta s^c (\xi,\mu_c^2)
,\nonumber \\
g^{\wum}_{1,c} (x,Q^2) &=& \Delta \bar s^c (\xi,\mu_c^2)
,\nonumber \\
g^{\wup}_{4,c} (x,Q^2) &=& -2 \xi \Delta s^c (\xi,\mu_c^2)
,\nonumber \\
g^{\wum}_{4,c} (x,Q^2) &=& 2\xi\Delta \bar s^c (\xi,\mu_c^2)
,\nonumber \\
g^{\wup}_{5,c} (x,Q^2) &=& -  \Delta s^c (\xi,\mu_c^2)
,\nonumber \\
g^{\wum}_{5,c} (x,Q^2) &=& \Delta  \bar s^c (\xi,\mu_c^2).
\end{eqnarray}
The resulting leading-order cross-section asymmetry (\ref{deltadef}) is 
the ratio of polarized to unpolarized effective strange-quark distributions:
\begin{equation}
A^{\wup}_{1,c} (x,Q^2) =  \frac{\Delta s^c (\xi,\mu_c^2)}{s^c (\xi,\mu_c^2)}
\; , \qquad
A^{\wum}_{1,c} (x,Q^2) = - \frac{\Delta \bar s^c (\xi,\mu_c^2)}
{\bar s^c (\xi,\mu_c^2)} \; .
\end{equation}
With $s^c (\xi,\mu_c^2)$ and $\bar s^c (\xi,\mu_c^2)$ being known from 
high-statistics measurements off unpolarized targets, a measurement 
of these asymmetries can be turned into a determination of 
the polarized strange quark and antiquark distributions. 

\begin{figure} 
\begin{center} 
\includegraphics[width=0.55\textwidth,clip,angle=-90]{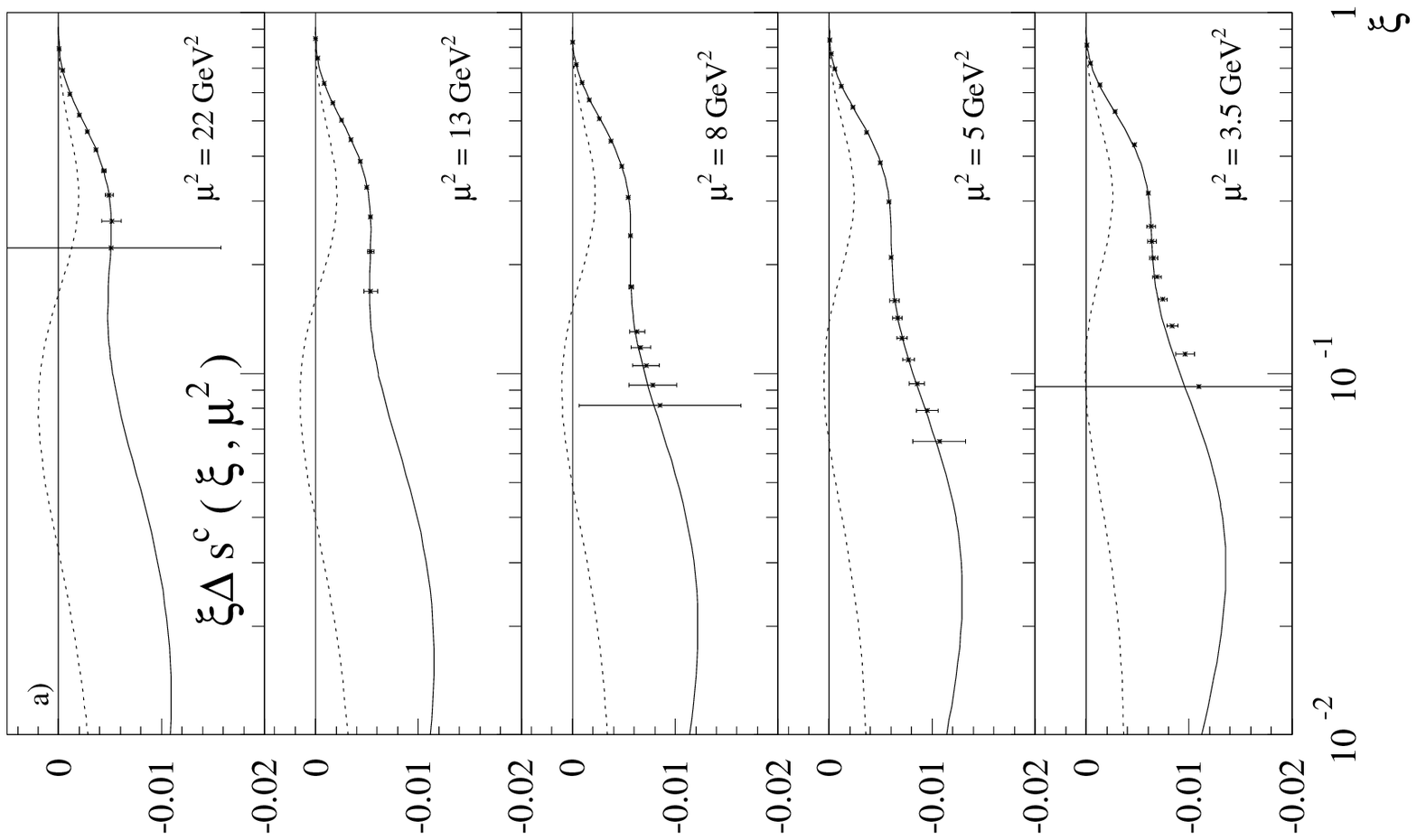} 
\includegraphics[width=0.55\textwidth,clip,angle=-90]{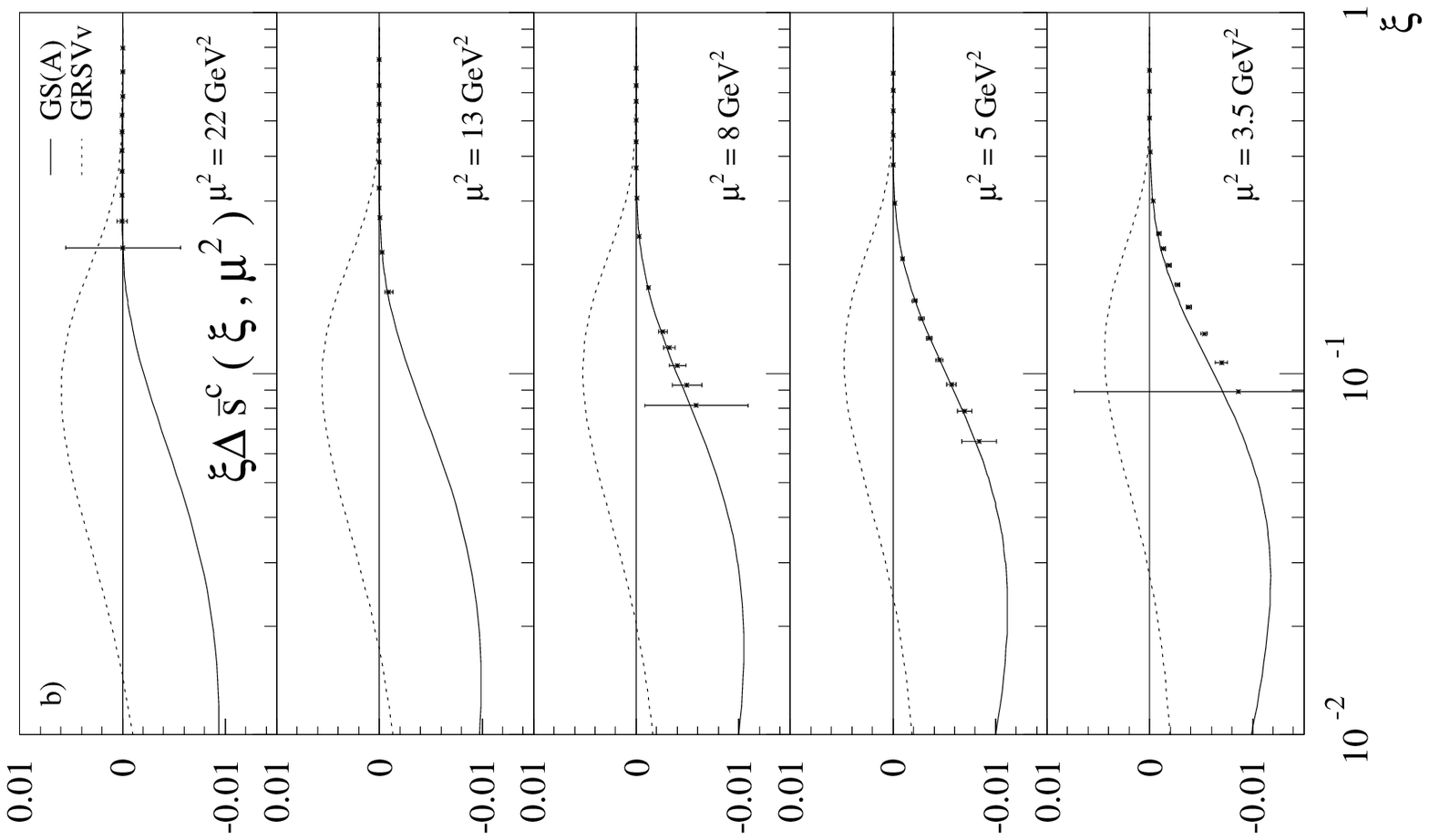} 
\includegraphics[width=0.55\textwidth,clip,angle=-90]{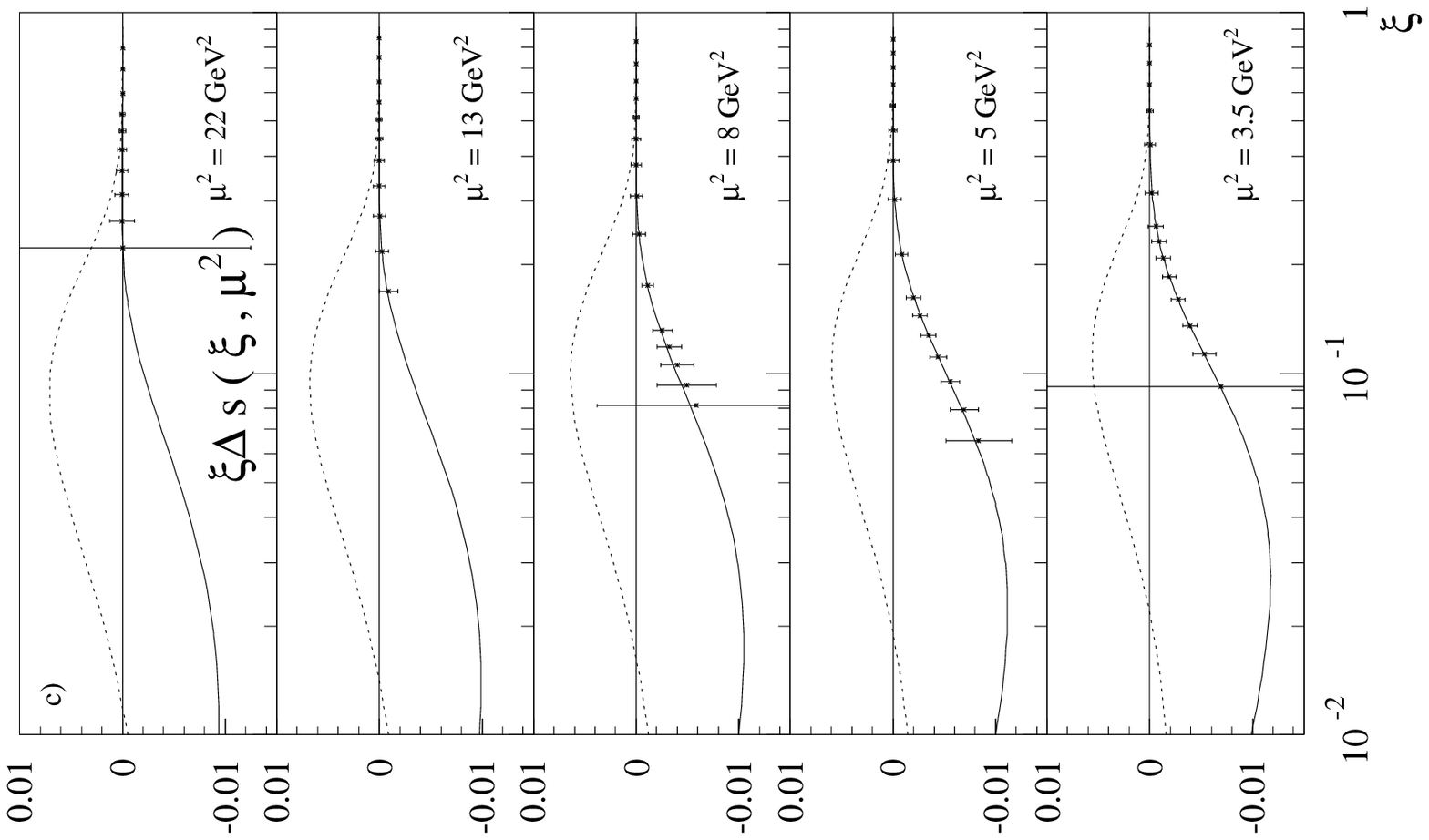} 
\end{center} 
\vskip -0.4cm 
\caption{Estimated statistical errors on the polarized strange and 
antistrange quark distributions, compared to the 
GS(A)~\protect\cite{Gehrmann:1996ag} and GRSVv~\protect\cite{Gluck:1996yr}
 parametrizations.
\label{fig:deltas}} 
\end{figure} 

The statistical error on an extraction of the polarized distributions 
from these asymmetries is given by~\cite{Ball:2000qd}: 
\begin{equation} 
\sigma_{\Delta s^c(\xi,\mu^2)} 
= F^{tgt}_{\nu,\bar \nu} \frac{s^c(\xi,\mu^2)}{\sqrt{2 N_{\nu,\bar \nu}}}\;,
\end{equation}
where $N_{\nu,\bar \nu}$ is the number of CC charm production 
events per target polarization. The estimate of 
statistical errors follows the analysis detailed above.
The target correction factors $F^{tgt}_{\nu,\bar \nu}$, which account for the 
target density, the incomplete target polarization and the dilution factor 
(cross-section-weighted fraction of polarizable to unpolarizable 
nucleons in the target), have to be re-evaluated. Taking into account the 
charm production cross sections off protons and isoscalar nucleons, we 
obtain $F^{tgt}_{\nu} = 3.3$ and $F^{tgt}_{\bar \nu}=2.5$. We assume 
100\% charm reconstruction efficiency. It is expected that the
experiments to be built for a \nufact\ will have efficiencies rather
close to this optimal choice.

In estimating the statistical errors, we use the same specifications as 
in the inclusive studies on polarized targets above: 
$E_{\mu} = 50$~GeV, $N_{\mu} = 10^{20}$ per  target polarization, 
decaying in a straight section of 100~m length, a 
proton target with
target density of 10 g/cm$^2$, a detector with radius of 50 cm positioned 
at  distance of 30~m from the end of the straight section. The cuts 
applied on the events are a minimum cut on the 
final-state muon energy of 3 GeV and minimum cut on the partonic
centre-of-mass energy equal to the charm-quark mass. We use the same 
bins in $x$ and $Q^2$ as in the inclusive studies 
(Fig.~\ref{fig:xqrange}), and 
compute the weighted mean values of $\xi$ and $\mu^2$ for each bin. 

Figures \ref{fig:deltas}a,b  show the expected statistical errors on 
 $\Delta s^c (\xi,\mu_c^2)$ and $\Delta \bar s^c (\xi,\mu_c^2)$. Since 
the polarized antidown-quark distribution is not expected to be 
substantially larger than the polarized antistrange 
one, one 
can determine 
$\Delta \bar s (\xi,\mu_c^2)\simeq\Delta \bar s^c (\xi,\mu_c^2)$. The 
effective polarized 
strange-quark distribution does, however, receive a significant contribution 
from the polarized down-quark distribution. 
Approximating the relative error on $\Delta d(x,Q^2)$ to be  
10\% over the full range in $x$ and $Q^2$ (Fig.~\ref{fig:deltad}), 
we can estimate 
the error on $\Delta s (\xi,\mu_c^2)$ extracted from $\Delta s^c 
(\xi,\mu_c^2)$. The result is shown in Fig.~\ref{fig:deltas}c. 
The comparison of Figs.~\ref{fig:deltas}b and 
\ref{fig:deltas}c
shows that a possible discrepancy between $\Delta s(\xi,\mu_c^2)$ and 
$\Delta \bar s(\xi,\mu_c^2)$ (as suggested for the unpolarized distributions 
in~\cite{Barone:2000yv})
could be detected at a level of about 20\%. 

\begin{figure} 
\begin{center} 
\includegraphics[width=0.35\textwidth,clip,angle=-90]{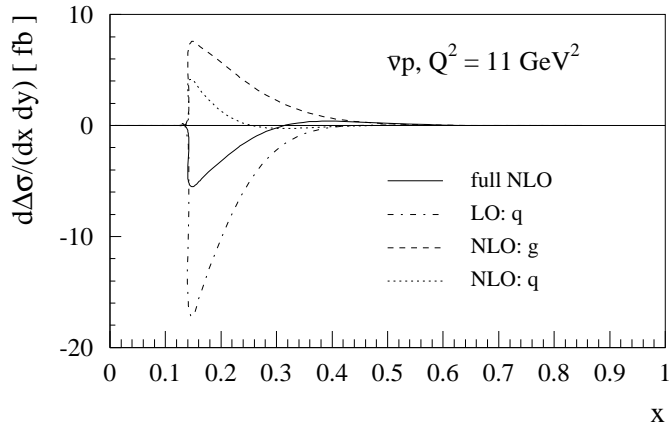} 
\end{center} 
\vskip -0.4cm 
\caption{Polarized heavy quark production cross section at NLO, 
using the polarized GRSVv~\protect\cite{Gluck:1996yr} parton 
distribution functions.}
\label{fig:nlosb}
\end{figure} 
A major uncertainty in the extraction of the polarized strange-quark 
distribution from charm-quark production arises from  
higher-order QCD corrections, consistent with the fact,
discussed at the end of Sect.~4.1, that
the singlet quark distribution is affected 
by large factorization scheme ambiguities.
 The 
NLO contribution from the boson--gluon fusion process to
heavy-quark production is proportional to the size of the polarized-gluon 
distribution, which is at present only constrained very loosely 
from the scale dependence of the inclusive polarized SFs. 
Figure~\ref{fig:nlosb} illustrates the relative magnitude of 
leading and NLO quark and gluon contributions for the
GRSVv polarized parton distribution.
It can be seen that the next-to-leading order gluon-induced subprocess
amounts to a 50\% correction for this distribution.  It follows that
the NLO error estimates of Figs.~\ref{fig:deltau},\ref{fig:deltad}
cannot be compared directly to the LO error estimates of
Fig.~\ref{fig:deltas}, which do not include the uncertainty from
higher order gluonic contributions.  For other parameterizations of
polarized parton distributions the effect is smaller (since the
strange quark distributions are in general assumed to be larger than
in GRSVv), amounting typically to a gluonic contribution of 25\%. Note
also that to NLO a finite renormalization is necessary in order to
relate quark distributions given in the \MS\ scheme (such as GRSVv)
with those of the AB--scheme shown in
Figs.~\ref{fig:deltau},\ref{fig:deltad}. This transformation is given
for first moments in Eq.~(\ref{abtoms}).
%It can be seen that 
%the NLO gluon-induced subprocess amounts to 
%a 50\% correction for this
%distribution. For other parameterizations of 
%polarized parton distributions, the effect is 
%smaller (since the strange-quark distributions are in general assumed to 
%be larger than in GRSVv), amounting typically 
%to a gluonic contribution of 25\%. 
% Notice that the GRSVv
%parton distributions are given in the \MS\ scheme, hence when
%comparing the errors of Fig.~\ref{fig:deltau},\ref{fig:deltad} 
%with those of Fig.~\ref{fig:deltas}, which apply to
%AB--scheme parton densities, the quark distributions should be
%transformed according to Eq.~(\ref{abtoms}).
%Notice that the errors given in 
%Fig.~\ref{fig:deltas} cannot be directly compared to those in 
%Fig.~\ref{fig:deltau} and \ref{fig:deltad}.
%The error estimates on the LO polarized strange
%quark distributions of Fig.~\ref{fig:deltas} do not take the
%uncertainty from higher order gluonic contributions (which can hardly
%be quantifed at present) into account. At NLO, a
%finite renormalization, similar to that done for first
%moments in \equ{abtoms}, should furthermore be performed to connect quark
%distributions obtained in the $\overline{{\rm MS}}$-scheme (such as
%GRSVv) to those in the AB-scheme.
%As outlined in Section~\ref{sec:psfform}, this
%ambiguity is however lifted once physical observables are computed
%consistently in a given scheme. 

As in the case of the results obtained form global fits of inclusive
SFs, discussed in the previous subsection, the
appearance of the gluon contribution at NLO poses the most significant
limitation to the extraction of the polarized strange densities, even
when using tagged-charm final states.  An accurate knowledge of the
polarized gluon density is therefore a mandatory ingredient for the
full statistical potential of the \nufact\ to be exploited.  The
COMPASS experiment will provide a measurement of the polarized gluon
distribution in the kinematic range relevant to the present
studies. To which extent this new knowledge will improve the prospects
for the extraction of the polarized strange component of the proton, 
will be known once the COMPASS results are available.

\section{MEASUREMENTS OF $\mathbf \alpha_{\sss S}$}
The value of the
strong coupling $\as$ is one of the fundamental
parameters of nature. There is almost no limit to our need
to determine it  with more and more precision.
The study of scaling violations of DIS SFs 
and the deviation from the quark--parton model prediction
of DIS sum rules, e.g. of the Gross--Llewellyn Smith (GLS)
sum rule \cite{GLS},
have provided in the past, and still provide,
an important framework for  measuring $\as$.
Good examples are the recent NLO
global analysis~\cite{Alekhin:2001ch} of
existing charged-lepton DIS data, giving 
$\as(M_Z)=0.1165 \pm0.0017 ({\rm stat+syst}) {+0.0026 \atop -0.0034}
({\rm theor})$ and the detailed 
NLO studies~\cite{Ball:1995qd}  of the 1993/94 HERA data for $F_2^p$ at small $x$ and large
$Q^2$, giving $\as(M_Z)=0.122 \pm 0.004 ({\rm exp}) \pm 
0.009 ({\rm theor})$.
Theoretical errors of the latter value of $\as(M_Z)$ are 
dominated by the renormalization and factorization scheme ambiguities,
and by ambiguities in the resummation of small-$x$ logarithms. The
former might be reduced after taking into account next-to-next-to-leading 
order (NNLO) perturbative QCD effects. Indeed, NNLO combined 
fits to the  charged-lepton DIS data 
~\cite{Santiago:1999pr} give  
smaller uncertainties.  The latter could be
reduced thanks to recent theoretical progress in the small-$x$
resummation~\cite{bfklnlo}. 

A  NNLO analysis of the
$\nu N$ DIS data of the CCFR collaboration
for $xF_3$  ~\cite{Seligman:1997mc}
gives $\as(M_Z)=0.118\pm 0.002 ({\rm stat}) \pm 0.005 ({\rm syst}) \pm 0.003
({\rm theor})$ ~\cite{KPS} (more detailed studies of these data
are now in progress ~\cite{KPS2}). This value
should be compared with  the independent NLO
extraction of $\as$  from the  $xF_3$ and $F_2$  data of the
same collaboration: $\as(M_Z)=0.1222\pm 0.0048({\rm exp})\pm 0.0040
({\rm theor})$~\cite{Alekhin:1999df}.
%with the help of the DGLAP method ~\cite{DGLAP}. 
Other available estimates of $\as(M_Z)$, including accurate
 NNLO results obtained from the analysis of LEP data and of
$\tau$ decays, can be found 
in recent extensive reviews~\cite{Bethke:2000ai,Hinchliffe:2000yq}. 
In this
section we will study the potential impact  of future SF
measurements at the \nufact~on the determination of $\as$.  The influence 
of the higher-twist (HT) terms, which become important in the
extraction of  $\as$ at relatively low energies, will be 
analysed as well.

\subsection{Determination of $\as$ and higher-twist terms from  QCD fits of
the SF data}
\label{sec:DGLAP}
To estimate the projected uncertainties on the
determination of $\as$ at the \nufact, we applied the 
same procedure as we used in the estimate of the uncertainties in PDFs (see
Section~\ref{sec:unpolsfres}).
The power corrections
of ${\cal O}(1/Q^2)$ were also included  into the generation of the
fake \nufact\ data and in the fits.
The target mass corrections (TMC) of ${\cal O}(M^2/Q^2)$
were taken into account following Ref.~\cite{Georgi:1976ve}.
The additional dynamical twist-4
non-perturbative contributions were parameterized in the additive form:
\ba
\label{eq:HT1}
F_2(x,Q^2)&=&F_2(x,Q^2)^{\rm LT,TMC}+H_2(x)\frac{1~{\rm GeV^2}}{Q^2}~, \\
\label{eq:HT2}
xF_3(x,Q^2)&=&xF_3(x,Q^2)^{\rm LT,TMC}+H_3(x)\frac{1~{\rm GeV^2}}{Q^2}~,
\ea
where $F_{2,3}^{\rm LT,TMC}$ are the results of NLO QCD calculations with
TMC included, and $H_{2,3}$ are parametrized 
at $x=0, 0.1,\dots, 0.8$ and linearly interpolated between these points. 
%This HT model is related to the one  defined in \equ{eq:HT}
%by the relation $H(x)=F^{\rm LT}(x,Q^2)h(x)$.
Since  the fitted functions depend on
$H_{2,3}$ linearly, the errors on the coefficients of $H(x)$
at $x=0, 0.1,\dots, 0.8$ do not depend on their central values.

The NLO fit
to the generated $F_2$ and $xF_3$ ``data'' returns a statistical
error $\Delta\as(M_Z)={0.00029}$. This is
much better than the statistical error on $\as$ obtained in the
global analysis of charged-lepton DIS data
\cite{Alekhin:2001ch} and it is 16 times smaller than the  one
obtained in the analysis of the CCFR  neutrino DIS
data of Ref.~\cite{Seligman:1997mc},
which used a model-independent description of the 
HT effects \cite{Alekhin:1999df}.

We verified that the extraction of the error is quite stable against
changes in the PDF parametrization. To obtain this result
we repeated the analysis,
generating central values of $F_2$ and $xF_3$ and using the parametrization
of these SFs obtained from the NLO fit to the CCFR
data given in Ref.~\cite{Alekhin:1999df}.
Even though the functional form and the number of $F_2$ and $xF_3$
parameters used in the two cases are quite different,
the difference in the errors on $\as$ is negligible with respect to the 
statistical accuracy of the individual fits.

We also carried out a NLO fit using only the $xF_3$ data.
This results in $\Delta\as(M_Z)=0.00074$,
which is about 2 times larger than the uncertainty  obtained
from the fit to  the combined $F_2$ and $xF_3$ data. 
It must be pointed out, however, that
the use of $F_2$ in the fit introduces
a strong correlation between the value of $\as$ and the sea and gluon
densities, leading to a potential source of further systematics.
In addition, the fit to $xF_3$ only has the advantage that
inclusion of higher-order QCD corrections
will be simpler, since
the NNLO corrections to the coefficient function of the DGLAP equation
are known ~\cite{Zijlstra:1992kj}.
Moreover, a model of the NNLO non-singlet (NS) splitting
function~\cite{vanNeerven:2000ca,vanNeerven:2000wp}
was also recently constructed, using the results of the explicit analytical
calculation of the NNLO corrections to the NS anomalous
dimensions, at fixed number of NS Mellin moments
\cite{Larin:1994vu,Larin:1997wd,Retey:2000nq}
and using additional theoretical information, contained in
Refs.~\cite{Gracey:1994nn,Blumlein:1996jp} (for more details see for instance
the review~\cite{Catani:2000jh}). For this reason the analysis 
of $xF_3$ may give useful information on $\as$ both at NLO and 
at NNLO.  

\begin{figure}
\begin{center}
\includegraphics[width=0.5\textwidth,clip]{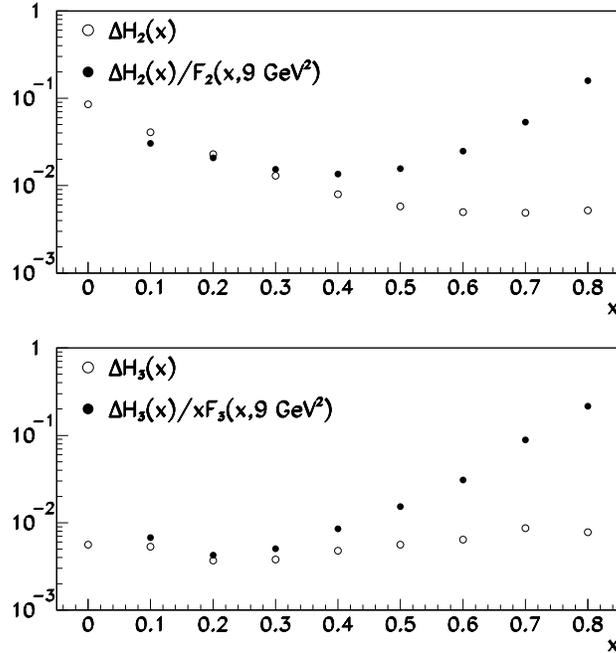}
\end{center}
\vskip -1cm
\caption{The errors on HT contributions accessible at the \nufact.}
\label{fig:htserrors}
\end{figure}

The expected precision of the \nufact\ data exceeds that of available
measurements of DIS SFs at moderate $Q^2$.
This may allow us to improve our knowledge of  the 
HT contributions $H_{2,3}$. The errors on the coefficients of the
functions
$H_{2,3}$, which can be obtained from the analysis of \nufact\ data,
are given in Fig.~\ref{fig:htserrors}.
These errors were obtained from the fits described above, alongside 
the errors on $\as$ (the simultaneous estimate of the HT and $\as$ errors
is very important in view of possible correlations between them). The errors
on the HT contributions are smaller by over one  order of magnitude
than those extracted from the CCFR data ~\cite{KPS,KKPS}.

The data obtained at the \nufact\ could then be used for the verification 
of the models describing the HT terms. 
One of these models is based on the application of the 
infrared renormalon (IRR) technique \cite{Beneke,BB}.
The advantage of this approach is that it connects the 
HT contributions with the leading twist ones.
For example, in the NS approximation  the IRR model of twist-4 
contributions has the following form:
\be
H_{2,3}(x)=A_2^{'}\int_x^1 dz C_{2,3}(z)F_{2,3}^{\rm LT}(x/z,Q)~.
\ee
Here  $C_{2,3}(z)$ are calculated in Ref.~\cite{IRRmodel},
and the parameter $A_2^{'}$ introduced there can be expressed as
\be
A_2^{'}=-\frac{2C_F}{\beta_0}\Lambda^2~,
\ee
where $C_F=4/3$  and $\beta_0$ is the first coefficient of the QCD
$\beta$-function.
(A similar definition was used in Ref.~\cite{Maul} in the 
comparison of the IRR model predictions for $F_{2,3,\rm L}$
with the data.)
The parameters $\Lambda_{2,3}^2$ for the IRR model can be extracted
from the \nufact\ data with the errors
$\Delta \Lambda_3^2={0.0030}~{\rm GeV}^2$ and 
$\Delta \Lambda_2^2={0.037}~{\rm GeV}^2$
(for comparison, the preliminary
results  of the combined analysis 
of the CCFR ~\cite{Seligman:1997mc} and the  JINR--IHEP~\cite{Sidorov:1999mk} 
collaborations data are 
$\Lambda_3^2=0.44\pm0.19~{\rm GeV}^2$ and
$\Lambda_2^2=0.91\pm0.77~{\rm GeV}^2$ ~\cite{Alekhin:2001zj})
\footnote{While completing our report we learned of the new $xF_3$
  data, obtained recently by the H1 Collaboration at HERA \cite{Adloff:2000qj}.
These data are related to rather high $Q^2$ region ( $Q^2\geq $1500~GeV$^2$).
We therefore expect no essential improvement of
our estimates from the inclusion of H1 data into these fits.}.

It is worth stressing that the results of the NNLO fits to the CCFR 
data~\cite{KPS,KKPS}, which use the NNLO QCD expressions for the Mellin
moments of $xF_3$ and the anomalous dimensions 
calculated in Refs.~\cite{Larin:1994vu,Larin:1997wd}, 
demonstrate the effect of shadowing of the twist-4 terms
by higher-order perturbative QCD corrections. 
It should also be stressed that the decrease of the size of the fitted
HT contributions at the NNLO level is confirmed independently
by the  DGLAP analysis of the combined $F_2$ charged-lepton data
made by the MRST collaboration~\cite{MRSTHO}, which incorporates
NNLO corrections to both the coefficient function~\cite{VZ}
and the  model of the splitting function~\cite{vanNeerven:2000ca}.
However, since Ref.~\cite{MRSTHO} did not assign errors to the
HT terms extracted at NLO and NNLO, we cannot decide
whether the effects observed in Refs.~\cite{KPS,KKPS}
arise from the incorporation of the NNLO corrections into DIS
fits, or whether they demonstrate a lack of precision of the analysed data.
The possible analysis of more precise data from the \nufact\ may
allow us to clarify this point. 
We evaluate that the correlation coefficients between $\as$ and $H_{2,3}$
are not so large: their maximal value is $\sim -0.7$ at $x \sim 0.5$.
This allows the unambiguous separation of the logarithmic-like and
power-like contributions to the Bjorken scaling violation.
In particular, it is possible to hope that a
clearer separation of the twist-4 effects from the perturbative QCD
contributions may be possible at the NNLO level. The detailed study of this
problem is rather intriguing.

\subsection{Determination of  $\as$ from  the Gross--Llewellyn Smith sum rule}
\label{sec:GLS}
The value of $\as$ can be also determined from the GLS integral
\ba
S_{\rm GLS}^N(Q^2) = \frac{1}{2}\int_0^1 dx\,\left(F_3^{\nu p}(x,Q^2)
+F_3^{{\nu} n}(x,Q^2)\right) \; .
\ea
At ${\cal O}(\as^3)$ and including 
the ${\cal O}(1/Q^2)$ corrections, the GLS integral for
$f=4$ massless active flavours is equal to~\cite{GLSQCD}:
\begin{equation}
S_{GLS}^N(Q^2)=3\left[1-\frac{\as(Q^2)}{\pi}-
3.25\left(\frac{\as(Q^2)}{\pi}\right)^2
-12.2\left(\frac{\as(Q^2)}{\pi}\right)^3
\right] - \frac{h}{Q^2}~.
\label{GLS}
\end{equation}
The GLS integral has been measured 
in a number of experiments (see for instance
Ref.~\cite{KSR} for a review).  Its $Q^2$ dependence was extracted
by combining the CCFR data for
$xF_3$~\cite{Seligman:1997mc} with those from 
CERN and IHEP experiments, at several energy bins~\cite{Kim:1998ki}.

In many other processes the theoretical $\as$ uncertainty is dominated by
the error due to the truncation of the higher-order perturbative QCD
corrections. 
Since the theoretical expression for the GLS integral is known
up to 3-loop $\as$ corrections, the scale- and the scheme-dependence
 ambiguities on the 
$\as$ extraction from the GLS sum rule 
can be minimized (see Ref.~\cite{ChK}).
The mentioned theoretical uncertainties can survive 
if the $Q^2$-region 
of the data used for the estimate of the integral is large 
and the \oatwo\ approximation needs to be used to interpolate 
data from  different $Q^2$ regions~\cite{KS}.
This problem can be avoided
if the data are split into relatively small 
$Q^2$ bins, as is expected at the \nufact.
An additional contribution to  the GLS sum rule comes from heavy
quarks ($c,s$).
The heavy-quark mass correction is known at \oatwo~\cite{BV}. 
Its effect is small at energies close to the threshold,
and is comparable in size with estimates of the massless 
$\oafour$ correction made with different methods~\cite{est}.
Together with the massless contributions,
the mass-dependent terms are therefore under control.
In the asymptotic regime, these affect the threshold
matching conditions
\cite{match}, and introduce an uncertainty of about
$6.5m_q$ in the choice of the matching point. 
This uncertainty leads to an additional theoretical ambiguity 
of approximately 0.002 on the value of $\as(M_Z)$ (see e.g. Ref.~\cite{KPS}).

\begin{table}
\begin{center}
\caption{The statistical errors on GLS integrals at different
$Q$ bins, obtained from the different data sets (I: generated data
for \nufact ~only; II: the same data for the \nufact ~combined with
the CCFR data of Ref.\cite{Seligman:1997mc}).}
\label{tab:glserrors}
\vspace*{0.1cm}
\begin{tabular}{|c|c|c|c|} \hline

$Q^2~[{\rm GeV}^2]$ & I &  II  \\\hline

1--2 &       0.0074 & 0.0073  \\\hline

2--3.5 &     0.0086 & 0.0084   \\\hline

3.5--7 &     0.013 & 0.013  \\\hline

7--14  &     0.028 & 0.021   \\\hline

14--28 &     0.11 & 0.039  \\\hline

28--200 &      -- & 0.054 \\ \hline
\end{tabular}
\end{center}
\end{table}

An important source of experimental uncertainty on the measured 
GLS integral is the error due to the extrapolation of 
$xF_3$ to the unmeasured high- and low-$x$ regions. 
This error can be large if the neutrino energy is limited, as is the case
for the 50 GeV \nufact\ option.
To estimate this error for different values of $Q^2$
we split the $xF_3$ data, used in the analysis of
Subsection~\ref{sec:DGLAP}, in several 
bins of $Q^2$, and then  generate random $xF_3$ values in each bin 
with the central values given by 
\begin{equation}
xF_3(x)=\frac{I_3}{A} x^a(1-x)^b~,~~~~~~~~~
A=\int_0^1 dx\,x^{a-1}(1-x)^b~.
\label{eqn:f3fit}
\end{equation}
The statistical errors are given by the study of Section~\ref{sec:F2F3}. 
The values of the parameters used for the generation
were chosen as $I_3=3$, $a=0.7$, $b=4$. 
Then Eq.~(\ref{eqn:f3fit}) is fitted to the generated data in each
$Q^2$ bin
with the parameters $a$, $b$, and $I_3$ set free. 
The uncertainty in the fitted value of $I_3$, which gives the
uncertainty on the GLS integral, accounts for the  
uncertainty in the extrapolation to the unmeasured $x$ regions.
The errors on the fitted values of $I_3$
are given in Table~\ref{tab:glserrors}.
One can see that at high $Q^2$ the  errors are quite large.
Combining the \nufact\ data with the CCFR measurements
of Ref.~\cite{Seligman:1997mc} one can obtain a significant improvement
in the precision of the GLS integral determination, 
since the two sets of data have a complementary $x$ coverage.
This is shown in the second
column of Table~\ref{tab:glserrors}, where the errors are
significantly smaller than those
obtained in Ref.~\cite{Kim:1998ki} from the analysis of the combined
CERN--FNAL--IHEP data with similar binning in $Q^2$. 

%It should 
%be noted that the used value of $a$ is in agreement 
%with the theoretical result  of Ref.~\cite{GET}, obtained 
%in the double logarithmic approximation, and investigated  in the case 
%of non-singlet SFs in Ref.~\cite{KL}. Moreover, the results 
%of the fits to the CCFR data, performed in Ref.~\cite{KPS}, are 
%supporting the results of the calculations of Ref.~\cite{GET} and in
%the leading order are in 
%agreement with the prediction of the $Q^2$-behaviour of the 
%parameter $b$, derived in Refs.~\cite{Korch,AB} (the $x\rightarrow 1$ 
%limit was also studied in the works of  Ref.~\cite{Parisi:1979se}.) 
%However, these QCD results do not include
%information about nuclear corrections, which can be important in
%the regions of low and high-$x$ values. The theoretical results of
%studying these effects will be described in the next section.

\begin{figure}
\begin{center}
\includegraphics[width=12cm,height=8cm,clip]{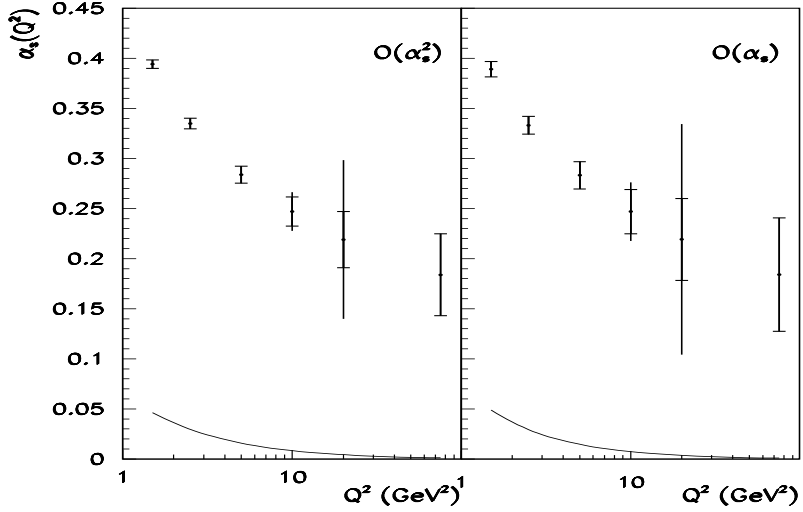}
\end{center}
\vskip -1cm
\caption{The outer error bars are determined by the GLS errors 
of column I of  Table \protect\ref{tab:glserrors}, 
while the inner  ones are fixed by the 
GLS errors of column II. 
The points at largest $Q^2$-bins are extracted from the CCFR data only.
The curve shows the 
errors on $\as$ due to the uncertainties of higher-twist contributions.}
\label{fig:alp}
\end{figure}

The estimates of the uncertainties on $\as$,
which can be obtained from the measurements 
of the GLS integral at a \nufact, are given in Fig.~\ref{fig:alp}.
The central points for $\as$ shown in this figure 
were calculated from the solutions of the 
2-loop and 3-loop renormalization group equations
with the boundary  value  $\as(M_Z)=0.118$. 
The error bars of the points  
were obtained by rescaling the statistical errors 
on $I_3$ given in  column II of  Table \ref{tab:glserrors}
by the factor of $dS_{GLS}/d\as$ in each $Q^2$ bin.    
The statistical error on $\as$ is smaller at ${\cal O}(\as^2)$ since 
$\vert dS/d\as\vert$ is larger at this order.
The errors  given in Fig.~\ref{fig:alp} are small enough  
to allow for the clear observation of the  $Q^2$ dependence of the QCD 
coupling constant measured in one single process
and the comparison with various 
theoretical predictions. 
%For example,  one can use the measurements from the \nufact\ to 
%verify  the findings 
%of Ref.~\cite{MSS}, namely that the application
%of the analytic perturbation theory approach of Ref.~\cite{ShS} (analogous
%to the dispersive approach of Ref.~\cite{DMW})  leads to a difference
%between the $Q^2$ evolution of the GLS sum rule and  the predictions
%of the standard perturbation theory.

In order to estimate the $\as(M_Z)$ precision accessible from the GLS
measurements, we fitted the expression of Eq.~(\ref{GLS}) to the points
of Fig.~\ref{fig:alp}.  The analysis was made both at ${\cal O}(\as)$
and at ${\cal O}(\as^2)$.  Since in our analysis the low-$Q^2$
data from the \nufact\ are used, an important source of the total
$\as$ error is related to the error in the HT parameter $h$ of
Eq.~(\ref{GLS}).  It should be stressed that, contrary to the
$x$-dependence of $H_3(x)$, the theoretical estimates of 
its first moment, related to the GLS sum rule, are
theoretically more solid.  In fact in this case we know the expression
for the local operator contributing to the dynamical ${\cal O}(1/Q^2)$
correction~\cite{ShV}. Moreover, there are several model-dependent
calculations of the value of this operator. The first one comes from
the QCD sum rules method~\cite{ShVZ}, implemented in its 3-point
function realization in the work of Ref.~\cite{BKol} and later on in
Ref.~\cite{RR}. Another estimate of the value of the twist-4
contribution to the GLS sum rule comes from the instanton vacuum model
\cite{BPW}. Within theoretical errors, the results of the
model-dependent calculations of Refs.~\cite{BKol,RR,BPW} are in
agreement.

To extract $\as(M_Z)$ with a model-independent treatment of the HT corrections,
we fitted  the expression of Eq.~(\ref{GLS}) with the parameter 
$h$ set free. 
At \oatwo, and using the analysis of the GLS data with 
the errors from column II of Table \ref{tab:glserrors}, 
we get the following results:
\be
\Delta \as(M_{\rm Z})={0.0035}\, , \quad \Delta h = {0.13}~{\rm
  GeV}^2~.
\ee
At \oas\ we get instead:
\be
\Delta \as(M_{\rm Z})={0.0039} \, , \quad \Delta h = {0.07}~{\rm
  GeV}^2~.
\ee
The cause of the only marginal improvement in accuracy when going to
higher order is the faster decrease of $S_{GLS}$ with $Q^2$, and the
consequent increased correlation of $\as$ with the HT term
$h/Q^2$.  The influence of the uncertainties of the high-twist correction
on the precision of the $\as$ determination at various $Q^2$ is
illustrated in Fig.~\ref{fig:alp}, where the value $\Delta h/Q^2$
rescaled with the factor $dS_{GLS}/d\as$ is given as well.  Notice
from Fig.~\ref{fig:alp} that this uncertainty weakly depends on the
perturbative approximation for the GLS sum rule used in the analysis.
One can also see that at small $Q^2$ the errors on $\as$ due to HT
uncertainties are increasing.  Since the values for $\as$ at large
$Q^2$ have large statistical errors, the related value of $\as(M_{Z})$
is mainly determined by the uncertainties of HT corrections and does
not change significantly from the ${\cal O}(\as)$ fit to the ${\cal
  O}(\as^2)$ one.

If one fixes $h$, the statistical
error on $\as(M_Z)$ in the ${\cal O}(\as^2$) fit to the data with
errors from column II of Table~\ref{tab:glserrors}
reduces to {0.00026}.  In this case, however,
the uncertainty on $\as$ due to the model dependence
of  $h$ \cite{BKol,RR,BPW}  is large (see e.g.~\cite{ChK}).
For  this reason, the error on $\as$
obtained by  considering the model-dependent estimates of the
HT contributions to the GLS sum rule  is essentially
the same as the one defined from the existing neutrino DIS data.
Therefore it seems more appropriate to analyse the  GLS sum rule 
data from the \nufact\ using the fit with the model-independent
definition of the HT contribution.

\subsection{Measurement of $F_1(x)$ and unpolarized Bjorken sum rule}
\label{sec:Bj}
\begin{figure}
\begin{center}
\includegraphics[width=0.5\textwidth,clip]{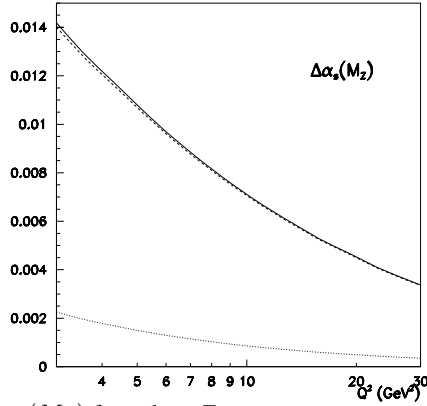}
\end{center}
\vskip -1cm
\caption{Theoretical errors  on $\as(M_Z)$
  from the \nufact\ measurement of the unpolarized Bjorken sum rule at
  different $Q^2$ values. The dotted line correspond to the
  contributions of the TMC uncertainty; the dashed one to the
  uncertainty due to the twist-4 contribution; the solid line is the
  combination of both.}
\label{fig:alpBj}
\end{figure}
While $F_1(x,Q^2)$ is known theoretically to be related to $F_2(x)$
via the Callan--Gross relation and its calculable higher-order
corrections, only recently have the experiments attempted a direct
measurement from the data.  Preliminary results on the determination
of $F_1^{\nu N}(x,Q^2)$ from the large $y=E_{had}/E_{\nu}$ behaviour
of $d^2\sigma{\nu N}/dxdy$ have been obtained by the CHORUS
collaboration at CERN \cite{Oldeman} and by the CCFR--NuTeV
collaboration at Fermilab \cite{CCFRF1}.  However, only a few points
for $F_1(x,Q^2)$ over a limited range of $x$ were extracted up to now.
As discussed above, the large statistics available at the \nufact\ 
will in principle allow a complete separation of the various SF
components, including in particular a measurement of $F_1(x,Q^2)$. An
example of the accuracy with which the individual components will be
extracted was given in Fig.~\ref{fig:F123err}. This improved knowledge
on $F_1$ can be used for different purposes, and in particular for an
independent measurement of $\as$.  This possibility is based on the
study of the unpolarized Bjorken sum rule, which was derived in
Ref.~\cite{Bjn}.  This old, but still experimentally untested NS
combination of neutrino DIS SFs, has the following form:
\be
S_{1}^{n-p} = \int_0^1 dx\,\left[F_1^{\nu n}(x,Q^2)-F_1^{\nu p}(x,Q^2)
\right] \; ,
\label{eqn:def}
\ee
which in terms of parton distributions can be expressed as 
\ba 
S_{1}^{n-p} =\int_0^1 dx\, \left[u(x,Q^2)-\overline{u}(x,Q^2)+d(x,Q^2)
-\overline{d}(x,Q^2)\right].
\label{eqn:part}
\ea
Taking into account the correction
of \oas\ 
calculated in Ref.~\cite{BBDM} and twist-4 terms, the theoretical expression
for $S_1^{n-p}$ reads
\be
S_{1}^{n-p}
 = 1-\frac{2}{3}\frac{\alpha_s}{\pi}+\frac{h_{Bj}}{Q^2} .
\label{eqn:bjn}
\ee
The massless corrections to
$S_{1}^{n-p}$ of \oatwo\
were calculated in Ref.\cite{Gorishnii:1984gs},
while the $\oacube$ contributions are  analytically
evaluated in  Ref.~\cite{Larin:1991zw}. The heavy-quark mass correction
to $S_{1}^{n-p}$ is  known from the  calculations
of Ref.~\cite{BV} and is comparable with the results of
existing $\oafour$ estimates~\cite{est}. 

The HT term in Eq.~(\ref{eqn:bjn}) is analogous to that of the GLS
sum rule. The value of 
$h_{Bj}$ is proportional to the  matrix element of a local twist-4 operator,
$h_{Bj}=-(8/9)\langle\langle \calO\rangle \rangle$ \cite{ShV}, with:
\be
2p_{\mu}\langle\langle \calO\rangle\rangle = \langle p|\calO_{\mu}|p\rangle~~,
\quad {\rm and} \quad
\calO_{\mu}=\overline{u}G_{\mu\nu}\gamma_{\nu}\gamma_5 u-
\overline{d}\tilde{G}_{\mu\nu}\gamma_{\nu}\gamma_5 d~~~,
\ee
where $\tilde{G}_{\mu\nu} =(\epsilon_{\mu\nu\alpha\beta}/2)G^{a}_{\alpha\beta}
(\lambda^a /2)$. 
The application of the 3-point function QCD sum rules
results in the following estimate: $\langle\langle \calO\rangle\rangle
=0.15\pm 0.07$ GeV$^2$ \cite{BKol},
where we take for the theoretical error the conservative estimate
of 50$\%$.

We constructed the \oas\ $Q^2$ evolution of $S_{1}^{n-p}$ (see
Eqs.~(\ref{eqn:def}),(\ref{eqn:part}) using the set of
parton distributions of Ref.~\cite{Alekhin:2001ch}, and taking into
account the twist-4 contribution, and estimated the theoretical error
on $\as(M_Z)$ extracted from this expression.  The behaviour of the
error $\Delta\as(M_Z)$ that can be obtained from the measurements of
the $S_{1}^{p-n}$ integral at different $Q^2$ is shown in
Fig.~\ref{fig:alpBj}.  This error is defined as
\be
\Delta\as(M_Z) = \Delta S_{1}^{n-p}(Q^2) \left[
\frac{dS_{1}^{n-p}(Q^2)}{d\as(Q^2)}
\frac{d\as(Q^2)}{d\as(M_Z)} \right]^{-1} \; ,
\label{eqner}
\ee
where $\Delta S_{1}^{p-n}$ is the error 
due to the different sources of theoretical uncertainties.
One of them comes from the error in the 
contribution of the TMC, which is related 
to the uncertainty in the existing sets of  PDFs.
We calculated this error using the uncertainties  
of PDFs from  Ref.~\cite{Alekhin:2001ch}
and convinced ourselves that it does not exceed
0.002 (see Fig.~\ref{fig:alpBj}).
At \oas\ $dS_{1}^{n-p}/d\as=2/(3\pi)$, and
the theoretical error due to the HT uncertainty is  
$(3\pi/2)\times0.07/Q^2$.
For the reference scales
$Q^2=4$~GeV$^2$ and $Q^2=$10~GeV$^2$,
we obtain
$\Delta^{HT}\as(M_Z)= 0.012$ and $\Delta^{HT}\as(M_Z)= 0.007$, respectively.
To keep a balance between the various sources of uncertainties
the corresponding statistical errors must not exceed these values.

From the theoretical point of view, the uncertainties of
Eq.~(\ref{eqner}) are consistent with those
of $\as(M_Z)$ extractions from the GLS sum rule value
at $Q^2=3$ GeV$^2$ \cite{ChK}, namely $\Delta^{HT}\as(M_Z)= 0.003$.
The latter one is a bit smaller because of  the differences between
the perturbative expressions for the GLS sum rule and the  unpolarized
Bjorken sum rule, and between the estimates of the twist-4 contributions
to these sum rules. 
It is therefore rather important to 
check the results for the high-twist contributions 
to $S_1^{n-p}$ obtained in Ref.~\cite{BKol}. Within the 
framework of the instanton model this question is now under
study\footnote{C.~Weiss, private communication.}. 

The experimental determination of the Bjorken unpolarized sum rule, which will
be possible at the \nufact, can therefore be considered as an additional
source for the determination of $\as$, provided the twist-4 contributions
are known with more precision.

Precise \nufact\ 
data on $F_1(x,Q^2)$ will also allow the measurement of the
$x$-dependence of the HT contribution to $F_1$, which in the IRR model of
Ref.~\cite{IRRmodel} is predicted to  coincide with the shape of the twist-4
contributions to $xF_3$. In view of the considerable interest given
to the analysis of the contributions of HT terms to different
quantities, it is quite desirable to study this prediction in detail,
using the experimental data for $F_1$.

To conclude this section we note that the measurement of the
unpolarized Bjorken sum rule requires the extraction of $F_1^{\nu
  p}(x,Q^2)$ from DIS on a hydrogen target and of $F_1^{\nu n}(x,Q^2)$
from DIS on a deuterium target.  The latter process necessitates the
analysis of nuclear corrections, especially in the small-$x$ region.
A detailed study of these problems, as well as the discussion of other
experimental alternatives, will be discussed in the next section.

\section{NUCLEAR EFFECTS IN DIS AT THE \nufact}
\label{sec:nuke}
%% MOTIVATIONS
There are two general motivations to study nuclear effects in DIS
experiments at the \nufact. First, nuclear physics of parton distributions
is of interest in itself, and the comparison of heavy-target data with
hydrogen and light-nuclei data (\eg deuterium) may give us new
insights into the structure of multi-quark systems.
%%From the practical point of view,
On the other side, an accurate knowledge of nuclear effects is
necessary in order to extract the SFs of a physical proton
and neutron from nuclear data. This applies in the first place to the
neutron, since available neutron targets are mainly nuclei.
Theoretical studies of nuclear effects, among other possible
applications, could also help in choosing the most appropriate neutron target.

%% WHAT IS KNOWN

\begin{figure}[ht]
\begin{flushleft}
\includegraphics[width=0.51\textwidth]{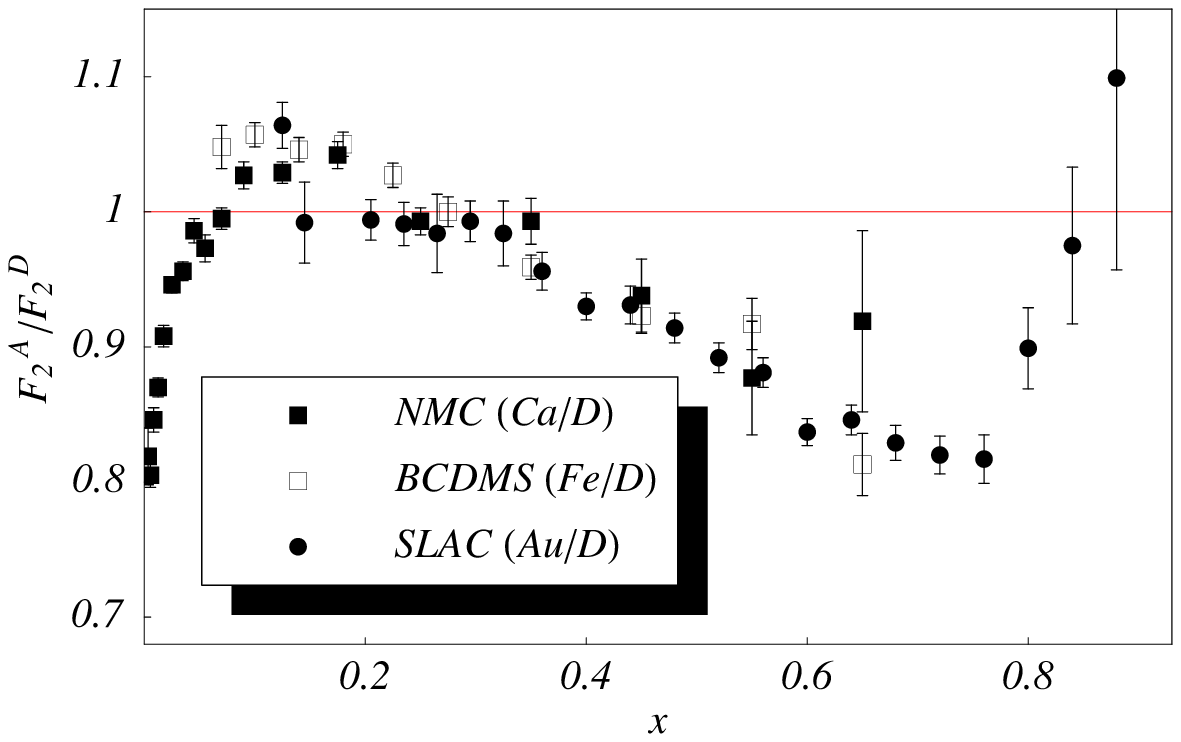}
\end{flushleft}
\vspace{-5.95cm}
%\vspace{-0.5\textwidth}
%\vspace{-1.cm}
\begin{flushright}
\includegraphics[width=0.51\textwidth]{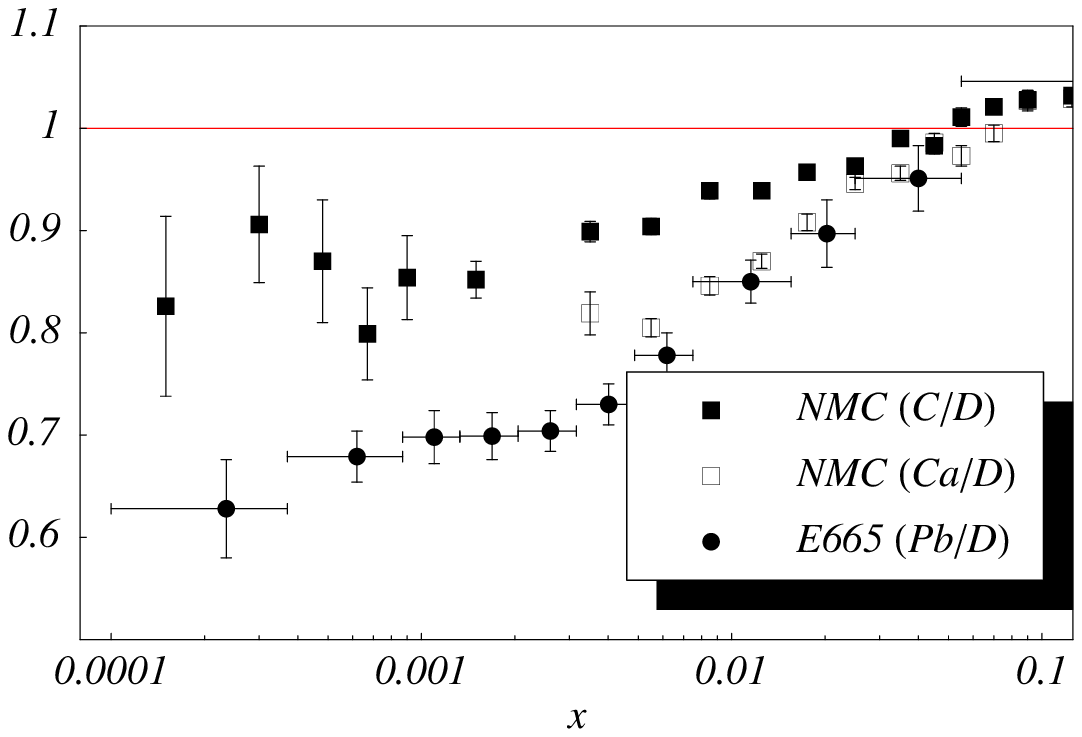}
\end{flushright}
\begin{center}
\caption{\label{emc_ratio}
  The $x$ dependence of heavy-target/deuteron SFs
  ratio as measured in DIS of muons and electrons off various nuclear
  targets and averaged over $Q^2$.  The left panel shows data from
  CERN and SLAC for different nuclear targets. The right panel focuses
  on the region of small $x$ and illustrates nuclear shadowing effect.
  Data points are from NMC \cite{nmc95}, BCDMS
  \cite{Benvenuti:1987az}, SLAC \cite{Gomez:1994ri} and and FNAL (E665
  collaboration \cite{Adams:1995is}) with only statistical errors
  shown.  Nuclear targets are specified in the brackets on the plot
  legends.  }
\end{center}
\end{figure}

DIS from different nuclear targets has been studied with
electromagnetic $\mu/e$ probes at CERN, SLAC, and FNAL (for a recent review
and references, see \cite{Arneodo:1994wf,Piller:2000wx}). It was
observed that heavy-target SFs differ substantially
from those of light nuclei in a wide kinematical region of $x$ and
$Q^2$. Figure~\ref{emc_ratio} presents a compilation of data on the
so-called EMC ratio (a traditional measure of the magnitude of nuclear
effects in DIS), $F_2^A/F_2^D$, where $F_2^A$ and $F_2^D$ are the
SFs per nucleon of a nucleus with mass number $A$ and
of deuterium, respectively. One passes through several distinct
regions with characteristic nuclear effects when going from small
to large $x$. At $x<0.1$ one observes a systematic reduction of the
nuclear SFs, the so-called nuclear shadowing. This is
illustrated in the right-hand panel of Fig.~\ref{emc_ratio}, showing the
EMC ratios on a logarithmic scale. A small enhancement appears there at
$0.1<x<0.3$, followed by a dip at $0.3<x<0.8$, which is usually
referred to as the `EMC effect', and finally an enhancement, which is
associated with nuclear Fermi motion.

%% WHAT ABOUT OTHER OBSERVABLES?

Experimental information about nuclear effects in other DIS observables,
such as the ratio of longitudinal to transverse cross sections or the
spin SF $g_1$, is available but scarse.
We note also that DIS off nuclear targets is characterized
by additional (new) SFs, which do not appear in DIS
off an isolated nucleon. As an example we refer to the tensor
SF $b_1$,
which is specific for spin-1 targets and appears in DIS on deuterium
(for a review and references see for instance~\cite{Piller:2000wx}).

%% WHAT ARE MECHANISMS FOR NUCLEAR EFFECTS?

\subsection{Nuclear shadowing}
Before we turn to the discussion of the DIS regime, it is useful to
discuss the low-$Q^2$ region away from scaling. Here the behaviour of
neutrino cross sections (SFs) is quite different from
that of charged leptons.  For the latter it is well known that the
longitudinal SF $F_L$, as well as $F_2$, vanish at low
$Q^2$, because of electromagnetic current conservation.  It was shown
long ago by Adler that, at low $Q^2$, CC neutrino
interactions are dominated by the axial current, and neutrino
cross sections can be expressed through PCAC in terms of pion cross sections
\cite{adler64}.  In contrast to charged-lepton scattering, $F^{\nu}_L$ is
finite, dominated by a pion pole for $Q^2$ at the pion mass scale, and
drives the neutrino cross section in this region.  Using the Adler
relation, Bell predicted nuclear-shadowing effects for neutrino
scattering similar to what is observed in pion--nucleus interactions
\cite{bell64}.  Going to larger $Q^2$ brings a finite contribution
from vector and axial-vector meson states, which have been discussed
in terms of an extension of the vector meson dominance model to vector
and axial-vector currents \cite{PS70,Kopeliovich:1993ym}.
Charged-current neutrino interactions with nuclear targets were studied in
bubble-chamber experiments \cite{Allport:1989vf}, where nuclear
shadowing was observed at low $Q^2$.

Most of the attempts to understand nuclear shadowing are based on the
space-time picture of DIS at small $x$ in the target rest frame, where
DIS is viewed as the process of interaction of the partonic (or
hadronic) component of the exchanged $\gamma^*$ or $W^*$ with the
target.  At small $x$ the typical propagation length of those states
exceeds the average distance between bound nucleons, and
coherent effects in the propagation of partons through the nuclear
medium are important.  Nuclear shadowing is usually explained by
multiple-scattering effects 
from bound nucleons \cite{Piller:2000wx}.

Nuclear shadowing in $F_2^\nu$ was calculated for both low and high
$Q^2$ regimes in terms of two different models in~\cite{Boros:1998mt}
in an attempt to match muon (NMC) and neutrino (CCFR),
data~\cite{Seligman:1997mc} on $F_2$ at small
$x$.
\footnote{This disagreement has recently been resolved
by CCFR/NuTeV~\cite{Yang:2000ju} who employed, among other things,  a
proper treatment of the charm mass threshold
effects~\cite{Baroneetal,Barone:2000yv}.
The ratio of the new $F_2$ values measured in
$\nu_{\mu}$ and $\mu$ scattering is
now in agreement with the NLO predictions, which use the massive charm
production scheme~\cite{Thorne:1998uu}  implemented in the
MRST parton distributions set \cite{Martin:1998sq}.}
It was found  that
nuclear shadowing in $F_2^\nu$ is similar (though slightly
smaller in magnitude) to that observed in muon-induced
reactions.

\begin{figure}[ht]
%\label{shadowing_kul}
\begin{center}
\includegraphics[width=0.7\textwidth]{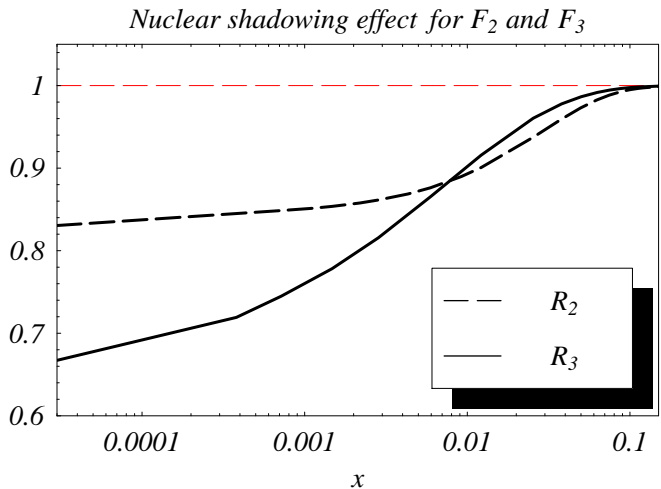}
\caption{\label{shadowing_kul}
  The ratios of a heavy target to the free nucleon SFs
  $R_2=F_2^A/F_2^N$ and $R_3=F_3^A/F_3^N$ calculated for the
  ${}^{56}$Fe nucleus in the region of small $x$ at
  $Q^2=10~\rm{GeV}^2$~\cite{Kulagin:1998wc} }
\end{center}
\end{figure}

At large $Q^2$ in the scaling regime both charged-lepton and
neutrino-induced  reactions are described by universal parton distributions.
Some observations on nuclear modifications of different combinations of parton
distributions can be made from existing charged-lepton DIS and
Drell--Yan data. Phenomenological constraints on the behaviour of
nuclear sea and valence quarks at small $x$ were discussed in
\cite{Frankfurt:1990xz,Kobayashi:1995ka,Eskola:1998iy}.  Explicit
evaluations of  nuclear effects in singlet and non-singlet 
combinations of parton distributions 
were performed in \cite{Kulagin:1998wc}, where nuclear
shadowing for $F_2^{\nu}$ and $xF_3$ was studied in terms of the
non-perturbative parton model of \cite{Landshoff:1971ff}, which was extended and
applied to nuclear targets in \cite{Kulagin:1994fz}.  It was found
that, while the shadowing effect in neutrino $F_2$ is similar to the
corresponding effect in charged-lepton DIS, 
the nuclear shadowing for $xF_3$ 
is enhanced with respect to that for $F_2$ in the region of small 
$x<0.01$ (see Fig.~\ref{shadowing_kul}).  
It was argued in \cite{Kulagin:1998wc} that
the underlying reason for the enhancement of nuclear shadowing
 for $xF_3$ is its negative $C$-parity.
In the small-$x$ region, $xF_3$ is determined by the
difference of effective quark and antiquark cross sections, and it is
known from the multiple-scattering   
theory that the double-scattering correction to
the difference of the cross sections is up to a factor of 2 larger
than the corresponding correction to their sum.

We note in this respect that a similar enhancement of nuclear
shadowing was predicted for the spin SF $g_1$
(see discussion in \cite{Piller:2000wx}), which involves the
differences of quark and antiquark distributions with helicities
parallel and antiparallel with respect to the helicity of the target.

\subsection{Nuclear effects at large $x$ }
The physics mechanisms that generate characteristic nuclear effects at
large $x$ are quite different from those that govern nuclear
shadowing at small $x$. At large $x$ the typical DIS time scale in
the laboratory reference frame is small with respect to an average
distance between bound nucleons. This allows us to assume that nuclear
DIS is dominated by incoherent scattering from bound nucleons.
It was found long ago that major nuclear effects here are due to
nuclear binding
\cite{Kulagin:1994fz,Akulinichev:1985ij,Kulagin:1989mu}, which leads
to a depletion of nuclear SFs at $x\sim 0.5$, and to
the Fermi motion \cite{west72}, which is responsible for the enhancement
at $x>0.7$.  These effects explain the bulk of the observed behaviour
of nuclear SFs at $x>0.2$, though detailed
understanding of this region is far from complete and further studies
of the reaction mechanism are required.
%%\cite{Kulagin:1994fz,Kulagin:1989mu}.

It is quite important to separate nuclear effects in a QCD analysis of
neutrino data. As an example we refer to the recent analysis of
higher-twist terms in the CCFR data on the $xF_3$ SF
\cite{Kulagin:2000yw}.  The theoretical and experimental situation
becomes less clear when going to the region of $x$ close to 1.  Here
we enter into the resonance region, and the notion of twist expansion
becomes less well defined.  Nuclear effects in SFs are
essential in this region, because the nucleon inelastic SFs
 must vanish as $x \; \to \; 1$. 
The impact of nuclear effects on HT terms in $xF_3$ was studied
in \cite{Kulagin:2000yw}, where it was shown that the consideration of 
nuclear effects at large $x$  somewhat decreases the magnitude of
dynamical HT terms extracted from CCFR data on iron target.
However, the results of calculations of nuclear corrections are
sensitive to the details of the nuclear structure input, \eg\ the behaviour of
the nuclear spectral function in the region of high excitation energy of
the residual nuclear system, which are not known yet.
%
%In any case, in order to study the possibility of extracting
%the $x$-shape of the dynamical  HT contributions
%it is highly desirable
%to use light nuclear targets in the experiments at the \nufact.
In order to minimize uncertainties associated with nuclear effects,
it is desirable to use light nuclear targets in the experiments at the \nufact.

We note also that nuclear SFs can extend beyond $x=1$,
the kinematical limit for scattering from a free nucleon.  Events with
$x>1$ have indeed been observed for $F_2$ in $\mu$ DIS from a carbon
target by the BCDMS collaboration \cite{Benvenuti:1994bb} and recently
by the CCFR collaboration in neutrino DIS from an iron target
\cite{Vakili:2000qt}. It is rather interesting to search for similar
events at the \nufact, where the statistics will be much higher.

The region of large $x$ will be explored in more detail in 
electron scattering experiments
at Jefferson Lab.
However the experiments at the \nufact\ at large
$x$ with different nuclear targets are challenging and could
significantly contribute to the field by providing a direct
measurement of the EMC effect for different parton combinations, such
as the $C$-even $F_2$ and $C$-odd $xF_3$.

\subsection{Nuclear effects in DIS sum rules}
As discussed in Section~\ref{sec:GLS}, the GLS sum rule is a convenient tool
to  extract $\as$ from neutrino data.
Neutrino data are usually collected on heavy nuclear targets;
therefore, it is of importance to separate contributions to the GLS
integral associated with nuclear effects.
We denote
$S_{\rm GLS}^A = S_{\rm GLS}^N + \delta S_{\rm GLS}$,
where $S_{\rm GLS}^A$ and $S_{\rm GLS}^N$ are
the GLS integral for the nucleus of $A$ nucleons and the isoscalar
nucleon respectively, and $\delta S_{\rm GLS}$ accumulates corrections
due to nuclear effects. There is a number of effects that
 can contribute to $\delta S_{\rm GLS}$.
Contributions due to nuclear binding, Fermi motion and off-shell
effects were discussed in \cite{Kulagin:1998vv}.
It was found that these effects cancel out in
the leading twist, and a tiny correction appears as a higher twist.
For example, 
$\delta S_{\rm GLS}^{\rm Fe} = -1.2\times 10^{-2}\mathrm{GeV}^2/Q^2$ and
$\delta S_{\rm GLS}^{D} = -1.9\times 10^{-3}\mathrm{GeV}^2/Q^2$
for
the iron and deuterium nuclei respectively.
These quantities are more than an order of magnitude
smaller than the corresponding QCD power correction estimated in ~\cite{BKol}.
However, as was discussed in \cite{Kulagin:1998wc}, nuclear
shadowing gives a finite and rather large negative correction
already in the leading twist.  In particular, a $\sim 4\%$
renormalization of the GLS sum rule due to nuclear shadowing
was found for the ${}^{56}$Fe nucleus,
$\delta_{\rm sh}S_{\rm  GLS}/S_{\rm GLS}^N = -0.035$,
at $Q^2=10$~GeV$^2$.

At this point we must mention that it is usually believed that the
GLS sum rule is not  renormalized by nuclear effects in the
leading in $1/Q^2$ order, since to this order the GLS integral counts
the baryon number of the target.  However, a negative sign of the
shadowing correction is also a generic feature of multiple scattering
theory.  It is therefore challenging to look for a dynamical mechanism that
would compensate a negative nuclear shadowing correction in the GLS sum
rule. Certainly more work is needed to clarify the status of the GLS
sum rule, as well as other DIS sum rules,  in  nuclear targets.

The Bjorken sum rule for the non-singlet combination of
the neutrino SFs $F_1$ was discussed in Section~\ref{sec:Bj}
as an alternative tool to extract $\as$.
Neutron data are necessary in order to measure this sum rule,
and it is clear that, since there exists no free
neutron target, nuclear data have to be used.
Combined hydrogen and deuterium data are usually used as a source of
information about the neutron.
The integrated difference between the hydrogen and
the deuteron SF
could then be used to extract
$S_1^{d-p} = S_1^{n-p}+\delta S_1$, where the last term incorporates
nuclear effects. Using the method described in \cite{Kulagin:1998vv},
it can be shown that the corrections due to nuclear binding and Fermi motion
 cancel out in the leading twist, similar to the GLS sum rule,
and the corresponding power $1/Q^2$ correction is small, of the same
order of magnitude as in the GLS sum rule.

%%Corrections due to nuclear shadowing and meson exchange currents in the deutron are currently under study.

Note also that the deuteron nucleus, in spite of a weak binding,
might not be a perfect source of information about neutron SFs, 
especially at very large or very small $x$.
It  was emphasized recently that DIS experiments with mirror
${}^3$He and ${}^3$H nuclei, which form an isotopic
doublet, could give information about the neutron SF practically
free from contamination from nuclear effects \cite{Afnan:2000uh}.
We then suggest that the direct measurement of the difference
$S_1({}^3\mbox{He})-S_1({}^3\mbox{H})$
could be a better source for the Bjorken sum rule
than the corresponding hydrogen--deuterium difference. 
The calculation of corrections to the Bjorken sum rule,
as well as to the more fundumental Adler sum rule, 
due to meson-exchange currents
in nuclei and nuclear shadowing, is under 
way\footnote{S.A.~Kulagin, work in progress.
During the course of writing this report we learned about 
the paper \cite{guzey01}, where the nuclear shadowing corrections
to the Gottfried sum rule for
${}^3\mbox{He}- {}^3\mbox{H}$ mirror nuclei were discussed.}.

%% ADVANTAGES OF NU FACTORY

In summary we note that  because of a larger number
of observables, which can be accessed with $\nu$ and $\bar \nu$ beams,
DIS studies at the \nufact\ can give unique information about
the structure of hadrons and nuclei, information
that is not accessible with $\mu/e$ machines.
In particular, a unique opportunity offered by the \nufact\
is a direct measurement of nuclear sea and valence quarks distributions 
in a wide kinematical region.

\section{ELECTROWEAK STUDIES AT THE $\mathbf{\nu}$-FACTORY}
Experiments with $\nu e^-$ and $\nu N$ have played a fundamental role in
establishing the SM. Before the start of LEP, the best determinations
of the electroweak mixing angle came from neutrino experiments, and
even now the results from neutrino DIS at NuTeV play an 
important role in  global analyses.

The characteristics of a \nufact\ are such that very precise 
tests of the SM and its extensions may be possible  
from  neutrino--electron scattering and neutrino induced DIS. 
As a  first approximation, the information available from these
experiments can be parametrized in terms of the uncertainty in the
determination of the sine of the Weinberg angle, $s_W^2\equiv
\sin^2\theta_W$. In fact, radiative corrections enter the two
processes in different ways and one should look at these experiments
as complementary measurements, akin to the present determinations of
$\sin^2\theta_{\mathrm{eff}}^{\ell}$ and $M_W$ at $e^+e^-$ and hadron
colliders. Moreover, a very precise low-energy
determination of $s_W^2$ would test 
a different variety of new-physics scenarios than usual colliders.
 
\subsection{$\mathbf{\nu e^-}$ scattering}
$\nu e^-$ scattering provides a particularly clean probe of the
electroweak coupling. There are several processes which contribute 
at a \nufact:
\ba
({\rm NC})&&\nu_\mu e^-\to \, \nu_\mu e^- \; , \ \ \
\bar{\nu}_\mu e^-\to \, \bar{\nu}_\mu e^- \; ,\\
({\rm NC+CC}) && \nu_e e^- \to \, \nu_e e^- \; ,\label{nccc}
\ \ \ 
\bar{\nu}_e e^- \to \, \bar{\nu}_e e^-,\ \ \  \bar{\nu}_\mu \mu^- \;, 
\;\dots \; ,\\
({\rm CC}) &&\nu_\mu e^-\to \, \nu_e \mu^- \; .
\label{cc}
\ea
Events originated by $\nu_\mu$ or $\bar{\nu}_e$ in a $\mu^-$ beam
without a muon in the final state
cannot be disentangled at a \nufact\  and must be considered together. 

In the case of $\nu e^-\to \; \nu e^-$ processes, 
numerical values for the total cross sections
are given by 
\be
\sigma(\nu e^- \to \; \nu e^-)= 
1.72 \times 10^{-41} {\rm cm^{-2}}\times E_\nu[\mathrm{GeV}]
\times \left[ g_L^2 + \frac13 g_R^2\right],
\ee
where $g_{L,R}= s_W^2$ or $g_{L,R}= \pm 1/2 +s_W^2$ according to the process. 

Despite  the very small cross section, the use of 
 a 2 ton dedicated, fully active target--detector  made of liquid CH$_4$
 \cite{bkbook} or of a 20 ton liquid argon \cite{Albright:2000xi} 
time projection chamber 
should provide around 10$^7$ $\nu e$ events/year with a $\mu^+$ beam and 
about half of it with  a $\mu^-$ beam. In this subsection
we use  default beam specifications, a 20 cm detector radius,  
$s_W^2=0.2314$, and  $\int L dt=8.6 \times  10^{46}$ cm$^{-2}$. 

The signal is a forward electron track with no hadronic activity and
energy above a threshold $E_{min}$. The transverse momentum of the
outcoming electron is very small, $p_t\sim \sqrt{m_e E_\nu}$.
In the configuration considered here
the transverse momentum due to the intrinsic divergence of the beam
is even smaller. 
The main source of background  is quasi-elastic $\nu_e N$ scattering, 
which can also produce a forward  electron, but  is characterized by $p_t\sim 
\sqrt{m_N E_\nu}$ and can be distinguished if the $p_t$ resolution of
the detector is  good. The signal-to-background ratio is expected
to be better than  5 at a 50 GeV \nufact\  \cite{Albright:2000xi}, leading to
a minor dilution of the sensitivities considered below. The $\mu^-$
beam has the advantage that quasi-elastic $\nu_e N$ scattering
produces positrons instead of electrons, so that this source of
background can be removed.

\begin{figure}[ht]\vspace{-.2cm}
\begin{center}
\includegraphics[width=0.45\textwidth,clip]{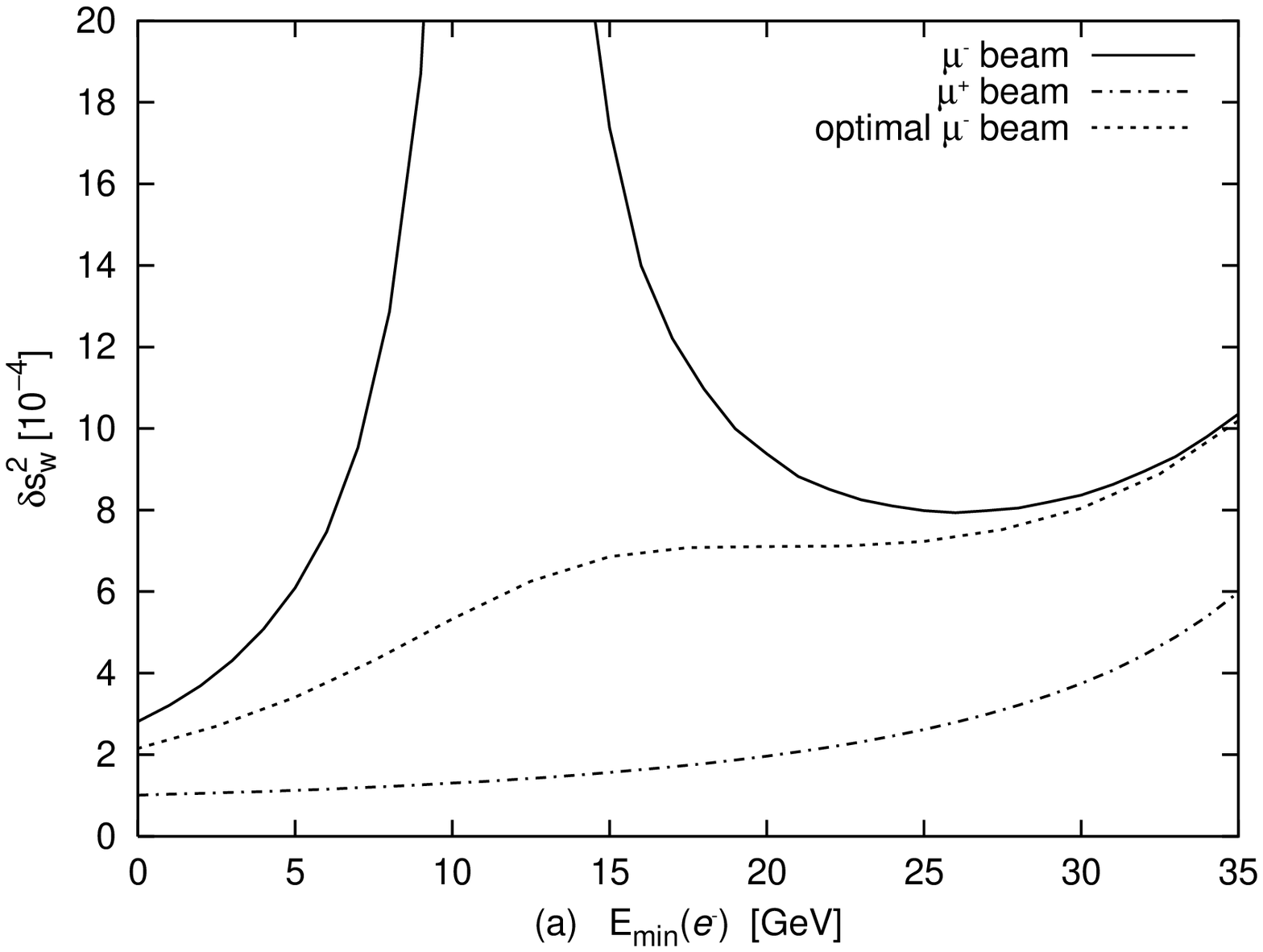} \hfill
\includegraphics[width=0.45\textwidth,clip]{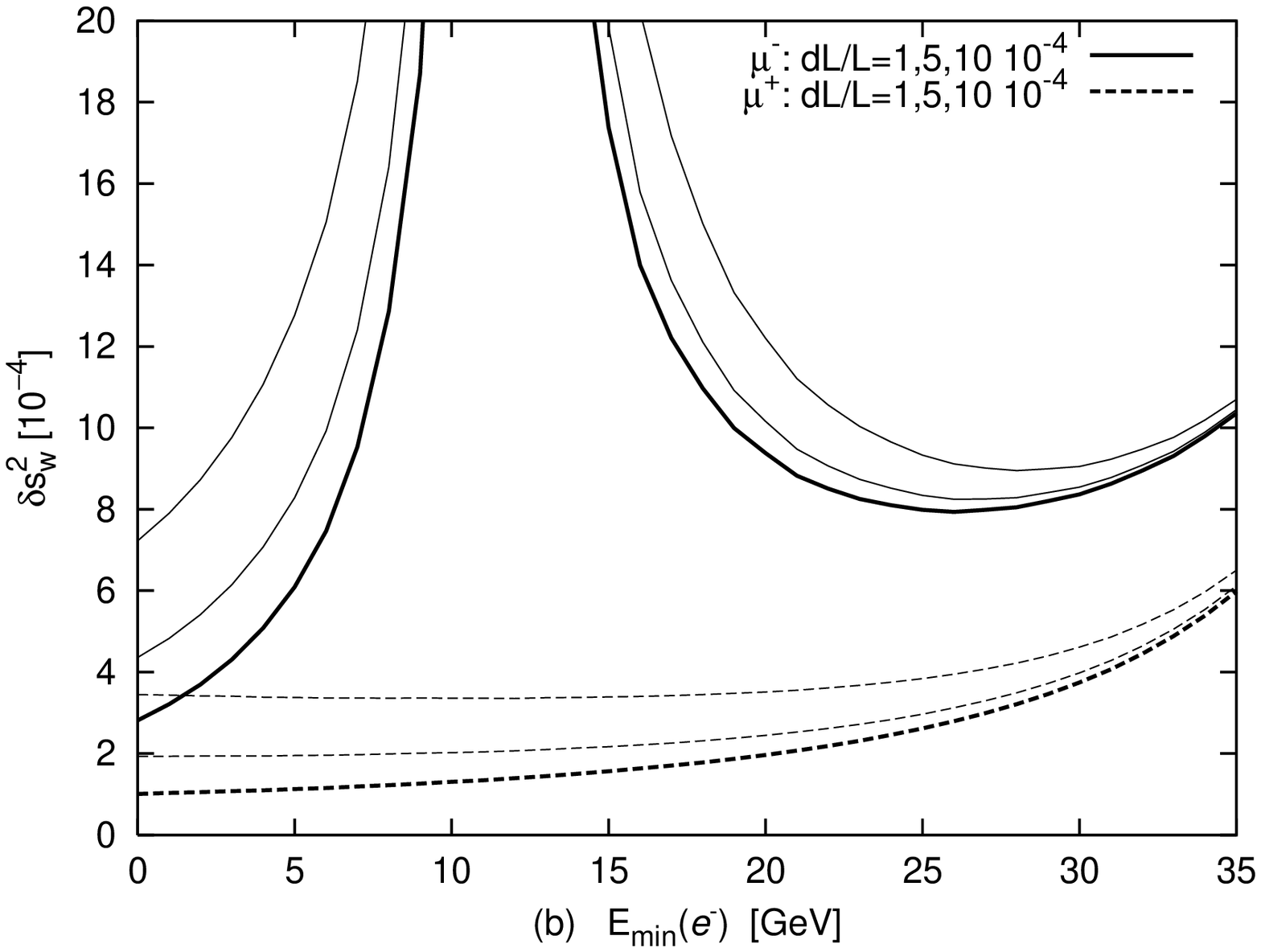} 
%\mbox{\epsfxsize=8cm\epsffile{gambino1.eps}}
%\hspace{-0.5cm}\mbox{\epsfxsize=8cm\epsffile{gambino2.eps}}
\end{center}
\vspace{-.6cm}
\caption{%\sf 
(a) statistical uncertainty (in units $10^{-4}$) in the
  extraction of $\sin^2\theta_W$ from $\nu e $ scattering as a
  function of the minimum electron energy.
(b) impact of luminosity measurement at the level of
  $10^{-3}$, 5 $10^{-4}$, 1 $10^{-4}$ on the same  $\sin^2\theta_W$
sensitivities.
\label{fig:gambino}}
\end{figure}

The statistical sensitivity to $s_W^2$ is shown in
 Fig.~\ref{fig:gambino}a  as a function of
$E_{min}$ for the $\mu^-$ and $\mu^+$ beams. When only
integrated cross sections are considered, the $\mu^+$ beam allows for a
superior $s_W^2$  resolution, very close to 1$\times 10^{-4}$, while  
the $\mu^-$ beam is much less sensitive
because of a cancellation among different terms. Assuming that a
measurement of the outgoing electron energy $E_e$ is possible, 
one can study the
$E_e$ dependence and achieve a better resolution in the $\mu^-$
case, $\delta s_W^2\approx 2\times 10^{-4}$. This  is close to having 
 an optimal observable. 
For small $E_{min}$, the sensitivity is a mild function of the threshold
energy.

A significant problem in $\nu e $ scattering is  the normalization of the 
cross sections.
As can be seen from Fig.~\ref{fig:gambino}b, in order to preserve 
the statistical sensitivity one
would need a luminosity determination at the level of 10$^{-4}$. 
The goal at a \nufact\  would be to
reach  a precision on the  flux of $10^{-3}$; this, however, seems
 inadequate.
 A realistic possibility is to normalize the $\nu e$ rate 
to $\nu \mu^-$, namely to the muon regeneration process (see
Eqs.~(\ref{nccc}),(\ref{cc})), which occurs for $E_\nu> m_\mu^2/2
m_e=10.8$ GeV with  cross sections given by 
\be
\frac{d \sigma}{dy}(\bar{\nu}_e e^-\to \mu^-\bar{\nu}_\mu)=
\frac{G_\mu^2}{\pi} s (1-y)\left(1-y+\frac{m_\mu^2}{s}\right); \ \ \ \ 
\frac{d \sigma}{dy}(\nu_\mu e^-\to \mu^-\nu_e)=
\frac{G_\mu^2}{\pi} \left(s-m_\mu^2\right).
\ee
Again, the  background from $\nu_\mu N$ scattering should be considered, 
but it can be drastically reduced  by a cut on the transverse momentum 
\cite{bkbook}.
This mechanism could 
provide a normalization accuracy of $3.6 \times 10^{-4}$,
but would be available only for the $\mu^-$ beam. 
On the other hand, it could serve to  calibrate  different luminosity
measurements.

We have also investigated the case in which the muon beams are
polarized and concluded that muon polarization would not add
significantly to the electroweak measurement. At best,
polarization asymmetries could help overcome the normalization problem.
With  70\% polarized muon beams the  statistical error is 
$\delta s_W^2\approx 3 \times 10^{-4}$ for the $\mu^-$ beam, and worse
for the $\mu^+$ beam.
On the other hand,  uncertainties in the polarization of the $\mu$ beam 
do affect the final precision. In order to maintain the final
precision at the level of $\delta s_W^2\approx 
10^{-4}$, it is necessary to know the
polarization of the muons to 0.05\% or better. 
 For this reason precision electroweak 
measurements require a storage ring design that minimizes these 
uncertainties. 

From the theoretical point of view, all one-loop QED and electroweak 
corrections are known \cite{nue1loop} and could be easily
implemented. The uncertainty from higher-order electroweak 
corrections can be certainly brought below $\delta s_W^2=1 \times 10^{-4}$ 
\cite{noi}.
QED corrections to the $e^-,\mu^-$ spectra are
relatively important and may need consideration of some higher order
effect. 
A major theoretical uncertainty is likely to come from the 
hadronic contribution to the $\gamma$--$Z^0$ amplitude, which must be
calculated at the relevant $Q^2\approx 10^{-3}{\rm  [GeV]} E_\nu$, a
region where perturbation theory cannot be applied. The
problem is analogous to --- but should not be confused with ---
the one of the hadronic contribution to the running of the electromagnetic
coupling, $\Delta\alpha_{hadr}$, which enters most electroweak tests.
$\Delta\alpha_{hadr}$ is needed to relate the fine-structure constant
$\alpha(q^2=0)$ to $\alpha(M_Z^2)$, which is relevant at the $Z^0$ pole
and for the electroweak corrections to muon decay 
as well as for low-energy NCs.
The latter, however, have an additional sensitivity to hadronic loops
in the $\gamma$--$Z^0$ mixing. This contribution can be calculated from
$e^+ e^-$ data using dispersion relations, SU(3) flavour symmetry,
 and perturbative QCD.
  The most recent estimate  \cite{Bahcall:1995mm}
 leads to $\delta s_W^2=5 \times 10^{-4}$. In view of recent 
progress~\cite{jegerlehner}, 
this can probably be reduced by a factor of 2 or more.
However, the use of SU(3) flavour symmetry implies a sizeable
ambiguity, which cannot be resolved by better data only. 
For what concerns  the uncertainty in 
 $\Delta\alpha_{hadr}$, it  should not be
considered a limiting factor for this experiment, as it affects
in the same way  most electroweak  observables. Moreover,  there has been and
there will be progress in its determination, and it is likely to play a
lesser role than the uncertainty from the $\gamma$--$Z^0$ mixing in $\nu
e$ scattering.

In conclusion, a total uncertainty on $s_W^2$ of about $2 \times 10^{-4}$
is probably achievable at a \nufact, in the case
where high-performance detectors are available, the 
polarization of the muons can be  controlled very precisely, and 
progress is achieved in estimating the hadronic effects. Higher precision
would require a substantial increase in luminosity
as well as major theoretical 
improvements, mainly  in hadronic physics.

\subsection{$\mathbf{\sin^2 \theta_W}$ from DIS}
Current electroweak analyses of $\nu N$ DIS (at NuTeV) are based on
the Paschos--Wolfenstein (PW) relation, 
\be
R^-=\frac{\sigma_{NC}({\nu}_\mu)-\sigma_{NC}(\bar{\nu}_\mu)}{
\sigma_{CC}({\nu}_\mu)-\sigma_{CC}(\bar{\nu}_\mu)}=\frac12 - s^2_W \; ,
\label{eq:PW}
\ee
which is designed to isolate the $u,d$ valence quark contributions 
that cancel out in the ratio
 and is therefore quite insensitive to the hadron
structure.  On the basis of about 1.3 (0.3) million $\nu_\mu$
($\bar{\nu}_\mu$) events, NuTeV has found \cite{nuTeV}
\be
s_W^2({\rm OS})=0.2253\pm 0.0019 ({\rm stat}) \pm 0.0010 ({\rm syst})
\pm 0.00025 (M_t)+0.0005 \ln\frac{M_H}{150{\rm GeV}} \; .
%\pm 0.0005(M_t,M_H)
\ee
As can be seen, if the on-shell definition $s^2_W({\rm OS})\equiv
1-M_W^2/M_Z^2$ 
is used, the indirect dependence of the result on the top and Higgs
masses through radiative corrections to \equ{eq:R} is relatively small. 
This is why the NuTeV result is often presented as an $M_W$
determination with $\delta M_W\approx 130$ MeV. This interpretation
can  be highly misleading 
when the experimental accuracy reaches ${\cal O}(10^{-4})$.

At a  \nufact, a striking improvement of the
statistical error can be expected, as well as  
the elimination of the main systematic problems of NuTeV
(uncontrolled $\nu_e$ beam contamination and NC event identification).
Whether the final $s_W^2$ sensitivity may reach the level of $10^{-4}$
or $\delta M_W\approx 5$ MeV depends, however, on many different factors.

If the detector can  identify primary $e^\pm$ in the final state,
the CC events originated by electron
neutrinos can be distinguished from NC events. One can then consider
ratios of NC/CC total cross sections \cite{bkbook}
\ba 
R^{\mu^-}=\frac{\sigma_{NC}(\nu_\mu) + \sigma_{NC}(\bar{\nu}_e)}{
\sigma_{CC}(\nu_\mu) + \sigma_{CC}(\bar{\nu}_e)}; \ \ \ \ 
R^{\mu^+}=\frac{\sigma_{NC}(\bar{\nu}_\mu) + \sigma_{NC}({\nu}_e)}{
\sigma_{CC}(\bar{\nu}_\mu) + \sigma_{CC}({\nu}_e)}
\label{eq:R}
\ea
Using our default beam specifications and $E_{lept}>3$ GeV, $E_h>1$
GeV, $s^2_W\equiv\sin^2\theta_W=0.225$,
we find $R^{\mu^-}\approx R^{\mu^+}=0.36$. Also the  
$s^2_W$ sensitivity of the two beams is very similar,
 $d R/d s^2_W\approx -0.5$.
Because of the   $10^9$ CC  and $3 \times 10^8$ 
NC events available for each beam, 
the statistical error on the $s^2_W$ determination from \equ{eq:PW}
is negligible, as shown  in Table~\ref{tab:pg1}.
However, one should take into account that $R^{\mu^\pm}$ have a higher
sensitivity to hadronic physics  than the PW relation.
As an illustration, we show in Table~\ref{tab:pg1} the sensitivity of
the $s_W^2$ determinations from $R^{\mu^\pm}$ on  present PDFs.
The values are  obtained by comparing  results for $R^{\mu^\pm}$ for which 
different sets of the CTEQ5 PDFs \cite{Lai:1997mg} (sets m, d, and hq)
have been employed. 
They are the result of a LO analysis based on 
present day information. Clearly, future improvements in our knowledge of
the hadron structure at the \nufact\  and
elsewhere, as well as a full NLO implementation of QCD radiative
corrections will  lower this uncertainty, but it seems
unlikely that they will bring it down to the level of the statistical error.

On the other hand, the PW relation is formally recovered in the combination
\be
(1+g\, r) \,R^{\mu^-}-( r+g)\, R^{\mu^+},
\label{eq:PW2}
\ee
where $g=\langle E_{\bar{\nu_e}}\rangle/\langle
E_{\nu_\mu}\rangle=0.857$ takes into account the
different mean energy of the neutrino and antineutrino beams,
$r=\sigma_{CC}(\bar{\nu})/\sigma_{CC}(\nu)$, and lepton universality
has been used.
Unfortunately, because of  the numerical closeness of $R^{\mu^-}$ and 
$ R^{\mu^+}$, this combination is not an efficient probe. Moreover
the $s_W^2$ sensitivities of $ R^{\mu^\pm}$ are also very similar and
 the improvement in the
PDF sensitivity is  paid for by the lower statistical sensitivity.
One possibility is to  fit a parameter 
$\alpha$ in $R^{\mu^-}-\alpha R^{\mu^+}$ in order to minimize the overall
$s_W^2$ uncertainty. For instance,  $\alpha\approx 0.8$ leads to 
 $\delta s_W^2({\rm PDF})\approx \delta s_W^2({\rm
  stat})\approx 0.0002$. 

\renewcommand{\arraystretch}{1.2}
\begin{table}[t]
\begin{center}
\begin{tabular}{|c|c|c|} 
\hline 
Observable & Stat. error & PDF 
 \\ \hline
$R^{\mu^-}$ & 0.4 & $\sim$ 12 
 \\ \hline
$R^{\mu^+}$ & 0.5 & $\sim$ 15 
\\ \hline
$R^{\mu^-}-0.8R^{\mu^+}$  & 2.2 & $\sim 2$ 
\\ \hline
$P$ & 4.9 & $\sim$ 4
\\ \hline 
%$\hat{R}^{\mu^-}$ & 1.0 & 222 
%\\ \hline
%$\hat{R}^{\mu^+}$ & 1.3 & 598
%\\ \hline
\end{tabular} 
\caption{Uncertainties on the $s^2_W$ determination from different
  observables in $\nu$ DIS in units $10^{-4}$
   (see text). \label{tab:pg1}}
\end{center} 
\end{table} 

Of course, one can construct other ratios of cross sections, also
combining the cross sections of the two beams in a way similar to
\equ{eq:PW}; however,
they tend to be less sensitive to $s^2_W$ and/or to depend even more
than $R^{\mu^\pm}$ on the hadronic structure. 
A possible exception could be the quantity
\be
P= \frac{\sigma_{NC}(\mu^-) - \sigma_{NC}(\mu^+)}{
\sigma_{CC}(\mu^-) - \sigma_{CC}(\mu^+)} \; ,
\label{eq:P}
\ee
which is modelled on the PW relation and has indeed a lower
sensitivity on hadronic physics (see Table~\ref{tab:pg1}). However, this
observable  is penalized
by  a lower $s_W^2$ resolution and requires the knowledge of 
the ratio $\sigma_{CC}(\mu^+)/\sigma_{CC}(\mu^-)$ --- and therefore of
the relative normalization of the two beams --- to better than $10^{-4}$. 

These high-precision measurements cannot be carried out without 
a high-performance tracking target. 
The detector and beam requirements  
 based on \eqs{eq:R} have been discussed in
\cite{bkbook}. The main ones are high electron/muon  detection efficiency,
 good primary lepton  identification, and precise momentum and energy
 measurements. A rough estimate \cite{bkbook}
of the corresponding uncertainties is $\delta s_W^2\approx 2 \times 10^{-4}$.
The ability  to have both $\mu^+$ and $\mu^-$ beams  would be an 
important asset, as it would allow the use of \equ{eq:PW2}. 

From a theoretical point of view, it will be necessary 
 to implement full NLO QCD
and QED corrections, to incorporate possibly complete two-loop 
electroweak corrections, to assess the importance of higher twists, and
to take into account charm production corrections. This is to be
contrasted with the obsolete and incomplete implementation of QCD and
electroweak corrections of the NuTeV analysis. On the other hand,
almost all theoretical tools are in principle already available,
including the dominant two--loop weak corrections \cite{noi}. 
Charm production, in particular, is likely to be a
significant source of uncertainty; it accounts for about 3\% of the 
total CC cross section and therefore even a 1\% error on the 
 overall charm yield would induce a $10^{-4}$ uncertainty on
 $R^{\mu^\pm}$ or $\delta s_W^2\approx 2\times 10^{-4}$.
An extrapolation from the CCFR measurements suggests
 $\delta s_W^2\sim 0.0003$ \cite{bkbook}.

If the outcoming electrons cannot be resolved and
CC events initiated by electron neutrinos are considered as NC events
--- a situation analogous to present-day $\nu N$ experiments ---
the only observable for the $\mu^-$ beam is
\be
\hat{R}^{\mu^-}= \frac{\sigma_{NC}(\nu_\mu) + \sigma_{NC}(\bar{\nu}_e)+
\sigma_{CC}(\bar{\nu}_e)}{\sigma_{CC}(\nu_\mu)} \; .
\ee
This observable and the analogous one for the $\mu^+$ beam have a
marginally lower statistical sensitivity and  a
much higher dependence on the SFs, $\delta
s_W^2> 0.02$, than $R^{\mu^\pm}$. 

In conclusion, efficient electron identification seems to 
be crucial for a very precise
determination of $s^2_W$ at the \nufact\ . The statistical
uncertainties are likely to be negligible in comparison with
theoretical and experimental 
systematic errors, which in turn are difficult to estimate at the moment.
Before a realistic assessment  of the potential of $\nu$ DIS at a \nufact\ 
for precision electroweak tests is possible,
several systematics will have to be understood;
most importantly the precision of the \nufact\
 measurement of the unpolarized SFs,
primary lepton identification, isospin--violating effects 
and charm production. 

\subsection{New physics through radiative corrections}
The very precise determination of the electroweak mixing angle
from $\nu e$  and $\nu N$ scattering at a \nufact\ would test
the SM at a level competitive with LEP, SLD and Tevatron measurements.
Experiments at the front end of a muon storage ring 
could therefore severely constrain many extensions 
of the SM, which potentially  affect the SM predictions
for neutrino scattering through virtual contributions.
If the  new physics is characterized by a high mass scale and it 
affects dominantly the 
two-point functions, its contributions can be parametrized in
a model independent way by the $S, T, U$ amplitudes introduced in
\cite{stu}. Similar strategies have been presented by a number of
authors (see for instance~\cite{others}).
Many models of physics beyond the SM can be studied at
least approximately in this simple way.

We use $M_H\approx100$~GeV and $\hat{s}_W^2=
\sin^2\theta_W(M_Z)_{\overline{\rm MS}}= 0.2315$  as reference SM
values and set $g =\langle E_{\bar{\nu_e}}\rangle/\langle
E_{\nu_\mu}\rangle=0.857$. 
As far as $\nu e$ scattering is concerned, in the case of 
the $\mu^-$ beam we can  consider the
ratio of electron- and muon-production total cross sections
\be 
r^-= \frac{\sigma(\nu e^-)}{\sigma(\nu\mu^-)} \; .
\ee
Virtual effects parametrized by $S$, $T$, and $U$ shift the measured value
of $r$ with respect to the SM prediction $r_0$ by 
\be
r^-=r^-_0\, (1 + 0.0055 \,S +0.0004 \,T) \; .
\ee
This observable is therefore very insensitive to the isospin
breaking parameter $T$, similarly to the weak charge in  
atomic parity violation experiments with cesium. 
It  could  allow for a very accurate determination
of the isospin conserving parameter $S$, within $\pm 0.1$; this is
roughly the  accuracy of the present global fit (see
Ref.~\cite{pdg2000},  p.105).
A similar observable can be constructed for the $\mu^+$ beam, provided 
the luminosity determination is based on a CC process
analogous to muon production. In this case the shift is 
\be
r^+=r^+_0 \,(1 + 0.0111 \,S -0.0100 \,T) \; ,
\ee
and implies a very different constraint in the $(S,T)$ plane.
Finally, the measured value of the Paschos--Wolfenstein relation (\ref{eq:PW})
is affected by the oblique parameters according to 
\be
R^-=R^-_0 \,( 1- 0.0136 \, S + 0.0253 \, T)
\ee
The different $\nu e $ and $\nu N$ measurements would therefore provide
complementary constraints on new physics.

\section{STUDIES WITH HEAVY QUARKS}
As has been pointed out in the past, and as shown in a previous
section, the \nufact\ can be seen as
a \nudis\ charm factory. A comprehensive review of the possible
physics goals of such a facility can be found in Ref.~\cite{bkbook}.
Here we shall explore in quantitative detail some
interesting example, drawing from the expertise available within the
Working Group. In particular we shall focus 
on two aspects of charm production in neutrino interactions:  
low-multiplicity processes such as the diffractive $D_s$ and the  
quasi-elastic charmed-baryon production.
We shall show that these 
 processes allow  a clear identification of the charmed  
hadron, and therefore a very good estimate of absolute decay branching  
ratios. Together with low systematic errors, these measurements can
provide, for instance, a precise measurement of $f_{D_s}$.
 
In the following we consider nuclear emulsions both as neutrino target
and tracking device. It is worth stressing that the use of nuclear
emulsions is limited by the overlapping of interactions. A density of
interactions of about $20$ per cm$^3$ is reasonable. On the other
hand, $10^7~\nu_\mu$ interactions are needed for the measurements we
will discuss in the following. Such a statistics could be obtained by
running the machine at low luminosity and taking data for a few years
(months) whether the experiment is located far from
($\mathcal{O}(1~\mbox{km})$) or close to ($\mathcal{O}(100~\mbox{m})$)
the neutrino source.

Our simulations include realistic estimates of the experimental
efficiencies and systematics.  It is important to point out, however,
that the methods discussed below can be used by electronic experiments
with a very good vertex detector as well, although with different
efficiencies and backgrounds.

\subsection{Direct evaluation of the $\mathbf \Lambda_c^+$ branching ratios} 
\label{chprod} 
\subsubsection{Model-independent extraction of 
$BR(\Lambda_c^+\rightarrow pK^-\pi^+)$ } 
\label{model} 
So far, only model-dependent extractions of $\Lambda _{c}^{+}$
branching ratios have been obtained, see~\cite{pdg2000}. A method,
based on the neutrino quasi-elastic charm production, for a
model-independent determination of most of the $\Lambda _{c}^{+}$  branching
ratios has 
been proposed in Ref.~\cite{lambda}. So far, two different methods to
extract $\Lambda _{c}^{+}$ branching ratios have been
used~\cite{pdg2000}. As discussed in Ref.~\cite{pdg2000} they rely on
different theoretical assumptions on $B$ physics, namely the $B$
branching ratios to $\Lambda_c$, and give results that are not in
quite a good agreement. Therefore, a model-independent 
determination of $\Lambda _{c}^{+}$ branching ratios would provide a
better theoretical understanding of the baryonic $b$-decays. For a
detailed discussion of this method, see Ref.~\cite{lambda}.

\subsubsection{Present knowledge, theoretical and experimental, of
  neutrino quasi-elastic charm production} 
The simplest exclusive charm-production reaction is the quasi-elastic 
process where a $d$ valence quark is changed into a $c$ quark, thus 
 transforming the target nucleon into a charmed baryon. Explicitly, the 
 quasi-elastic reactions are 
  
\ba
&&
\nu_\mu n\rightarrow\mu^- \Lambda_c^+(2285) \;, 
\nu_\mu n\rightarrow\mu^- \Sigma_c^{+}(2455)\;, 
\nu_\mu n\rightarrow\mu^- \Sigma_c^{*+}(2520)\;,
\\
&& \nu_\mu p\rightarrow\mu^-\Sigma_c^{++}(2455)\;, 
\nu_\mu p\rightarrow\mu^- \Sigma_c^{*++}(2520)\;. 
\ea
For a detailed theoretical review of neutrino quasi-elastic charm 
production, see Refs.\cite{mode1}--\cite{kova}. 
 
 The predicted cross sections, assuming a neutrino energy of $10$~GeV and 
according to different authors, are shown in Table~\ref{tab:crosspred}.  As we 
can see, these predictions can even differ by one order of 
magnitude. If we express the total quasi-elastic charm production rate with  
respect to 
deep-inelastic CC interactions, it ranges from $0.33\%$ 
to $9.0\%$ (see Table~\ref{tab:crosspred}).  From Fig. 3 of
Refs. \cite{mode2} and \cite{kova} we note that the total 
cross section in both models is almost flat for neutrino energies 
above $8$~GeV. 
  
\begin{table}[tbp] 
\small 
\begin{center} 
\begin{tabular}{||c|c|c|c|c|c||} 
\hline 
$\sigma(10^{-40}~\mbox{cm}^2)\backslash$ Model & F.R.~\cite{mode1} &
S.L.~\cite{mode2}  
& A.K.K.~\cite{mode3,mode4,mode5} & A.G.Y.O.~\cite{mode6} & K.~\cite{kova} \\ 
\hline\hline 
$\nu_\mu p\rightarrow\mu^-\Sigma_c^{++}$ & $0.2(0.030\%)$ &
$9.0(1.3\%)$ & $8.0(1.2\%)$ & $1.0(0.15\%)$ & $3.0(0.45\%)$  \\ 
\hline 
$\nu_\mu p\rightarrow\mu^-\Sigma_c^{*++}$ & $0.6(0.089\%)$ &
$16.0(2.4\%)$ & $10.0(1.5\%)$ & $0.6(0.089\%)$ & -  \\ 
\hline 
$\nu_\mu n\rightarrow\mu^-\Lambda_c^{+}$ & $1.0(0.15\%)$ &
$23.0(3.4\%)$ & $41.0(6.1\%)$ & $3.0(0.45\%)$ & $5.0(0.74\%)$  \\ 
\hline 
$\nu_\mu n\rightarrow\mu^-\Sigma_c^{+}$ & $0.1(0.015\%)$ & $5.0(0.74\%)$ & - & 
$0.6(0.089\%)$ & $1.5(0.22\%)$   \\ 
\hline 
$\nu_\mu n\rightarrow\mu^-\Sigma_c^{*+}$ & $0.3(0.045\%)$ & $8.0(1.2\%)$ & - & 
$0.3(0.045\%)$ & - \\ 
\hline\hline 
Total &  $2.2(0.33\%)$ & $61.0(9.0\%)$ & $59.0(8.9\%)$ & $5.5(0.82\%)$
& $9.5(1.41\%)$  \\ 
\hline\hline 
\end{tabular} 
\caption{Predicted quasi-elastic charm-production cross section assuming a 
neutrino energy of $10$~GeV. In brackets the quasi-elastic charm rate with  
respect to deep-inelastic interactions is given.} 
 \label{tab:crosspred} 
\end{center}  
\end{table}
The statistics of neutrino quasi-elastic charm events collected by 
bubble chamber and emulsion experiments is rather poor. The measured 
cross sections, obtained by a reanalysis that uses the latest results 
on $\Lambda_c^{+}$ branching ratios~\cite{pdg2000}, are shown in 
Table~\ref{tab:qemeas}. 
\begin{table}[tbp] 
\small 
\begin{center} 
\begin{tabular}{||c|c|c|c|c|c||} 
\hline 
Experiment & $\sigma_{\Sigma_c^{++}}$ & $\sigma_{\Sigma_c^{*++}}$ &
$\sigma_{\Lambda_c^{+}}$ & $(\sigma_{\Lambda_c^+} + \sigma_{\Sigma_c^+}
+ \sigma_{\Sigma_c^{*+}})$ &  $(\sigma_{\Sigma_c^{++}} +
\sigma_{\Sigma_c^{*++}})$ \\  
 & $(10^{-40}~\mbox{cm}^2)$ &  $(10^{-40}~\mbox{cm}^2)$ &  $(10^{-40}~\mbox{cm}^2)$ &
 $(10^{-40}~\mbox{cm}^2)$ &  $(10^{-40}~\mbox{cm}^2)$ \\  
\hline\hline 
\cite{sigmac++} & $(2.3^{+2.7}_{-1.6})$ & & & & \\ 
\hline 
\cite{ammosov} & $(2.3\pm2.0)$ & $(4.5\pm4.0)$ & & & \\ 
\hline 
\cite{son} & & & $(0.9^{+0.9}_{-0.7})$ & & \\ 
\hline 
\cite{e531} & & & $(3.7^{+3.7}_{-2.3})$ & & \\ 
\hline  
\cite{armenise} & & & &  $(38.3\pm23.1)$ &  $<8.8$\\ 
\hline\hline 
Average & $(2.3\pm1.5)$ & $(4.5\pm4.0)$ &  $(1.1\pm0.8)$ &
$(38.3\pm23.1)$ &  $<8.8$ \\ 
\hline\hline 
\end{tabular} 
\caption{Summary of the experimental measurements of quasi-elastic
  charm-production cross sections. } 
\label{tab:qemeas} 
\end{center} 
\end{table}
Despite  the large statistical error, these measurements are clearly 
inconsistent with the predictions of Refs.~\cite{mode2}--\cite{mode5}, 
while the agreement is fair for the ones in 
Refs.~\cite{mode1,mode6,kova}. The average value of the cross sections 
predicted by Refs.~\cite {mode1,mode6,kova} has been used  to get a 
rough estimate of the expected number of events, given in  
Table~\ref{tab:crossused}. 
 
\begin{table}[hbtp] 
\small 
\begin{center} 
\begin{tabular}{||c|c|c|c||} 
\hline 
Reaction & $\sigma(10^{-39}~\mbox{cm}^2)$ & $\mathcal{R}(\%)$ & Expected events \\ 
\hline\hline 
$\nu_\mu p\rightarrow\mu^-\Sigma_c^{++}$ & 0.14 & $0.14$ & 14000 \\ \hline 
$\nu_\mu p\rightarrow\mu^-\Sigma_c^{*++}$ & 0.06 & $0.06$ & 6000 \\ \hline 
$\nu_\mu n\rightarrow\mu^-\Lambda_c^{+}$ & 0.3 & $0.3$ & 30000 \\ \hline 
$\nu_\mu n\rightarrow\mu^-\Sigma_c^{+}$ & 0.07 & $0.07$ & 7000 \\ \hline 
$\nu_\mu n\rightarrow\mu^-\Sigma_c^{*+}$ & 0.03 & $0.03$ & 3000 \\ \hline 
All & 0.6 & $0.6$ & 60000 \\ \hline\hline 
\end{tabular} 
\caption{Quasi-elastic charm-production cross section and its contribution 
to the total CC neutrino cross section. The last column shows the 
expected number of events, assuming a starting sample of $10^7$ 
CC neutrino-induced events. } 
\label{tab:crossused} 
\end{center} 
\end{table} 
 
\subsubsection{Description of the method} 
\label{metho} 
\begin{figure}[ht] 
\begin{center} 
\rotatebox{0}{\ \resizebox{0.7\textwidth}{!}{\ \includegraphics{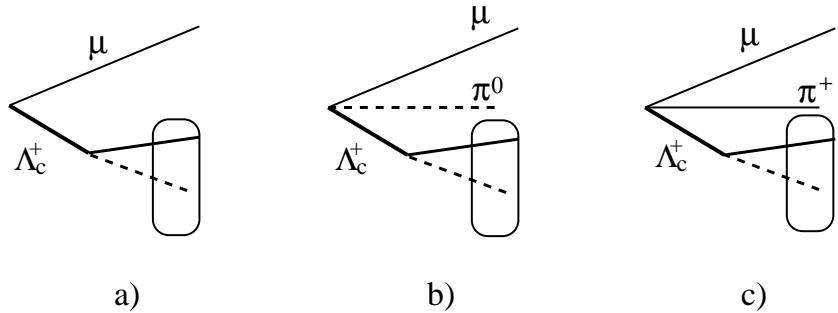} }} 
\end{center} 
\caption{Topology of the quasi-elastic neutrino-induced charm events in the 
case of the reactions a) $\nu_\mu n\rightarrow\mu^-\Lambda_c^+$, b) $\nu_\mu 
n\rightarrow\mu^-\Sigma_c^+(\Sigma_c^{*+})$ and c) $\nu_\mu 
p\rightarrow\mu^-\Sigma_c^{++}(\Sigma_c^{*++})$, with subsequent
$\Sigma_c$ decay  into $\Lambda_c$.  The particles inside the 
box represent the $\Lambda_c^+$ decay products.} 
\label{fi:topo} 
\end{figure}

The method consists in identifying the $\Lambda_c$ by means of the peculiar 
topology of the quasi-elastic reaction. As shown in Fig.~\ref{fi:topo}, 
the only charged particles produced in this process are the muon and
the short-lived  particle ($\Lambda_c$) 
except for the reaction Fig.~\ref{fi:topo}c) where an additional $\pi$ is produced. Therefore 
there is only a small contamination of these events from deep-inelastic charm 
production where a charmed hadron is produced (faking a $\Lambda_c$)
and the topology is quasi-elastic-like. 

In this way, since we only use topological information, the
$\Lambda_c$ identification  
is model-independent. The relative contamination of $D^+$ and $D^+_s$ 
from deep-inelastic events, $\varepsilon_{fake}$, can be dealt with as 
a relative systematic error on the branching ratios. We assume the 
relative systematic error to be $\varepsilon_{fake}+3\sigma_{fake}$ where 
$\sigma_{fake}$ is the error on $\varepsilon_{fake}$ . 

The normalization to determine the $\Lambda_c^+$ absolute branching
ratios is simply given by the number of events with a vertex topology
consistent with Fig.~\ref{fi:topo}. No model-dependent information is
used to define the normalization. The little knowledge we have about
the quasi-elastic charm-production cross section, which is
model-dependent 
unless measured, plays a role only in the evaluation of the
deep-inelastic contamination, namely the systematic error. In fact, in
the evaluation of the contamination, the ratio $\mathcal{R}$ between
charm quasi-elastic and standard deep-inelastic production appears.
It is worth noting that, even if the ratio $\mathcal{R}$ had an
uncertainty of $500\%$\footnote{Nevertheless, this is not the case.
  From Table~\ref{tab:qemeas} we see that $\mathcal{R}$ is measured
  with an accuracy better than $100\%$.}, the relative systematic
error on the branching ratios would be $\sim7.2\%$
(see~\cite{lambda}).

% 
%%%%%%%%%%%%%%%%%%%%%%%%%%%%%%%%%%%%%%%%%%%%%%%%%%%%%%%%%%%%%%%%%%%%%%%%%%%%%%% 
% Sensitivity of perfect detector 
%%%%%%%%%%%%%%%%%%%%%%%%%%%%%%%%%%%%%%%%%%%%%%%%%%%%%%%%%%%%%%%%%%%%%%%%%%%%%%% 
% 
 
\subsubsection{Measurement accuracy} 
 
\label{sensi} 
 
\begin{table}[tbp] 
\small 
\begin{center} 
    \begin{tabular}{||c|c|c|c|c||} 
      \hline 
      Channel & PDG BR~\cite{pdg2000} & $\Delta BR$ &$\Delta BR$  &
      $\Delta BR$ \\  
       &  & ($\frac{\Delta\mathcal{R}}{\mathcal{R}}=10\%$) &
      ($\frac{\Delta\mathcal{R}}{\mathcal{R}}=100\%$)&
      ($\frac{\Delta\mathcal{R}}{\mathcal{R}}=500\%$)\\  
      \hline 
      \hline 
      $\Lambda_c^+\rightarrow p K^-\pi^+$ & $(5.0{\pm}1.3)\%$ &
      $({\pm}0.09{\pm}0.04)\%$& $({\pm}0.09{\pm}0.09)\%$&
      $({\pm}0.09{\pm}0.4)\%$ \\ 
      \hline 
      $\Lambda_c^+\rightarrow \Lambda \mu^+\nu_\mu$ &
      $(2.0{\pm}0.7)\%$ & $({\pm}0.06{\pm}0.01)\%$&
      $({\pm}0.06{\pm}0.04)\%$& $({\pm}0.06{\pm}0.1)\%$ \\ 
      \hline 
      $\Lambda_c^+\rightarrow \Lambda e^+\nu_e$ & $(2.1{\pm}0.7)\%$ &
      $({\pm}0.06{\pm}0.01)\%$&  $({\pm}0.06{\pm}0.04)\%$&
      $({\pm}0.06{\pm}0.1)\%$ \\ 
      \hline 
      \hline 
    \end {tabular} 
\caption{Statistical and systematic accuracy achievable in the 
  determination of the $\Lambda_c^+$ absolute branching ratios,
  assuming a collected statistics of $10^7~\nu_\mu$
  CC events, as a function of the relative error on
  $\mathcal{R}$. The central values are taken from Ref.~\cite{pdg2000}.}
\label{tab:brscena1} 
\end{center} 
\end{table} 
 
The expected accuracy on the determination of the $\Lambda_c^+$
branching ratios as a function of the relative error on $\mathcal{R}$,
the quasi-elastic charm production cross section relative to the
deep-inelastic one, is shown in Table~\ref{tab:brscena1}. To compute
the expected number of events in each decay channel we use the central
values (shown in Table~\ref{tab:brscena1} together with their errors)
given by the Particle Data Group~\cite{pdg2000}. From this table we
can see that, even assuming a very large (unrealistic) systematic error,
the achievable accuracy on the branching ratios makes the
discrimination between the methods discussed in Ref.~\cite{pdg2000}
still possible.

% 
%%%%%%%%%%%%%%%%%%%%%%%%%%%%%%%%%%%%%%%%%%%%%%%%%%%%%%%%%%%%%%%%%%%%%%%%%%%%%%% 
% Diffractive production of Ds* 
%%%%%%%%%%%%%%%%%%%%%%%%%%%%%%%%%%%%%%%%%%%%%%%%%%%%%%%%%%%%%%%%%%%%%%%%%%%%%%% 
 
\subsection{Direct evaluation of $\mathbf D_s$ branching ratios and
  $\mathbf f_{D_s}$ measurement} 
The experimental knowledge on leptonic $D_s$ decays is very 
little. Currently, the branching ratios for $D_s\rightarrow l\nu$ 
decays have been estimated by the PDG~\cite{pdg2000} to be 
$BR(D_s\rightarrow \mu\nu) = (4.6\pm1.9)\times10^{-3}$ and 
$BR(D_s\rightarrow \tau\nu) = (7\pm4)\times10^{-2}$. These large 
uncertainties translate into a large uncertainty on the extraction of 
the decay constant $f_{D_s}$. 
 
A method that would allow at the \nufact\ the extraction of most of the $D_s$
branching ratios, and consequently of $f_{D_s}$, by means of purely
leptonic decays, has been proposed in Ref.~\cite{dspap};
the expected accuracy would be better
than $5\%$. Once $f_{D_s}$ will
be measured with such an accuracy, one will feel more confident about
extrapolating to the decay constants in the $B$ system, $f_{B}$ and
$f_{B_s}$, which are crucial quantities for the quantitative
understanding of $B^0_{(s)}$--$\bar{B}^0_{(s)}$ oscillations and the
extraction of $V_{td}$ ($V_{ts}$) from them.
 
\subsubsection{Topology of neutrino-induced diffractive charm events 
and background} 
In $D_s^{(*)}$\footnote{In the following, $D_s^{(*)}$ means either
$D_s$ or $D_s^{*}$.  The same notation is also used for $D$ mesons.}
diffractive production, only a muon is produced at the interaction
point (primary vertex), besides the charmed meson. Therefore, these
events are characterized by a peculiar topology: two charged tracks at
the primary vertex, one of them being a short-lived particle.
 
Neutrino-induced quasi-elastic charm events are characterized by the 
topologies shown in Fig.~\ref{fi:topo}. Therefore, they are
similar to the diffractive ones, but with a cross section twice 
as large. Since antineutrinos cannot induce quasi-elastic charm 
production, while diffractive production is the same for both $\nu$ and 
$\bar{\nu}$, in the following we will consider only $\bar{\nu}$ 
beams. We will then make the assumption that all the events with the 
above topology are due to $D_s^{(*)}$ diffractive production. In this case 
we will wrongly classify some of the charmed hadrons produced 
in deep-inelastic interactions. 
 
The charm production in $\bar{\nu}$ interactions and the event topology 
have been studied by using the HERWIG event generator~\cite{herwig}, 
as well as an event generator based on JETSET~\cite{jetset} and 
LEPTO~\cite{lepto}, assuming  the energy dependence of the charm fractions 
reported in~\cite{bolton}\footnote{We assumed that charm fractions 
are equal for both $\nu$ and $\bar{\nu}$.}. The average charm
fractions, convoluted with the neutrino spectrum, are $F_{\bar{D}^0} = 
61\%$, $F_{{D}^-} = 26\%$, $F_{{D}^-_s} = 7.3\%$ and 
$F_{{\Lambda}^-_c} = 5.7\%$.

The signal kinematics has been studied by using an event generator developed within the CHORUS Collaboration~\cite{chorus_ds}.

The contamination to the diffractive sample, which 
comes from deep-inelastic events, can be written as 
 
\begin{equation} 
\varepsilon_{fake} = \frac{\sigma(\bar{\nu}_\mu N\rightarrow\mu^+ C X)}{ 
\sigma(\bar{\nu}_\mu N\rightarrow\mu^+X)}{\times}\frac{1}{\bar{\mathcal{R}}}{\times} 
 (F_{D^-}+F_{\Lambda_c^-}) \times f_{fake}{\times}\varepsilon_{kin} 
\label{eq:eprio} 
\end{equation} 
where  
\be
  \bar{\mathcal{R}}\equiv\frac{\sigma(\bar{\nu}_\mu 
  N\rightarrow\mu^+D_s^{(*)-} N)}{\sigma(\bar{\nu}_\mu 
  N\rightarrow\mu^+X)}\,.
\ee

We take the charm production in $\bar{\nu}$ interaction to be
$3\%$~\cite{conrad}. By using the charm fractions discussed above,
$(F_{D^-}+F_{\Lambda_c^-})=31.7\%$. The factor $f_{fake} = (6.0\pm
0.1)\%$ is the fraction of deep-inelastic charmed events faking the
diffractive topology; $\varepsilon_{kin}=(0.4\pm0.2)\%$ gives the
efficiency of kinematical cuts as explained in
Ref.~\cite{dspap}. Finally we get $\varepsilon_{fake}\simeq0.04\%$.
 
The currently measured $\bar{\mathcal{R}}$ has an error of about
$15\%$ (see Ref.~\cite{dspap}), which affects the relative error on
$\varepsilon_{fake}$. As described in Ref.~\cite{dspap}, a kinematical
analysis allows a good $D_s$ detection efficiency ($\sim 73\%$) with a
small background. The expected contamination and its error are given
in Table~\ref{tab:erro2}.
 
Another possible source of irreducible background is the diffractive
production of $D^{(*)-}$ (see Ref.~\cite{dspap}). It is suppressed by
a factor $\left |  V_{cs} / V_{cd} \right|^2\sim20$ and only $30\%$
of the $D^{*-}$ decay into a charged charmed meson ($BR(D^{*-}\rightarrow
D^{-} ) = 0.323\pm0.006$~\cite{pdg2000}). On the other hand, all the
diffractively produced $D^{-}$ are background: we assume that $D^-$
are half of the diffractive sample~\footnote{This assumption has been
driven by the NuTev results: $\sigma(\nu_\mu N\rightarrow\mu^- D_s N)
= (1.4\pm0.3)$~fb/nucleon, $\sigma(\nu_\mu N\rightarrow\mu^- D_s^* N) =
(1.6\pm0.4)$~fb/nucleon~\cite{dsprod_nutev}.}. Finally, we get
$\varepsilon_{D^*}=(3.3\pm0.8)\%$.

From the numbers given above, it turns out that the little knowledge we
have about the diffractive charm production cross-section plays a role
only in the evaluation of the deep-inelastic contamination, namely a
term of the systematic error. Even if the ratio $\bar{\mathcal{R}}$
had an uncertainty of $100\%$\footnote{Nevertheless, this is not the
case. We recall that the Big Bubble Chamber Neutrino
Collaboration~\cite{dsprod} and NuTeV~\cite{dsprod_nutev} combined
analysis gives an accuracy of about $15\%$ for $\bar{\mathcal{R}}$.},
the relative systematic error on the branching ratios would be $\sim
0.04\%$ (see Table~\ref{tab:erro2}).  We want to stress that, since
the $\varepsilon_{D^{(*)}}$ contribution is dominant, the overall
systematic uncertainty does not depend at all on the
$\bar{\mathcal{R}}$ accuracy. Writing the relative systematic error
from deep-inelastic events as $\varepsilon_{fake}+3\sigma_{fake}$,
where $\sigma_{fake}$ is the error on $\varepsilon_{fake}$, the
overall relative systematic uncertainty is
\be 
\varepsilon_{sys} = 
\sqrt{(\varepsilon_{fake}+3\sigma_{fake})^2 +
  \varepsilon_{D^{(*)}}^2 } = (3.3\pm0.8)\% \; ,
\ee
which is dominated by the $\varepsilon_{D^{(*)}}$ term. 

\begin{table}[tbp]
\begin{center}
{\small
\begin{tabular}{||c|c|c|c||}
\hline $\Delta\bar{\mathcal{R}}/\bar{\mathcal{R}}$ &
$\sigma_{fake}/\varepsilon_{fake}$ &
$\varepsilon_{fake} (\%)$ & $\varepsilon_{D^{(*)}} (\%)$ \\ 
\hline\hline 
$15\%$ & $23\%$ & $0.037{\pm}0.009$ & $3.3\pm0.8$ \\ 
\hline 
$30\%$ & $34\%$ & $0.04{\pm}0.01$ & $3.3\pm0.8$ \\
\hline 
$50\%$ & $53\%$ & $0.04{\pm}0.02$ & $3.3\pm0.8$ \\ 
\hline 
$100\%$ & $101\%$ & $0.04{\pm}0.04$ & $3.3\pm0.8$ \\ 
\hline
\end{tabular}
}
\end{center}
\caption{The relative and absolute error on $\varepsilon_{fake}$ as
  a function of the relative error on $\bar{\mathcal{R}}$. In the fourth column 
the systematic uncertainty due to the $D^{(*)}$ contamination is reported. }
\label{tab:erro2}
\end{table}
 
\subsubsection{Description of the method} 
\label{metho2} 

An almost pure sample of $D_s^-$ from diffractive events, with a small 
contamination of $D^-$ and $\Lambda_c^-$ produced in deep-inelastic 
events and in diffractive $D^{(*)-}$ production, can be built by using
diffractive $D_s^{(*)}$ production from antineutrinos. 
The normalization to determine the $D_s$ 
absolute branching ratios is given by the number of events with 
a vertex topology consistent with one $\mu$ plus a short-lived 
particle. No model-dependent information is used to define the normalization.

It is worth noting that the contamination of $D^-$ and $\Lambda_c^-$
 events does not affect the $D_s\rightarrow\tau$
channel. Indeed, such events would present a unique topology with
two subsequent kinks. An event with a double kink has been observed recently
in CHORUS (see Ref.~\cite{chorus_ds}).

\subsubsection{Measurement accuracy at a neutrino factory} 
At present there are no experiments with both an adequate spatial 
resolution to fully exploit the diffractive topology and a sufficient 
antineutrino-induced CC event statistics. Therefore, the 
method proposed in this paper could only be exploited with the above-mentioned 
detector exposed at a \nufact. 

Let us assume that $10^7$ $\bar{\nu}$~CC events are collected into an emulsion
target and that the detection efficiency is about $73\%$ for the
$D_s$ decays. By assuming a vertex
location efficiency of about $50\%$\footnote{This efficiency accounts
for the electronic detector reconstruction and the automatic location
of the event vertex inside the emulsions.} and assuming a $\bar{\nu}$
diffractive production rate of $6.2\times10^{-3}$/CC event, we expect to
detect a number of $D_s$ equal to $N_{D_s} = 10^7\times
6.2\times10^{-3}\times0.73 \times 0.5\simeq 2.3\times10^4$.

The expected accuracy on the determination
of the $D_s$ branching ratios is shown in Table~\ref{tab:brds} for a few 
channels, together with the current status. To compute the expected number of
events in each decay channel we have used the central values (shown in
Table~\ref{tab:brds} together with their errors) given by the
Particle Data Group~\cite{pdg2000}.

\begin{table}[tbp]
\begin{center}
{\small
%\begin{sideways}
    \begin{tabular}{||c|c|c||}
      \hline
      Channel & PDG BR~\cite{pdg2000} & New method \\
      \hline
      \hline
      $D_s\rightarrow \mu\nu$ & $(4.6 \pm 1.9)\times10^{-3}$ & 
$({\pm}0.55{\pm}0.15)\times10^{-3}$ \\
      \hline
      $D_s\rightarrow \tau\nu$ & $(7 \pm 4)\%$ & 
$({\pm}0.17{\pm}0.23)\%$  \\
      \hline
      $D_s\rightarrow \phi l\nu$ & $(2.0 \pm 0.5)\%$ & 
$({\pm}0.08{\pm}0.07)\%$ \\
      \hline
      \hline
    \end {tabular}
%    \end{sideways}
}
\end{center}
\caption{Statistical and systematic accuracy achievable in the
  determination of the $D_s$ absolute branching ratios, assuming a
  collected statistics of $10^7~\bar{\nu}_\mu$ CC events. 
The central values are  taken from Ref. \cite{pdg2000}.}
\label{tab:brds}
\end{table}

As discussed in Ref.~\cite{richman} the leptonic branching ratios are
proportional to the decay constant. Therefore, by using the measured
branching ratios given in Table~\ref{tab:brds}, $f_{D_s}$ can be
extracted. If we collect $10^7$ $\bar{\nu}_{\mu}$ CC interactions we
get $f_{D_s} = 288\pm4(\mathrm{stat})\pm5(\mathrm{syst})$~MeV,
where the central value is taken from Ref.~\cite{pdg2000}.

\subsection{Theoretical estimates for $\mathbf \nu$-induced exclusive
  $\mathbf D_s$ production}
\begin{figure}[th]
\centering
\includegraphics[width=0.45\textwidth,clip]{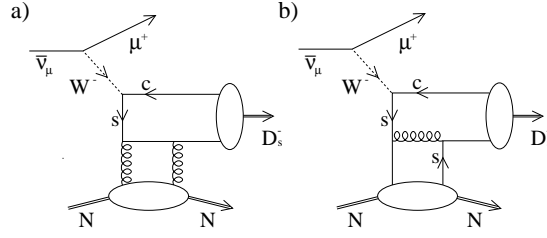}
\caption{Two of the five contributing diagrams.}
\label{graph}
\end{figure}

The observation of exclusive $D_s$ production at the \nufact, through
the process
$\bar\nu_{\mu} + N \to \; \mu^+ + N + D_s^-$,
is also of interest for the study of the production mechanism within QCD
and opens a new possibility to study the nucleon structure.
There exists a QCD factorization theorem~\cite{factor,Rad}, which states
that the amplitude for hard exclusive
meson-production processes can be written as a convolution of a skewed parton
distribution (SPD), a distribution amplitude, and a hard part.
This theorem has recently been applied to the investigation of
electroproduction of single light mesons \cite{mesonprod} and meson
pairs~\cite{maxben}.
Motivated by the possible implications for the
\nufact, the formalism has been extended to CC-induced
 processes\footnote{B.~Lehmann-Dronke and A.~Sch\"afer, in
  preparation.}. We summarize here the main results
of this work, whose validity is limited to values of $Q^2$
large compared to $-t$ and to the squared masses of the involved
particles.
The differential cross section of the leptoproduction process is given
by
\be
\frac{\rmd\sigma}{\rmd\xBj\rmd Q^2\rmd t}
=
\frac{e^2}{4(4\pi)^3\sin^2\theta_W}\frac{\xBj}{Q^2(Q^2+M_W^2)^2}
\biggl(1-\frac{Q^2}{2\xBj\,p\cdot l}\biggr)
\sum_{s'}|T|^2\;,
\ee
where $l$ is the neutrino momentum and $T$ is the amplitude
for the subprocess
\be
W^{-^\ast}_L(q) + N(p)\to D_s^-(q')+N(p')\;.
\ee
At leading order, $T$ is obtained from the sum of three
diagrams involving a gluon SPD (Fig.~\ref{graph}a
and diagrams obtained by an interchange of the order of the vertices)
and two diagrams with a contribution of the (polarized and unpolarized)
strange quark SPD (Fig.~\ref{graph}b plus one
diagram with a changed order of vertices).
The relative Feynman diagrams are convoluted  with the
gluon and $s$ quark SPD and with the
distribution amplitude $\Phi_{D_s^-}(z)$ for the
$D_s^-$ meson ($z$ being the fraction of the $D_s^-$ momentum carried by the
strange quark).

For a numerical estimate of the cross section we adopt the
distribution amplitude from~\cite{Bauer}:
\be
\Phi_{D_s^-}(z)=
N_{D_s}\sqrt{z(1-z)}\exp\bigg[-\frac{m_{D_s}^2}{2\omega^2}z^2\bigg] \; ,
\ee
where $\omega=1.38\,\mathrm{GeV}$ was
obtained in~\cite{omega} as the best fit for the $D$ meson,
and $N_{D_s}$ has to be chosen to satisfy the sum rule
\be
\int_0^1 \rmd z\,\Phi_{D_s^-}(z)=f_{D_s}\label{DA}
\ee
For a comparison, we shall also use the
asymptotic form of the $D_s$ distribution amplitude:
\be
\Phi_{D_s^-}(z)=6f_{D_s}z(1-z)\;.
\label{DA2}
\ee
In both cases we use
$f_{D_s}=270\,\mathrm{MeV}$, namely the average of the results
obtained so far in lattice calculations \cite{fDs}.
The gluon and quark SPD's are parameterized as in \cite{maxben},
combining Radyushkin's model \cite{Rad,Rad99} with the
parameterizations of the usual (forward) parton distributions of
Ref.~\cite{MRS}.

\begin{figure}
\centerline{\includegraphics[width=0.5\textwidth,clip]{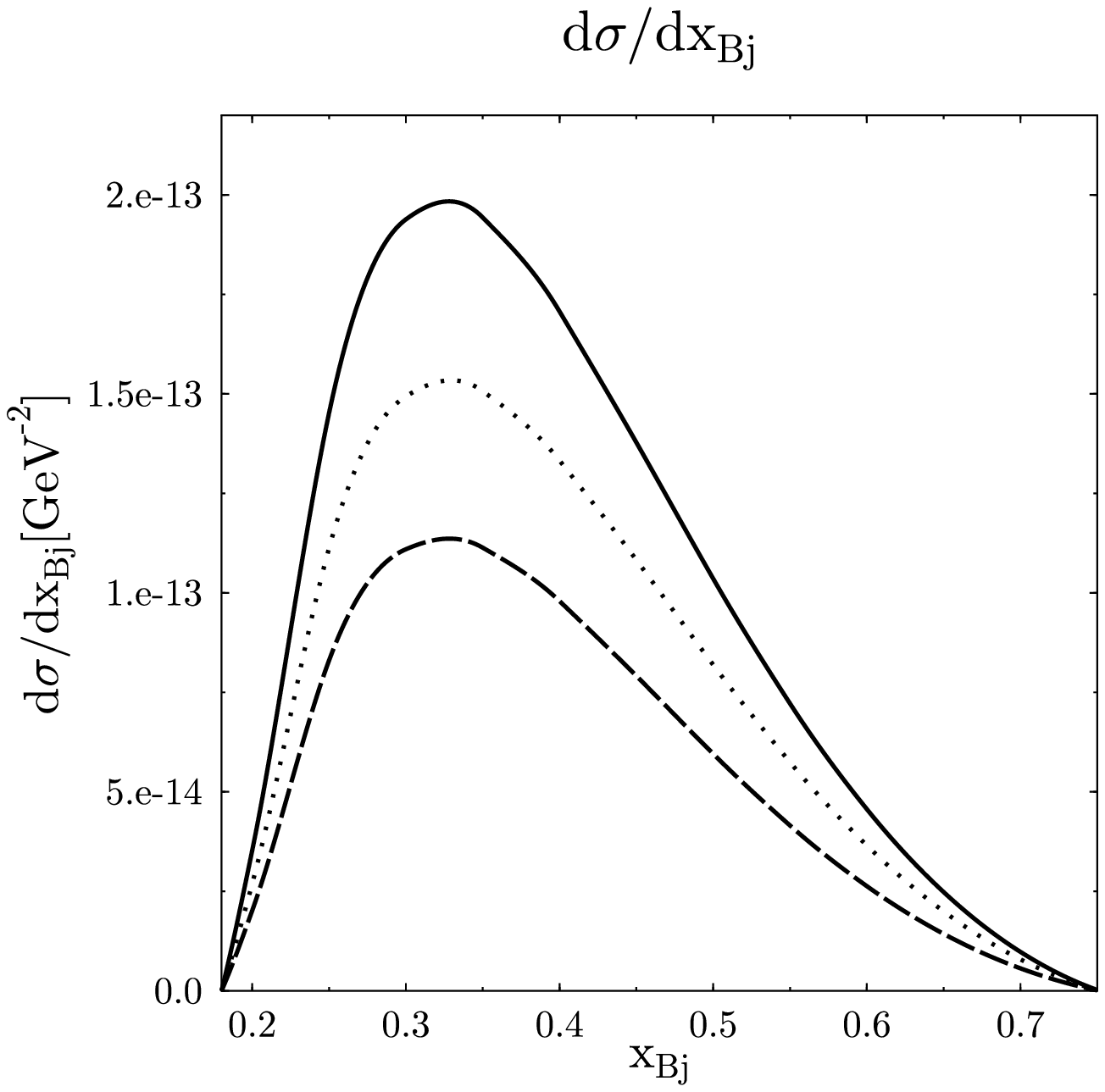}
\hskip 1cm
\includegraphics[width=0.5\textwidth,clip]{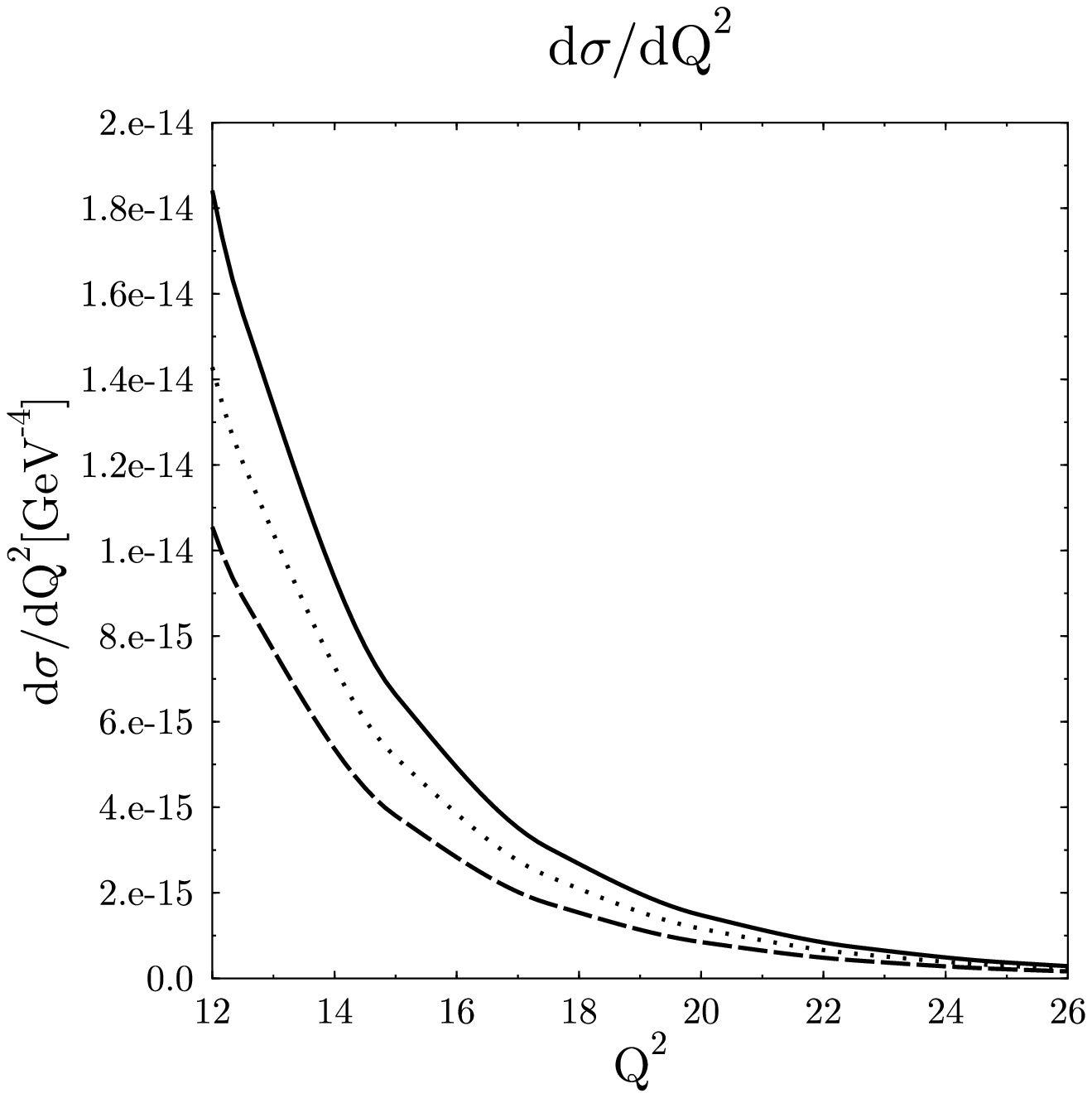}}
\caption{The differential cross section for exclusive $D_s^-$ production
  as a function of $\xBj$ or $Q^2$ respectively. The dotted lines show
  the contribution stemming from the gluon SPD. The results obtained
  for the asymptotic form of the distribution amplitude $\Phi_{D_s^-}$
  are plotted with dashed lines.}
\label{result}
\end{figure}

Figure~\ref{result} shows the results obtained for the differential
cross sections $\rmd\sigma/\rmd\xBj$ and $\rmd\sigma/\rmd Q^2$ where
$t=(p-p')^2$ has been integrated over the interval
$t_{\mathrm{min}}=m_N^2\xBj^2/(1-\xBj)<-t<2\,\mathrm{GeV}^2$, assuming
for simplicity a fixed neutrino energy $E_{\bar\nu}=34\,\mathrm{GeV}$. For the
plot of the $\xBj$-dependence $Q^2$ has been integrated from
$12\,\mathrm{GeV}^2$ to the upper bound given by the constraint
$y<1$, with $y:=p\cdot q/p\cdot l=Q^2/(2\xBj p\cdot l)$.
The plot of $\rmd\sigma/\rmd Q^2$ is based on the $\xBj$-integrated
cross section, integrated between $\xBj=0.18$ and $\xBj=0.75$ and taking into
account the same kinematical constraint.
The solid lines correspond to the form of $\Phi_{D_s^-}$ given in
\equ{DA}, while the dashed lines correspond to the
asymptotic form of \equ{DA2}.
The dotted lines are obtained by setting to zero
the strange quark SPD,
proving the dominance of the gluon contribution, and
suggesting that this process is a potential probe for the
measurement of the gluon SPD.

Integrating over all variables $Q^2$, $\xBj$, and $t$ over the same
kinematical region gives a value for the total cross section of
$\sigma=5.6\times 10^{-14}\,\mathrm{GeV}^{-2}
=2.2\times 10^{-5}\,\mathrm{pb}$.
Uncertainties of this rough estimate result from the
current limited knowledge of the SPDs, by the 20\% uncertainty on
$f_{D_s}$, and by the lack of knowledge about the exact form of the
meson-distribution amplitude, as illustrated in
Fig.~\ref{result}. It is worth noting, however, that experiments of the
kind discussed here can provide much more precise 
information on the $D_s$ decay constant independently of
the exact cross section.
As discussed in the previous subsection the relatively high production rate
of $D_s$ mesons allows us to determine $f_{D_s}$ by
measuring the $D_s$ branching ratios and its total width \cite{dspap}.
In spite of the small cross section, the huge integrated luminosities of
$\int L\rmd t>10^{9}\,\mathrm{pb}^{-1}$ available
at the \nufact\ would lead to samples of the order of
$10^4$ events.

The analysis shows that the large rates, and the extended $Q^2$ range
available, will allow a more accurate determination of the
gluon SPD and help to better test the theory of these
exclusive processes.
The available statistics may even allow a study of higher-twist
corrections to the leading $Q^2$ behaviour, probing the limits of the
factorization theorem.

\section{$\mathbf \Lambda$ POLARIZATION IN NEUTRINO DIS} 
Many experiments have reported~\cite{wa21}--\cite{nomad}
 the observation of longitudinal
polarization of $\Lambda$ baryons produced in neutrino 
DIS on an unpolarized target. 
Measurements in the
current fragmentation
region (CFR) give information on distribution and fragmentation
functions, while measurements in the target fragmentation region  (TFR)
allow to access  fracture
functions~\cite{tv}. 
In both cases the measurement of $\Lambda$~ polarization provides a 
sensitive way for studying perturbative and non-perturbative spin phenomena. 

\subsection{$\mathbf \Lambda$ polarization in the current fragmentation region}
\def\beq{\begin{equation}}
\def\eeq{\end{equation}}
\def\bea{\begin{eqnarray}}
\def\eea{\end{eqnarray}}
\newcommand{\bm}[1]{\mbox{\boldmath $#1$}}
\newcommand{\Lup}{\Lambda^\uparrow} 
\newcommand{\Ldown}{\Lambda^\downarrow} 
\newcommand{\bfk}{\mbox{\boldmath $k$}}
The polarization of spin 1/2 baryons inclusively produced in polarized
DIS in the CFR may be useful
to obtain new information on polarized distribution and fragmentation 
functions. A lot of theoretical attention has been dedicated to 
the self-revealing polarization of $\Lambda$'s and other hyperons 
\cite{noi1}--\cite{noi2}. Most papers, with the exception of Refs.~\cite{kbh}, 
\cite{bjk} and \cite{mssy}, do not consider weak interaction contributions, 
since there is no available experimental information.

The NOMAD collaboration has published some interesting results
\cite{nomad} on the $\Lambda$ polarization in $\nu_\mu$ CC 
interactions; more data, sensitive to $\gamma/Z$ interference effects,
 might be available from high-energy NC
processes at HERA; 
more complete information is however only expected 
from experiments at the \nufact.

These experiments are a unique source of new data, owing to the natural 
neutrino polarization and to the selected couplings of $W$'s to pure helicity 
states. We will be able to study in detail processes like
\bea
\nu     \, p  \, \to \, \ell^-   \, \Lambda^{\uparrow} + X && \quad 
\bar\nu \, p  \, \to \, \ell^+   \, \Lambda^{\uparrow} + X \quad 
{\mathrm{ (CC)}} \nonumber \\
\nu     \, p  \, \to \, \nu      \,\>\> \Lambda^{\uparrow} + X &&\quad 
\bar\nu \, p  \, \to \, \bar \nu \,\>\> \Lambda^{\uparrow} + X \quad 
{\mathrm {(NC)}} \nonumber 
\eea
where the proton $p$ may or may not be polarized, depending on the 
experimental setup, whereas neutrinos are obviously always polarized 
($\lambda _\nu = -1/2$, $\lambda _{\bar \nu} = +1/2$). Notice that though 
the formulae we present here hold for proton targets, they can be easily 
modified into analogous expressions valid for polarized and unpolarized 
neutrons. We consider CC and NC processes separately.
The explicit expressions for the polarization of the final $\Lambda$'s in 
terms of elementary dynamics, quark distribution and fragmentation functions 
show how these experiments can provide precious information on distribution 
and fragmentation functions which are still far from being well known. 

\subsubsection{Charged current neutrino processes, $\nu p \to \ell \Lup X$}
For the neutrino-initiated processes, the longitudinal polarizations 
$P_{[\nu,\ell]}$ and $P_{[\bar \nu,\ell]}$ for any spin $1/2$ baryon $B$ 
($\Lambda$'s and $\bar \Lambda$'s for instance) produced in CC
DIS processes are defined as 
\be
P_{[\nu,\ell]} (B) = 
\frac{d\sigma^{\nu p \to \ell ^- B_+ X} - 
      d\sigma^{\nu p \to \ell ^- B_- X}}
{d\sigma^{\nu p \to \ell ^- B_+ X} + 
      d\sigma^{\nu p \to \ell ^- B_- X}} \label{pnul}
\ee
and
\be
P_{[\bar \nu,\ell]} (B) = 
\frac{d\sigma^{\bar \nu p \to \ell ^+ B_+ X} - 
      d\sigma^{\bar \nu p \to \ell ^+ B_- X}}
{d\sigma^{\bar \nu p \to \ell ^+ B_+ X} + 
      d\sigma^{\bar \nu p \to \ell ^+ B_- X}} \, , \label{pbnul}
\ee
where $B_{\pm}$ denotes a baryon $B$ with helicity $\pm$.
 
In the most general case, when also the proton $p$ is polarized -- and we
denote by a superscript $S$ its spin state -- we obtain at leading
twist in the QCD factorization theorem:
\be
P^{(S)}_{[\nu,\ell]} (B) = 
-\frac{\sum_{i,j} [ (d_j)^{(S)}_- d\hat\sigma_{--}^{d_j \to u_i} 
\Delta D_{B/u_i} - (\bar u_i)^{(S)}_+ d\hat\sigma_{-+}^{\bar u_i \to \bar d_j}
\Delta D_{B/\bar d_j}]}
{\sum_{i,j} [ (d_j)^{(S)}_- d\hat\sigma_{--}^{d_j\to u_i} D_{B/u_i} +
(\bar u_i)^{(S)}_+ d\hat\sigma_{-+}^{\bar u_i \to \bar d_j} D_{B/\bar d_j}]}
\label{Pnu}
\ee
and 
\be
P^{(S)}_{[\bar \nu,\ell]} (B) =
-\frac{\sum_{i,j} [ (u_i)^{(S)}_- d\hat\sigma_{+-}^{u_i \to d_j} 
\Delta D_{B/d_j} -
(\bar d_j)^{(S)}_+ d\hat\sigma_{++}^{\bar d_j \to \bar u_i} 
\Delta D_{B/\bar u_i}]}
{\sum_{i,j} [ (u_i)^{(S)}_- d\hat\sigma_{+-}^{u_i \to d_j} D_{B/d_j} +
(\bar d_j)^{(S)}_+ d\hat\sigma_{++}^{\bar d_j \to \bar u_i} D_{B/\bar u_i}]}
\,,
\label{Pantinu}
\ee
where $(q)^{(S)}_{\pm}$ stands for the number density
(distribution function) of quarks $q$ with helicity $\pm$ inside a proton 
with spin $S$, whereas $q_{\pm}$ alone will refer, as usual, to a proton with 
helicity~+. For the flavours we use the notation $u_i = u,c$ and 
$d_j = d,s$. The polarized fragmentation functions are defined in
terms of fixed-helicity fragmentation functions as
\be
\Delta D_{B /q} \equiv D_{B_+/q_+} - D_{B_-/q_+} = D_{B_-/q_-} - D_{B_+/q_-} 
\, ,
\ee
and $d\hat\sigma_{--}^{d_j \to u_i}$ stands for the 
$d\hat\sigma/dy$ cross section for the elementary interaction 
$\nu \, d_j \to \ell \, u_i$, where $\nu$ and $d_j$ having negative
helicities.  

The above polarizations depend on the usual DIS variables $x$, $z$ and $y$; 
apart from the $Q^2$ dependence of distribution and fragmentation functions
due to the QCD evolution, there is a kind of factorization 
in the dependence on 
the three variables, in that the distribution functions depend on $x$,
the fragmentation functions on $z$ and the SM dynamics on $y$.
If convenient, and according to experimental setups, numerator and denominator
of Eqs.~(\ref{Pnu}) and (\ref{Pantinu}) can be integrated over some variables.
 
Performing the sum over flavours in the numerators and denominators, 
neglecting $c$-quark contributions and inserting the elementary dynamics 
expressions, Eqs.~(\ref{Pnu}) and (\ref{Pantinu}) give,
for longitudinally polarized ($\pm$ helicity) protons
\be
P_{[\nu,\ell]} ^{(\pm)} (B;x,y,z) = 
-\frac{[d_\mp + R\,s_\mp]\, \Delta D_{B/u} - 
(1-y)^2 \,\bar u_\pm \,[\Delta D_{B/\bar d} + R\, \Delta D_{B/\bar s}]}
{[d_\mp + R\,s_\mp] \, D_{B/u} +
(1-y)^2 \,\bar u_\pm \, [D_{B/\bar d} + R\, D_{B/\bar s}]} \; ,
\label{Pnu+-}
\ee
and 
\be
P_{[\bar \nu,\ell]} ^{(\pm)} (B;x,y,z) = 
\frac{[\bar d_\pm + R\, \bar s_\pm]\, \Delta D_{B/\bar u} - 
(1-y)^2 \,u_\mp \, [\Delta D_{B/d} + R\, \Delta D_{B/ s}]}
{[\bar d_\pm + R\, \bar s_\pm] \, D_{B/\bar u} +
(1-y)^2 \, u_\mp \, [D_{B/ d} + R\, D_{B/s}]} \,,
\label{Pantinu+-}
\ee
where $R\, \equiv \sin ^2 \theta_c/\cos ^2 \theta _c \simeq 0.056$. 

In the simpler case in which the proton is unpolarized, we replace
$q_{\pm}$ by $q/2$, and Eqs.~(\ref{Pnu+-}) and (\ref{Pantinu+-})
become respectively 
\be
P_{[\nu,\ell]} ^{(0)} (B;x,y,z) = 
-\frac{[d + R\,s] \, \Delta D_{B/u} - 
(1-y)^2 \, \bar u \, [\Delta D_{B/\bar d} + R\, \Delta D_{B/\bar s}]}
{[d + R\,s] \, D_{B/u} +
(1-y)^2 \, \bar u \, [D_{B/\bar d} + R\, D_{B/\bar s}]} \; ,
\label{Pnu0}
\ee
and 
\be  
P_{[\bar \nu,\ell]} ^{(0)} (B;x,y,z) =  
 \frac{[\bar d + R\, \bar s] \, \Delta D_{B/\bar u} - 
(1-y)^2 \, u [\Delta D_{B/ d} + R\, \Delta D_{B/ s}]}
{[\bar d + R\, \bar s] \, D_{B/\bar u} +
(1-y)^2 \, u [D_{B/ d} + R\, D_{B/s}]} \, \cdot
\label{Pantinu0}
\ee
The formulae given above hold for any baryon and antibaryon with spin $1/2$; 
further simplifications are possible when a $\Lambda$ baryon
(and, more in general, a baryon rather than an antibaryon) is
produced: in this case, in the 
kinematical regions characterized by large values of $x$ and $z$
one can neglect terms that contain both $\bar q$ distributions (in a proton)
and $\bar q$ fragmentations (into a $\Lambda$) as they are both 
small. Then one simply has:  
\be
P_{[\nu,\ell]} ^{(\pm)} (\Lambda;z) \, \simeq \, 
P_{[\nu,\ell]} ^{(0)} (\Lambda;z) \,\simeq \,  
-\frac{\Delta D _{\Lambda/u}}{D_{\Lambda/u}} \; , \label{PnuL}
\label{pnlu}
\ee
\be
P_{[\bar \nu,\ell]} ^{(\pm)} (\Lambda;z) \, \simeq 
\, P_{[\bar \nu,\ell]} ^{(0)} (\Lambda;z) \,\simeq \,
- \frac{\Delta D _{\Lambda/d} + R\,\,\Delta D _{\Lambda/s}}{D_{\Lambda/d} 
+ R\,\, D _{\Lambda/s}}\,, \label{PantinuL}
\label{pnblds}
\ee
and the polarizations, up to QCD evolution effects, become functions of 
the variable $z$ only, since any other term apart from the fragmentation 
functions cancels out.

Equations~(\ref{PnuL}) and (\ref{PantinuL}) relate the values of the 
longitudinal polarization $P(\Lambda)$ to a quantity with a clear physical 
meaning, \ie\ the ratio $\Delta D_{\Lambda/q}/D_{\Lambda/q}$; 
this happens with weak CC interactions -- while it cannot happen 
in purely electromagnetic DIS \cite{noi2} -- due to the selection of the 
quark helicity and flavour in the coupling with neutrinos. A measurement of 
$P(\Lambda)$ offers new direct information on the fragmentation process.     

Weak CCs couple to pure helicity states, and do not transfer
transverse polarization from quarks to final baryons; therefore 
transversely polarized protons do not add 
any information. However, one might
still have -- in analogy to what happens in unpolarized $N$--$N$ 
interactions --
final $\Lambda$'s with transverse (with respect to the production plane)
polarization. This can only originate in the fragmentation process
of an unpolarized quark; recently, new polarizing fragmentation functions
have been introduced~\cite{noi4} to describe such an effect:
\be   
\Delta^N D_{\Lup/q}(z, \bfk_{\perp}) = 
\hat D_{\Lup/q}(z, \bfk_{\perp}) - \hat D_{\Ldown/q}(z, \bfk_{\perp})  
\label{pff}
\ee
where $\bfk_{\perp}$ is the transverse momentum of the $\Lambda$ with respect
to the fragmenting quark momentum. A measurement of {\it transverse} $\Lambda$
polarization would give a direct measurement of such a new function:
\be   
P_{[\nu,\ell]}^{(0)}(\Lambda; z, k_\perp) \simeq \frac
{\Delta^ND_{\Lup/u}}{D_{\Lambda/u}} \, \cdot
\label{polt}
\ee

\subsubsection{Neutral current neutrino processes, $\nu p \to \nu \Lup X$} 
In analogy to the previous paragraph, the 
longitudinal polarizations of the produced baryon $B$ is given by
\be
P_{[\nu ,\nu]}(B) = 
\frac{d\sigma^{\nu p \to \nu B_+ X} - 
      d\sigma^{\nu p \to \nu B_- X}}
{d\sigma^{\nu p \to \nu B_+ X} + 
      d\sigma^{\nu p \to \nu B_- X}}\,,
\ee
and 
\be
P_{[\bar \nu ,\bar \nu]}(B) =  
\frac{d\sigma^{\bar \nu p \to \bar \nu B_+ X} - 
      d\sigma^{\bar \nu p \to \bar \nu B_- X}}
{d\sigma^{\bar \nu p \to \bar \nu B_+ X} + 
      d\sigma^{\bar \nu p \to \bar \nu B_- X}} \>\cdot
\ee

For the numerator and denominator of $P_{[\nu, \nu]}(B)$ and 
$P_{[\bar \nu, \bar \nu]}(B)$ separately, one obtains, for a generic spin 
state $S$ of the proton ($C \equiv \sin^2\theta_W/3$):
\bea
N_{[\nu,\nu]}^{(S)} (B) \!\!&=&\!\! 
 \sum \,\! _{_j} \left\{ \left[ 
(u_j)_+^{(S)} \, (1-y)^2 \, 16\,C^2 -
(u_j)_-^{(S)} \, (1 - 4\,C)^2 \right] \Delta D_{B/u_j}  
\right . \nonumber \\ 
&& \mbox{} \hspace{0.7cm} + \left. \left[ 
(d_j)_+^{(S)} \, (1-y)^2 \, 4\,C^2 -
(d_j)_-^{(S)} \,(1 - 2\,C)^2 \right] \Delta D_{B/d_j} 
\right .\nonumber \\ 
&& \mbox{} \hspace{0.7cm} + \left. \left[ 
(\bar u_j)_+^{(S)} \,(1-y)^2 \, (1 - 4\,C)^2 -
(\bar u_j)_-^{(S)} \, 16\,C^2 \right] \Delta D_{B/\bar u_j} 
\right .\nonumber \\ 
&& \mbox{} \hspace{0.7cm} + \left.\left[ 
(\bar d_j)_+^{(S)} \,(1-y)^2 \, (1 - 2\,C)^2 -
(\bar d_j)_-^{(S)} \, 4\,C^2 \right] \Delta D_{B/\bar d_j} 
\right\}
\eea
\bea
D_{[\nu ,\nu]}^{(S)} (B) \!\!&=&\!\!  
\sum \,\! _{_j} \left\{ \left[ 
(u_j)_+^{(S)} \,(1-y)^2 \, 16\,C^2 +
(u_j)_-^{(S)} \,(1 - 4\,C)^2 \right] D_{B/u_j}  
\right . \nonumber \\ 
&& \mbox{} \hspace{0.3cm} + \left. \left[ 
(d_j)_+^{(S)} \,(1-y)^2 \, 4\,C^2 +
(d_j)_-^{(S)} \,(1 - 2\,C)^2 \right] D_{B/d_j} 
\right .\nonumber \\ 
&& \mbox{} \hspace{0.3cm} + \left. \left[ 
(\bar u_j)_+^{(S)} \,(1-y)^2 \, (1 - 4\,C)^2 +
(\bar u_j)_-^{(S)} \, 16\,C^2 \right] D_{B/\bar u_j} 
\right .\nonumber \\ 
&& \mbox{} \hspace{0.3cm} + \left.\left[ 
(\bar d_j)_+^{(S)} \,(1-y)^2 \, (1 - 2\,C)^2 +
(\bar d_j)_-^{(S)} \, 4\,C^2 \right] D_{B/\bar d_j} 
\right\}\
\eea
and 
\bea
N_{[\bar \nu ,\bar \nu]}^{(S)} (B) \!\!&=&\!\!  
 \sum \,\! _{_j} \left\{ \left[ 
(u_j)_+^{(S)} \, 16\,C^2 -
(u_j)_-^{(S)} \,(1-y)^2\,(1 - 4\,C)^2 \right] \Delta D_{B/u_j} 
\right . \nonumber \\ 
&&\mbox{} \hspace{0.7cm}+\left. \left[ 
(d_j)_+^{(S)} \, 4\,C^2 -
(d_j)_-^{(S)} \,(1-y)^2\,(1 -2\,C)^2 \right] \Delta D_{B/d_j} 
\right .\nonumber \\ 
&& \mbox{} \hspace{0.7cm}+ \left. \left[ 
(\bar u_j)_+^{(S)} \, (1 - 4\,C)^2 -
(\bar u_j)_-^{(S)} \,(1-y)^2 \, 16\,C^2 \right] 
\Delta D_{B/\bar u_j} \right .\nonumber \\ 
&&\mbox{} \hspace{0.7cm}+ \left.\left[ 
(\bar d_j)_+^{(S)} \, (1 - 2\,C)^2 -
(\bar d_j)_-^{(S)} \, (1-y)^2\, 4\,C^2 \right] 
\Delta D_{B/\bar d_j} \right\}
\eea
\bea
D_{[\bar \nu ,\bar \nu]}^{(S)} (B) \!\!&=&\!\!  
\sum \,\! _{_j} \left\{ \left[ 
(u_j)_+^{(S)} \, 16\,C^2 +
(u_j)_-^{(S)} \,(1-y)^2\, (1 - 4\,C)^2 \right] D_{B/u_j}  
\right . \nonumber \\ 
&& \mbox{} \hspace{0.3cm}+ \left. \left[ 
(d_j)_+^{(S)} \, 4\,C^2 +
(d_j)_-^{(S)} \,(1-y)^2\, (1 - 2\,C)^2 \right] D_{B/d_j} 
\right .\nonumber \\ 
&& \mbox{} \hspace{0.3cm}+ \left. \left[ 
(\bar u_j)_+^{(S)} \,(1 - 4\,C)^2 +
(\bar u_j)_-^{(S)} \,(1-y)^2 \, 16\,C^2 \right] 
D_{B/\bar u_j} \right .\nonumber \\ 
&& \mbox{} \hspace{0.3cm}+ \left.\left[ 
(\bar d_j)_+^{(S)} \, (1 - 2\,C)^2 +
(\bar d_j)_-^{(S)} \, (1-y)^2 \, 4\,C^2 \right] D_{B/\bar d_j} 
\right\}\>.
\eea

In the case of $\Lambda$ (or any baryon, rather than antibaryon) production, 
a simple expression for its longitudinal polarization $P$ can be 
obtained by neglecting the antiquark contributions and the terms proportional 
to $\sin^4 \theta_W$. For longitudinally polarized protons in this 
approximation we have
\be
P_{[\nu ,\nu]}^{(\pm)} (\Lambda) \simeq
- \frac{\sum_j \; \left\{ (u_j)_\mp \, (1 - 8\,C) \Delta D _{\Lambda/u_j} +
(d_j)_\mp \, (1 - 4\,C) \Delta D _{\Lambda/d_j} \right \} }
{\sum_j \; \left\{ (u_j)_\mp \, (1 - 8\,C) D _{\Lambda/u_j} +
(d_j)_\mp \, (1 - 4\,C) D _{\Lambda/d_j} \right \} } \,,
\label{pnnlap}
\ee
whereas for unpolarized protons, where $q_\pm$ is replaced by
$q/2$, we obtain
\be
P_{[\nu ,\nu]}^{(0)} (\Lambda) \simeq  
- \frac
{\sum_j \; \left\{ u_j \, (1 - 8\,C) \Delta D _{\Lambda/u_j} +
d_j \, (1 - 4\,C) \Delta D _{\Lambda/d_j} \right \} }
{\sum_j \; \left\{ u_j \, (1 - 8\,C) D _{\Lambda/u_j} +
d_j \, (1 - 4\,C) D _{\Lambda/d_j} \right \} } \>\cdot
\label{pnnla0}
\ee

\subsubsection{Present knowledge on $\mathrm \Lambda$ fragmentation functions
  and numerical results}
The study of $\Lambda$ polarization gives direct access to new fragmentation
functions of quarks into $\Lambda$; it is thus worth looking at the present 
knowledge of these functions, both unpolarized and polarized. 

{\it Unpolarized} $\Lambda$ fragmentation functions are determined by fitting 
$e^+e^- \to \, \Lambda + \bar \Lambda + X$ experimental data, which are
sensitive  only to singlet combinations such as $(D_{\Lambda/q} +
D_{\Lambda/\bar q})$.    It is therefore impossible to separate the
fragmentation functions relative  to $\Lambda$'s from those for
$\bar\Lambda$'s in a model-independent way.  Furthermore, flavour
separation is not possible without appropriate initial
assumptions. For example, one can assume $SU(3)$ flavour symmetry,  as
in Ref.~\cite{vog}; this leads to the simple approximation 
\be
D_{\Lambda/u} = D_{\Lambda/d} = D_{\Lambda/s} = 
D_{\bar\Lambda/\bar u} = D_{\bar\Lambda/\bar d} 
= D_{\bar\Lambda/\bar s}\,.
\ee
Other parameterizations rely on different assumptions: for instance, 
$SU(3)$ flavour and $SU(6)$ spin-flavour symmetry breakings lead in 
Ref.~\cite{blt} to  $D_{\Lambda/s} > D_{\Lambda/u} 
= D_{\Lambda/d}$, and in Ref. \cite{ind} to
$D_{\Lambda/s} \gg D_{\Lambda/u} = D_{\Lambda/d}$.

Other examples could be given, but at this stage much more 
stringent data are greatly needed for the determination of $\Lambda$ 
fragmentation functions.

For {\it polarized}  $\Lambda$ fragmentation functions the situation is also
problematic. In fact, polarized $\Lambda$ fragmentation functions 
are obtained by fitting $\Lambda$ polarization at LEP, that is only
sensitive to non-singlet combinations such as 
$\Delta D_{\Lambda/q} - \Delta D_{\Lambda/\bar q} = 
\Delta D^{val}_{\Lambda/q}$. 
In this case we have direct information on the valence contributions to the 
polarized $\Lambda$ fragmentation function, and the sea contributions are 
generated through evolution, for each flavour, starting from a given initial 
scale. It is therefore possible to determine polarized fragmentation 
functions for $\Lambda$ and $\bar \Lambda$ separately. Unfortunately, flavour 
separation is very difficult in the polarized case, because
experimental data currently
 available cannot even discern between remarkably different input 
models adopted for the valence contributions: for example, the ratios
$C_q(z) \equiv \Delta D_{\Lambda/q}(z)/D_{\Lambda/q}(z)$ range from
values such as 
$C_s =1$, $C_u = C_d = 0$ in the non-relativistic quark model to 
$C_s =0.6$, $C_u = C_d = -0.2$ in Ref.~\cite{bj}, to 
$C_s = C_u = C_d = z^{\alpha}$ in one of the scenarios of Ref.~\cite{vog}. 
The parameterizations of sea and gluon fragmentations have to be guessed 
by mere assumptions.

\begin{figure}[htb]
\centerline{\hspace{4cm}
\includegraphics[width=0.5\textwidth,clip,angle=-90]{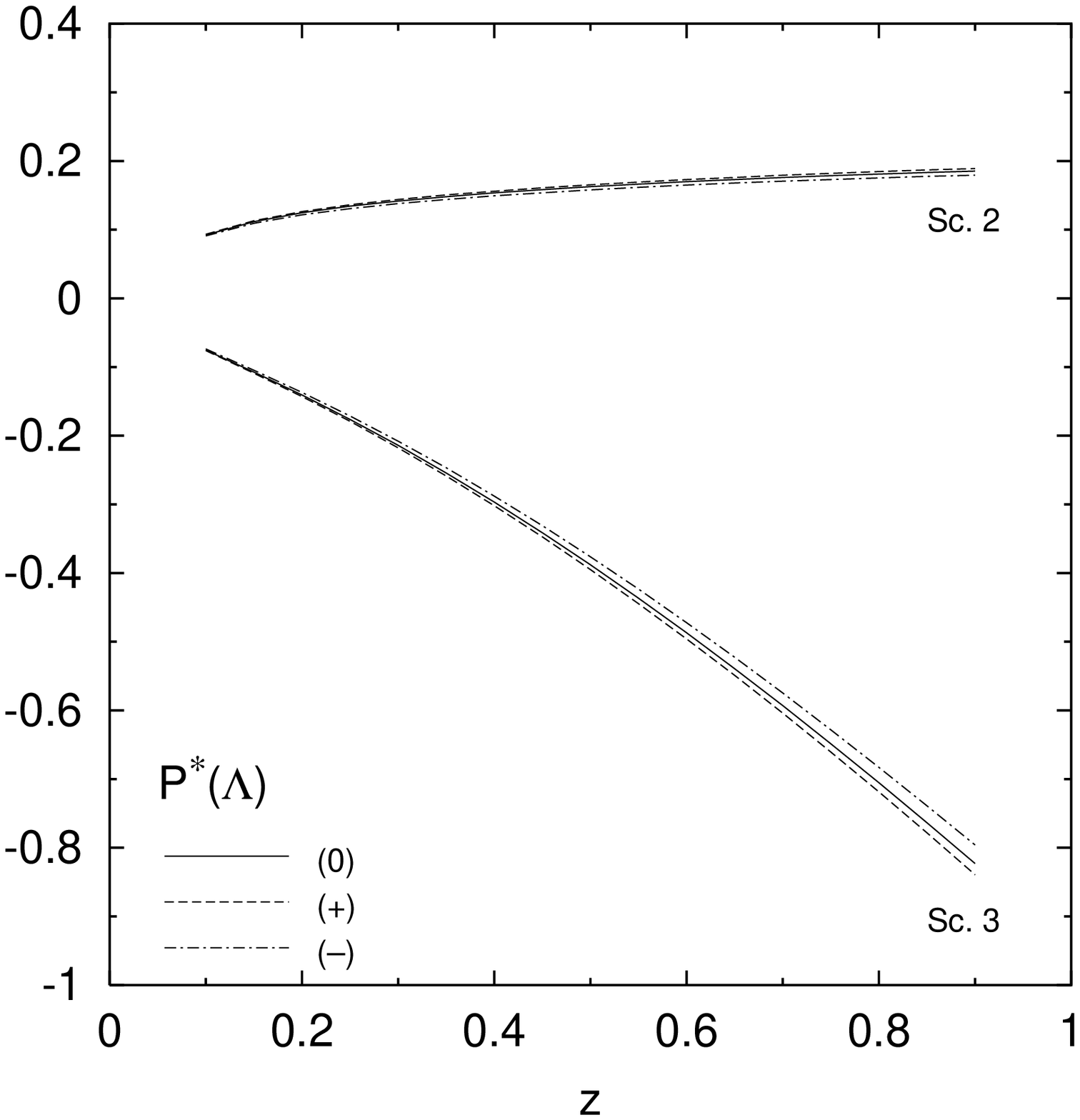} \hspace{-3.5cm}
\includegraphics[width=0.5\textwidth,clip,angle=-90]{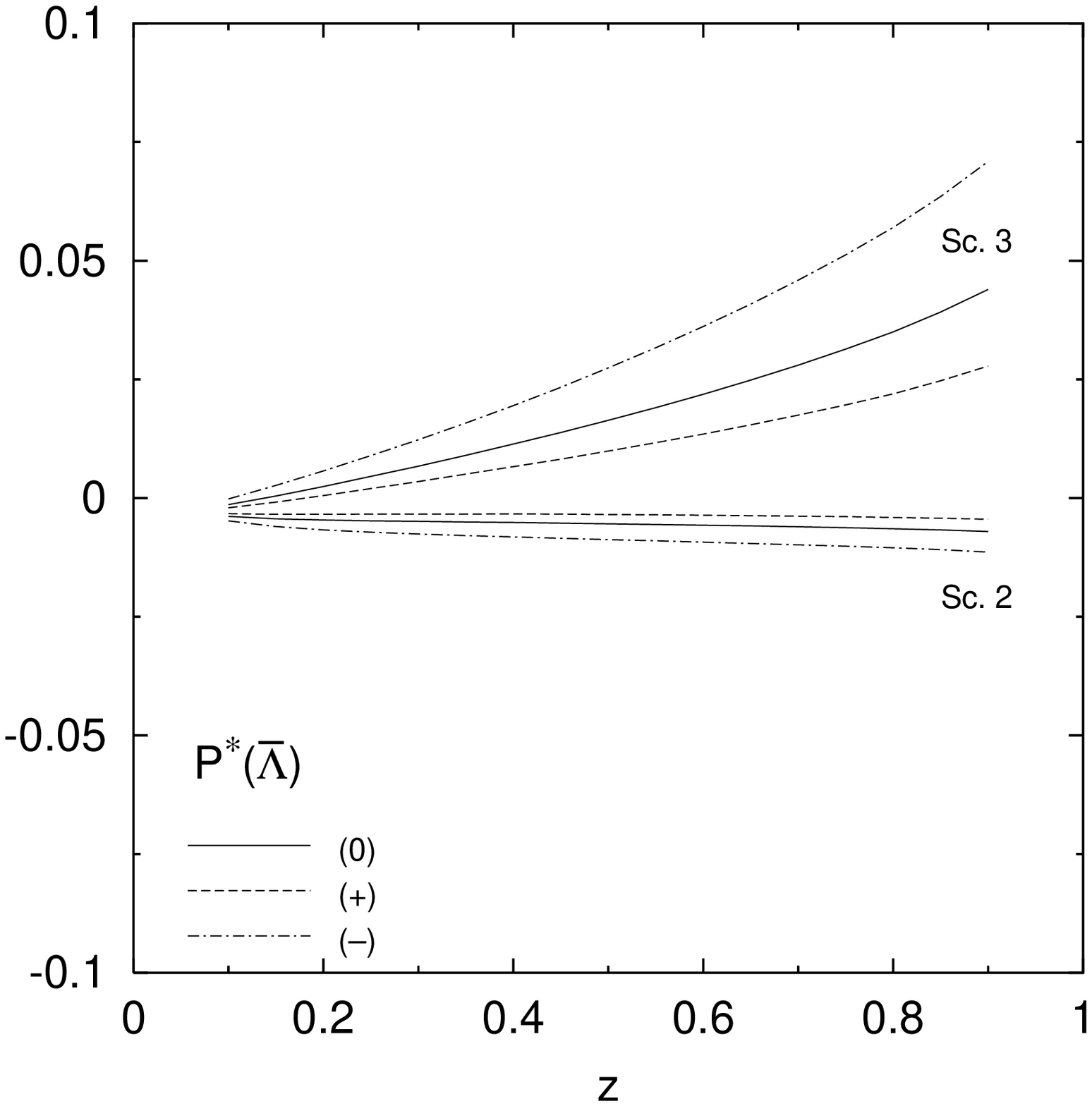} }
\caption{
$P^*(\Lambda)$ (left) and $P^*(\bar \Lambda)$ (right) for the process
$\nu_\mu\,p\to\mu^-\,\Lambda^{\uparrow}\,X$ as a function of $z$, as
predicted by using scenarios 2 and 3 for the polarized
$\Lambda$, $\bar\Lambda$ fragmentation functions of Ref.~\protect\cite{vog}.
Results for unpolarized (solid) and longitudinally polarized
($\lambda=+$: dashed; $\lambda=-$: dot-dashed) proton target are shown.
Kinematical conditions typical for the \nufact\  have been considered:
$E_\nu=30$ GeV, $0.01\leq x\leq 0.7$, 1 GeV$^2\leq Q^2\leq 100$ GeV$^2$,
$W^2 \geq 4$ GeV$^2$.}
\label{fig:abdm}
\end{figure}

The most consistent and model-independent fragmentation functions that
can be derived from $e^+e^-$ data are $(D_{\Lambda/q} + D_{\bar\Lambda/q})$
and, separately, $\Delta D_{\Lambda/q}$ and $\Delta D_{\bar\Lambda/q}$. 
These do not allow a computation of $P(\Lambda)$ and $P(\bar \Lambda)$.
The lack of knowledge of unpolarized fragmentation functions 
for separate $\Lambda$ and $\bar \Lambda$ could be partially overcome by 
introducing new measurable quantities:
\be
P^*(\Lambda) = \frac{d\sigma(\Lambda ^+) - d\sigma(\Lambda ^-)}
{d\sigma(\Lambda + \bar\Lambda )} = \frac{1}{1+T} \, P(\Lambda),
\ee 
\be
P^*(\bar\Lambda) = \frac{d\sigma(\bar\Lambda ^+) - 
d\sigma(\bar\Lambda ^-)} {d\sigma(\Lambda + \bar\Lambda )} = 
\frac{T}{1+T} \, P(\bar\Lambda) \,,
\ee 
where $T=d\sigma(\bar\Lambda)/d\sigma(\Lambda)$. 
These two quantities can be computed using the singlet unpolarized 
fragmentation functions, and they are simply 
related to $P(\Lambda)$ and $P(\bar\Lambda)$ through the factor $T$,
which can be measured; notice that one always has $P^* \le P$.
In Fig.~\ref{fig:abdm} we show, as an example, the expected values of  
$P^*(\Lambda)$ and $P^*(\bar\Lambda)$ obtained for typical
kinematical values of the planned \nufact\  experiments, adopting
two different sets of fragmentation functions from Ref.~\cite{vog}. 
Note that, in the chosen kinematical region, 
one expects $T \ll 1$ at large $z$ so that $P^*(\Lambda) \simeq P(\Lambda)$,
while $P(\bar\Lambda)$ could be sensibly larger than $P^*(\bar\Lambda)$
and in fact comparable in size with $P(\Lambda)$.
 
Let us finally emphasize the impact that neutrino semi-inclusive DIS (SIDIS)
 data would have on our knowledge 
of both nucleon distribution functions and $\Lambda$ fragmentation functions.
Once the $\Lambda$ and $\bar \Lambda$ polarizations in neutrino and 
antineutrino SIDIS off a polarized target will be measured, 
a carefully combined study of equations like (\ref{Pnu+-}),
 (\ref{Pantinu+-}), (\ref{pnlu}), (\ref{pnblds}), (\ref{pnnlap}),
 (\ref{pnnla0})  would not 
only allow the extraction of new information about polarized and unpolarized  
$\Lambda$ and  $\bar \Lambda$ fragmentation functions, but also a cross check
of our knowledge on nucleon polarized distribution functions. This combined 
program would offer an invaluable insight in the internal spin structure 
of hadrons in terms of their elementary constituents.

\subsection{Models for $\mathbf \Lambda$  polarization in the target
fragmentation region}
To describe the $\Lambda$~polarization in the TFR one has to model
the fracture function, namely the probability of finding a
parton $q$ in the target nucleon and a final hadron with given
momentum and polarization. This problem was addressed in only few
models, which we discuss below.

\subsubsection{$\mathbf SU(6)$ Quark--diquark model}
Longitudinal polarization of $\Lambda$'s produced in DIS was first
considered in ~\cite{bi}. After kicking out a 
left-handed $u$ or $d$ quark from an unpolarized nucleon, we have
the following relative probabilities for the polarization states of
the remnant diquark:

\begin{eqnarray}
\nu p:\;p\ominus d^\uparrow &\Rightarrow& \nonumber
\frac{1}{36}[2(uu)_{10}+4(uu)_{1-1}]\\ \nonumber
\nu n:\;n\ominus d^\uparrow &\Rightarrow&
\frac{1}{36}[9(ud)_{00}+(ud)_{10}+2(ud)_{1-1}]\\ 
\bar{\nu} p:\;p\ominus u^\uparrow &\Rightarrow&
\frac{1}{36}[9(ud)_{00}+(ud)_{10}+2(ud)_{1-1}]\\ 
\bar{\nu} n:\;n\ominus u^\uparrow &\Rightarrow& 
\frac{1}{36}[2(dd)_{10}+4(dd)_{1-1}], \nonumber
\label{eq:remdiq}
\end{eqnarray}
%ak
where, for example, $(uu)_{10}$ denotes a $uu$-diquark with total spin 
$S=1$ and spin projection $S_z=0$.
%ak 
It is assumed that during recombination with the unpolarized $s$ quarks the
diquark does not changes its polarization. Since in the naive quark
model (NQM)
the polarization of $\Lambda$'s is equal to the $s$-quark polarization,
the $\Lambda$'s that are directly produced will be unpolarized. However, the
final state $\Lambda$'s may also be produced indirectly via
electromagnetic decay of $\Sigma^0$ or strong decay of $\Sigma^*$'s. In
both cases, the non-strange diquark changes its spin from 1 to 0, while
the strange quark retains its polarization ~\cite{ashli}. Using the
$SU(6)$ wave functions of octet and decuplet baryons, we obtain for
the $\Lambda$ polarization:
\begin{eqnarray}
P_\Lambda^{\nu p}&=&P_\Lambda^{\bar{\nu} n}\simeq -0.55\\
P_\Lambda^{\nu n}&=&P_\Lambda^{\bar{\nu} p}\simeq -0.05,
\label{eq:lpolpn}
\end{eqnarray}
yielding, for an isoscalar target,
\begin{eqnarray}
P_\Lambda^{\nu (p+n)}=P_\Lambda^{\bar{\nu}(p+n)}\simeq -0.30.
\label{eq:lpoliso}
\end{eqnarray}

\subsubsection{Meson-cloud model}
Some non-perturbative features of the nucleon structure, such as
the deviation from the QCD-parton-model-inspired  Gottfried
sum rule, can be explained in the framework of the meson-cloud model.
The pion-cloud model provides a natural explanation  of the
isospin symmetry breaking in the unpolarized proton sea. In the case of 
polarized DIS, the scattering on the lowest $\Lambda\,K$ and $\Sigma\,K$ 
component of the nucleon wave function  provides a possible mechanism 
leading to a violation of the Ellis--Jaffe sum rule.
The polarization of $\Lambda$'s produced in the TFR in this model has
been considered in \cite{melth95}.
It appears that $\Lambda$~ polarization is almost 100\% anticorrelated
with the target polarization. It is thus expected to be 0 for an 
unpolarized target.

\subsubsection{Polarized-intrinsic-strangeness model}
\begin{figure}[ht]
\begin{center}
\includegraphics[width=0.75\textwidth,clip]{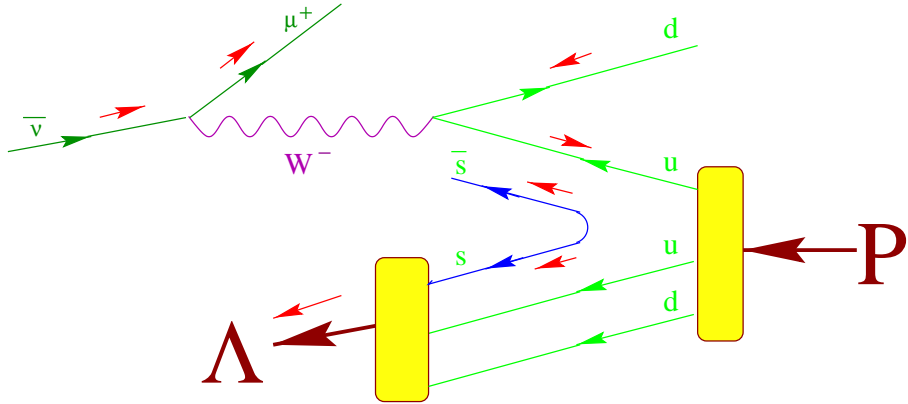} 
\end{center}
\caption{Dominant diagram for $\Lambda$ production in the
target fragmentation region due to the scattering on a valence $u$-quark.
Each small arrow represents the longitudinal polarization
of the corresponding particle.}
\label{fig:nubpr}
\end{figure}

The polarized-intrinsic-strangeness model (for a review, see ~\cite{ekks})
qualitatively reproduces experimental features of the $\phi$ production in
$\bar{p}N$ annihilation. The model is based on  the following 
major observations. First, the fact that the masses of pseudoscalar mesons 
are small with respect to the
typical hadronic scale can be attributed to the existence of effective
strong attraction in the $J^{PC}=0^{-+}$ channel.  
Secondly, from phenomenological
analyses of the quark condensates in the framework of QCD sum rules,
it is known that the vacuum density of strange--antistrange quark
pairs is comparable to the density of $u$ and $d$ quarks.
It is natural to assume that the polarized constituent quark can
contain an $\bar s s$ pair with the vacuum quantum numbers
corresponding to a
$^3P_0$ state. Hence, in the polarized nucleon, the spin of $\bar s$ will 
be antiparallel to the valence quark spin, $S_z(\bar s)=-1/2$ for
$S_z(q_v)=1/2$. 
In Ref.~\cite{ekk} there was considered the case (in the following
referred to as ${\bf A}$) of an angular momentum
projection of the $\bar s s$ pair $L_z(\bar s s)=+1$ ($S_z(\bar s s)=-1)$.
In this case any $s$ quark in the target fragment
should have {\it negative} longitudinal polarization, so that the
longitudinal $\Lambda$ polarization should also be {\it negative}, see
Fig.~\ref{fig:nubpr}.

In the quark--parton model of deep inelastic
$\nu$ or $\bar \nu$ scattering, the net longitudinal
polarization $P_{s}$ of the remnant $s$ quark is given by
\begin{equation}
P_{s}=\frac{\sum_q C_{s\,q}N_q -
\sum_{\bar q} C_{s\,\bar q}N_{\bar q}}{N_q+N_{\bar q}},
\label{eq:ps}
\end{equation}
where $N_q$ ($N_{\bar q}$) is the total number of events 
in which a quark (antiquark) is struck, and $C_{s \,q}$ is the
spin-correlation coefficient. The antiquarks contribute with
a negative sign because their CC weak interactions are
right-handed. The final $\Lambda$ polarization $P_\Lambda=D_F\,P_s$,
where $D_F$ is a dilution factor that describes the spin transfer
during hadronization.

Let us consider also the scenario ${\bf B}$, where both projections 
$L_z(\bar s s)=+1$ and $L_z(\bar s s)=0$ of the 
$\bar s s$-pair angular momentum contribute with equal probabilities. 
$L_z(\bar s s)=0$ means neglecting the transverse motion of the
$\bar s s$ pair ~\cite{ck}. 

The correlation of the remnant $s$-quark polarization with that of any
other struck sea quark ($q_{sea} \neq \bar s$) depends on whether they come 
from the same parent constituent quark. If they do, which might be the 
dominant case, then a strong spin-correlation is expected 
(case {\bf a}). 
Otherwise, the correlation should be reduced (case {\bf b}).
Then for the spin-correlation coefficients we have
\begin{eqnarray}
\nonumber
{\bf A:} \quad C_{sq_{val}}&&=-1,\;C_{s \bar s}=1 \\ \nonumber
%ak
{\bf B:} \quad C_{sq_{val}}&&=-\frac{1}{3},\;C_{s \bar s}=\frac{1}{3}\\ 
%ak
{\bf Aa:} \quad C_{sq_{sea}}&&=1, \\ \nonumber
{\bf Ba:} \quad C_{sq_{sea}}&&=\frac{1}{9}, \\ \nonumber
%ak
{\bf Ab, Bb:}  C_{sq_{sea}}&&=0. \\ \nonumber
%ak
\label{eq:scenar}
\end{eqnarray}
The results for the remnant $s$-quark polarization 
and the predictions of {NQM} are
presented in the Table~\ref{tab:restfr} together with the measured $\Lambda$~
polarization. 
\begin{table}[ht]
\begin{center}
\caption{$\Lambda$ polarization in the TFR of (anti)neutrino
SIDIS. Model predictions in boldface lie within $\pm 1\sigma$ of the
  experimental data.}
%\label{tab:results}
\begin{tabular}{|c||c|c|c|c|c|c|} \hline
Experiment (Reaction)&Data&{Aa}&{Ab}&{Ba}&{Bb}&{NQM}\\
\hline \hline
WA21~\cite{wa21} ($\nu_{\mu} - p$)&-0.29 
$\pm$ 0.18&-0.51&-0.75&-{\bf 0.22}&-{\bf 0.25}&-0.55\\\hline
WA21~\cite{wa21} ($\bar{\nu}_{\mu} - p$)&-0.38
$\pm$ 0.18&-0.85&-0.92&-{\bf 0.30}&-{\bf 0.31}&0.03\\\hline
WA59~\cite{wa59} ($\bar{\nu}_{\mu} - Ne$)&-0.63
$\pm$ 0.13&-0.82&-0.91&-0.29&-0.30&-0.30\\\hline
E632~\cite{e632} ($\nu_{\mu} - Ne$)&-0.43
$\pm$ 0.20&-0.70&-0.84&-{\bf 0.27}&-{\bf 0.27}&-0.30\\\hline
NOMAD~\cite{nomad} ($\nu_{\mu} - C$)&-0.21
$\pm$ 0.04&-0.59&-0.80&-{\bf 0.24}&-{\bf 0.27}&-{\bf 0.30}\\\hline
NOMAD~\cite{nomad} ($\nu_{\mu} - ``p''$)&-0.29
$\pm$ 0.06&-0.54&-0.77&-{\bf 0.23}&-{\bf 0.26}&-0.55\\\hline
NOMAD~\cite{nomad} ($\nu_{\mu} - ``n''$)&-0.16
$\pm$ 0.05&-0.61&-0.81&-0.25&-0.27& 0.03\\\hline
\end{tabular}
\label{tab:restfr}
\end{center}
\end{table}
The data on longitudinal $\Lambda$ polarization 
from the NOMAD experiment~\cite{nomad}
have the  best statistical accuracy, and we will base the conclusions
mainly on comparisons with these data.
As one can see from Table~\ref{tab:restfr}:
\begin{itemize}
\item the predictions of NQM are not so different from 
  the NOMAD data on  the isoscalar (mainly carbon) target, 
  but contradict the NOMAD $p$ and $n$ and the WA21 data;
\item the meson-cloud model predicts zero polarization and is in
  contradiction with all data;
\item the best description of the NOMAD data is achieved in the 
polarized-intrinsic-strangeness 
model with scenarios {\bf Ba} and {\bf Bb}, provided 
  that $D_F\approx 1$. 
%ak
  The remnant $s$-quark polarization is higher in absolute value for the 
  {\bf Aa} and {\bf Ab} scenarios.
  If one allows a large depolarization during
  hadronization, $D_F \approx 0.4$-$0.5$, then the scenarios {\bf Aa} and 
  {\bf Ab} can also provide a fair description of all data.
%ak**** WHERE DOES THIS CONCLUSION COME FROM? ****
\end{itemize}
It is possible to distinguish between scenarios {\bf A} and {\bf B} of
the last model by measuring the $\bar \Lambda$~polarization in the TFR of the 
neutrino SIDIS. One should expect that $P_{\bar \Lambda} \approx P_\Lambda$ 
in case {\bf A} and $P_{\bar \Lambda} \approx 3 P_\Lambda$ in
case {\bf B}. 

The models predict different polarizations for
intrinsic $s$- and $\bar s$-quark sea:
\begin{itemize}
\item NQM:  $\Delta s = \Delta\bar{s} = 0$;
\item meson-cloud model~\cite{ints}:$\Delta \bar{s}\approx 0,\,\Delta s < 0$;
\item intrinsic-strangeness-model {\bf A}:
  $\Delta s \approx \Delta\bar{s} < 0$;
\item intrinsic-strangeness-model {\bf B}:
  $\Delta s \approx 1/3;\; \Delta\bar{s} < 0$;
\end{itemize}
In principle, these predictions can be independently tested by measuring the
asymmetries of strange-particle production in the current
fragmentation region of SIDIS or of (anti)neutrino DIS on
a polarized target.

 One could try to
improve the naive quark model by taking into account the $SU(6)$
symmetry breaking as it was done in~\cite{blt}. In the meson-cloud 
model one can expect that the contributions from higher possible
fluctuations with vector meson $K^{+*} \Lambda$ will lead to non-zero 
$\Lambda$~polarization. 
However, the estimates of ref.~\cite{ints} show that the 
relative probability of this state is small with respect to 
$K^{+} \Lambda$  (less than 10\%). Moreover they predict a 
positively-correlated $s$-quark spin in a polarized proton.

\subsection{Discussion}
The study of $\Lambda$ polarization in the CFR
 allows us to understand
the spin transfer from the quarks to the $\Lambda$. Due to the natural
helicity and flavour selection of neutrino couplings, we can precisely
 single out specific quark contributions: this information cannot
be obtained from the usual lepton-initiated DIS. The information on polarized
and unpolarized fragmentation functions into $\Lambda$ which is available
from LEP data is scarce and uncertain; also data from NOMAD are far from
being decisive in fixing the features of quark-spin transfer. The
\nufact\  data will induce a big improvement in our understanding
of fragmentation processes.

In the target fragmentation region, one can study a new phenomenon,
namely,
the polarization transfer
from the lepton to the final hadron. 
The models discussed here are the first attempts to describe this
effect. 
The different models of spin transfer from a polarized quark to a polarized
$\Lambda$ are able to describe the existing LEP data 
but give different predictions for (anti)neutrino DIS 
%ak(**** ANY REF HERE? ***)
(see, for example, \cite{kbh,mssy} and~\cite{naumov} where comparison 
with NOMAD data are presented).
%ak

Finally, we would like to mention that the best existing data on
$\Lambda$~polarization, which come from the NOMAD experiment, are based on
the analysis of about one million DIS events. The statistics at
the \nufact\ is expected to be a hundred times higher, thus providing
ten times better statistical accuracy. This will provide a more detailed
study of the $\Lambda$~polarization dependence on kinematic variables and
allow better comparisons with different model predictions.

\section{SEARCHES FOR NEW PHYSICS}
We now review briefly some prospects in searching for new physics using
the intense beams from a neutrino factory.
\subsection{A search for $\mathbf Z^{\prime}$ in
  muon--neutrino-associated  charm  production} 
In many extensions of the SM the presence of 
an extra neutral boson, $Z^{\prime}$, is invoked. A precision study of 
weak NC-exchange processes involving only second-generation 
fermions is still missing. A search for $Z^{\prime}$ in 
muon--neutrino-associated charm production has recently been
proposed~\cite{zprime}.  This process only involves $Z^{\prime}$ 
couplings with fermions from the second generation. It is interesting
because  an exotic $Z^{\prime}$ with stronger coupling to the $I_3 = 1/2$ 
component of weak isospin doublets could still give measurable effects at 
neutrino factories, unlike LHC experiments, which are only sensitive 
to the $Z^{\prime}$ coupling to  charged leptons ($I_3 = -1/2$). 

We briefly review the method and the application to neutrino factories
using an {\it  ideal detector}. For a detailed discussion of this
method, see Ref.~\cite{zprime}. 
 
\subsubsection{The process} 
\label{sec:contact} 
At $Q^2 \ll M_Z^2$  
the NC effective Lagrangian ruling the associated charm  
production induced by $\nu_\mu$  and including the new physics term is  
given by~\cite{zprime} 
\begin{equation} 
{\cal L}_{T}^{\nu c \overline{c}} =  
- \frac{G_F}{\sqrt{2}} 
\overline{\nu}_\mu \gamma ^{\alpha }(1-\gamma _{5})\nu_\mu~ 
\overline{c}\gamma_\alpha \left[\epsilon_V(c) - \epsilon_A(c) \gamma_5 
\right]c~~~, 
\label{eq:T} 
\end{equation} 
where 
\begin{eqnarray} 
\epsilon_V(c)&=& \epsilon_L(u)+\epsilon_R(u)    
+ \left(M_Z^2 \over M_{Z'}^2\right) (\eta_L + \eta_R ) \nonumber\\ 
&\equiv &  \epsilon_V(u)+ \left(M_Z^2 \over M_{Z'}^2\right)  
\eta_V = \left[ 1+ \left(M_Z^2  \over M_{Z'}^2\right) x \right] 
\epsilon_V(u)~~~, 
\label{epsv}\\ 
\epsilon_A(c)&=& \epsilon_L(u)-\epsilon_R(u)  
+ \left(M_Z^2 \over M_{Z'}^2\right) 
(\eta_L - \eta_R ) \nonumber\\ 
&\equiv&  \epsilon_A (u) + \left(M_Z^2  
\over M_{Z'}^2\right) \eta_A = \left[1+ \left(M_Z^2  \over M_{Z'}^2\right)  
y\right] 
\epsilon_A(u)~~~, 
\label{epsa} 
\end{eqnarray} 
and the parameters $x$ and $y$ give the departure of the couplings from 
SM predictions. 
 
\subsubsection{Description of the method} 
\label{method}  
 
The peculiar topology of the associated charm production in $\nu_{\mu}$  
NC interactions is exploited:  
two charmed hadrons in the final state. Consequently, there are no  
other physical processes that may mimic it.  
Experimentally we are sensitive to the ratio  
\begin{equation} 
R = \frac{\sigma_{c \bar{c}}^{NC}}{\sigma^{CC}} 
\end{equation} 
which can be written as the product  
\begin{equation} 
\label{eq:rapporto} 
R = \frac{\sigma_{c \bar{c}}^{NC}(Z^{0} + Z^{\prime})} 
{\sigma_{c \bar{c}}^{NC}(Z^{0})}  
\times \frac{\sigma_{c \bar{c}}^{NC}(Z^{0})}{\sigma^{CC}} = r \times f 
\end{equation} 
where $\sigma_{c \bar{c}}^{NC}(Z^{0})$ is the cross section of the 
associated charm-production process in $\nu_{\mu}$ 
interactions in the absence of the  
$Z^{\prime}$ boson, $\sigma_{c \bar{c}}^{NC}(Z^{0}+ Z^{\prime})$ includes 
the  
contribution of the new neutral boson, and $\sigma^{CC}$ is the $\nu_{\mu}$  
DIS CC cross section. 
 
From Eq.~(\ref{eq:rapporto}) it is clear that the relevant information about  
the $Z^{\prime}$ comes from the ratio $r$, 
which is unity in absence  
of the $Z^{\prime}$. 
In the following we assume a $50$~GeV mono-energetic  
$\nu_{\mu}$ beam  
\footnote{ 
The results achievable with a real neutrino spectrum of mean  
energy $\langle E_{\nu}\rangle$ are  rather well reproduced  by  
using a simple mono-energetic beam  
with energy equal to $\langle E_{\nu}\rangle$.}.  
Under this assumption, by using the HERWIG simulation program to compute the 
the ratio $f$, we get $ f = (1.25 \pm 0.01) \times 
10^{-4}$.

\begin{figure}[ht] 
  \begin{center} 
              \resizebox{0.5\textwidth}{!}{ 
\includegraphics{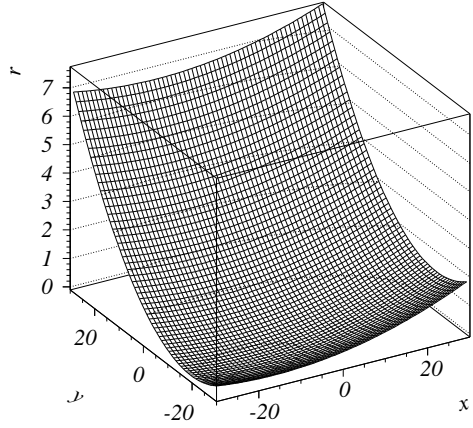}} 
        \caption{\small The ratio $r$ is plotted by assuming $M_{Z^{\prime}} = 
500~{\mathrm GeV}/c^2$. } 
   \label{fig:r50b} 
  \end{center} 
\end{figure}

If we parameterize the ratio $r$ in terms of the $x$, $y$ and 
$M^2_{Z'}$ variables defined in Section \ref{sec:contact},   
the most general expression we obtain is: 
\begin{equation} 
r(x,y,M^2_{Z'}) =  
1 +  \left(500  \over M_{Z'}\right)^2 ( A_1 y + B_1 x)  
 +  \left(500  \over M_{Z'}\right)^4  
(A_2 y^2 +  B_2 x^2 + C_1 xy) . 
\label{eq:fitfor} 
\end{equation} 
Fitting the data from the calculation with the previous function,   
we obtain the following values of the coefficients: $A_1 = 0.1$, $A_2
= 0.003$, $B_1 = 0.02$, $B_2 = 0.0007$ and $C_1 = -0.0002$.  
The fit is valid in the $[-30, 30]$ range for both $x$ and $y$ variables.  
Figure~\ref{fig:r50b} shows the fitted function $r$ for $M_{Z^{\prime}} = 
500~\mbox{GeV/c}^2$.
The number of observed events, $N_{S}$, can be written as 
 \begin{equation} 
N_{S} = N_{c \bar{c}} \cdot \frac{\varepsilon_S}{\varepsilon_B} \cdot  r 
\; ,
\label{eq:events} 
\end{equation} 
where $N_{c \bar{c}}$ is the number of observed events without the  
$Z^{\prime}$ effect, $\varepsilon_S$ and $\varepsilon_B$ are  
the reconstruction efficiencies for the events with and 
without a $Z^{\prime}$, respectively.  
 
\subsubsection{Measurement accuracy} 
\begin{figure}[hbtp] 
  \begin{center} 
              \resizebox{0.9\textwidth}{!}{ 
\includegraphics{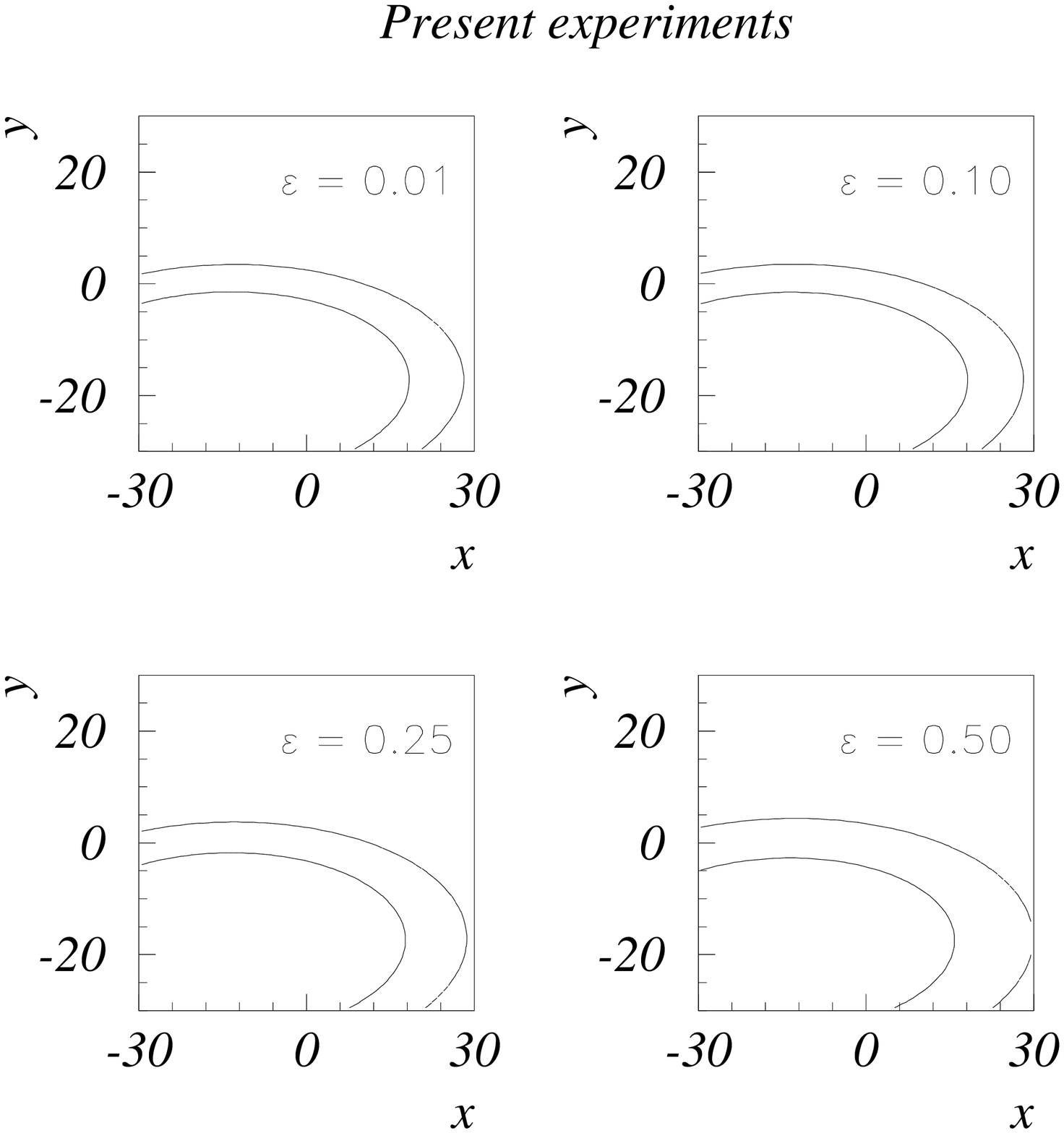}
\includegraphics{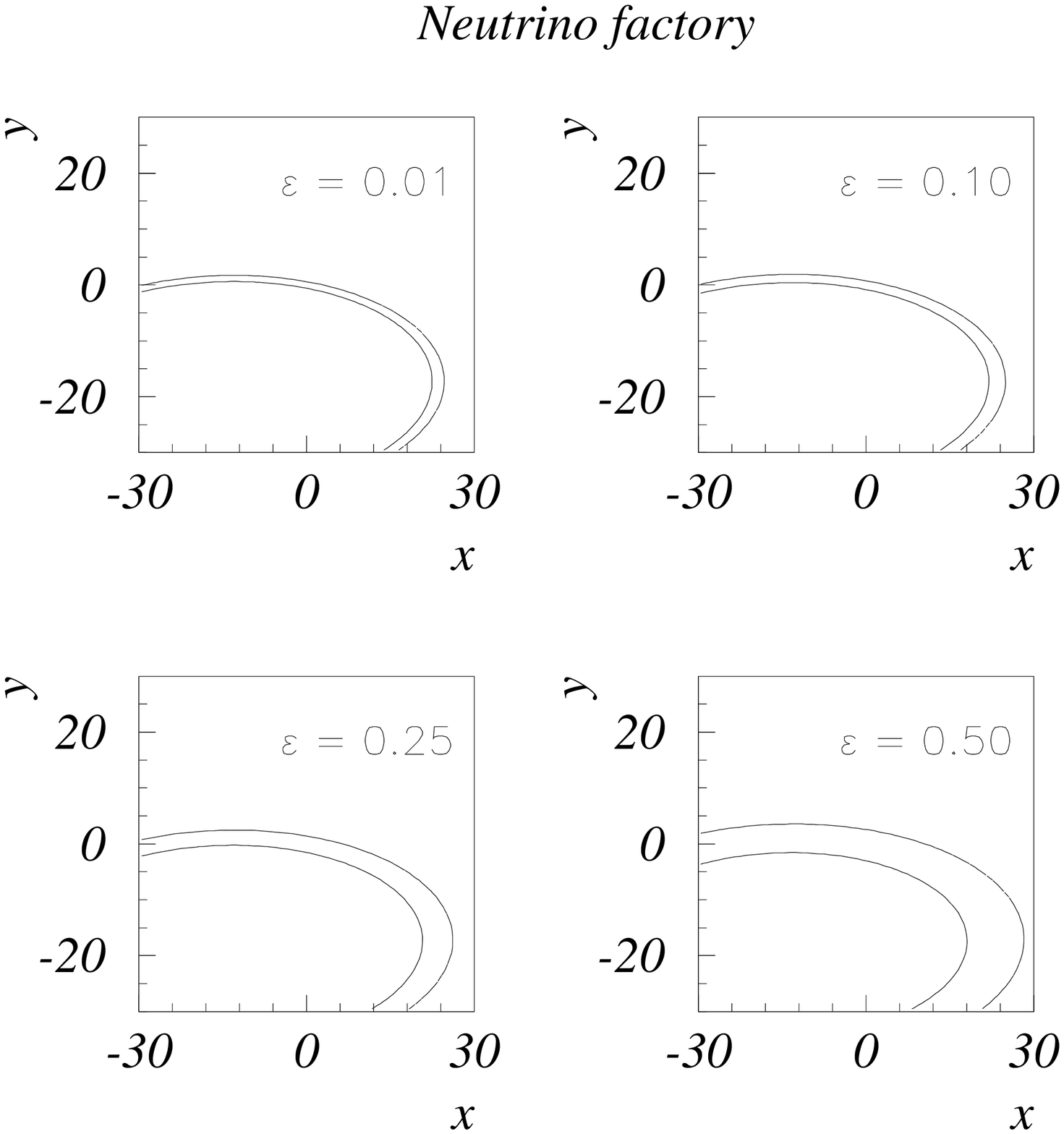}} 
        \caption{\small The sensitivity plots for the $x$ and $y$
          variables at $M_{Z^{\prime}}$ = 500 GeV$/c^2$ are shown.
          $\varepsilon$ indicates the systematic error. $10^7$
          CC interactions are assumed to be collected at
          a \nufact\  (right plot). The sensitivity achievable
          with present experiments is also considered (left plot). }
   \label{fig:cont50ev} 
  \end{center} 
\end{figure}

Once  the charmed particles have been tagged, the $Z^{\prime}$ effect 
would show up as an excess/defect of doubly charmed events in NC 
interactions. 
From \equ{eq:fitfor} we can argue that for `large' $Z^{\prime}$ 
couplings, i.e.~$x$ $\mbox{and}$ $y> 20$, we can get an enhancement of 
the associated charm production of about a factor 7. 
On the other hand, if we do not observe any excess/defect, we can put a 
limit on the $x$ and $y$ parameters. As an example we report in  
Fig.~\ref{fig:cont50ev} the sensitivity plot at 95\% C.L. 
for the $x$ and $y$ variables  at $M_{Z^{\prime}}$ = 500 GeV$/c^2$.  
Different systematic errors are assumed from 1\% to 50\%.  
 
The allowed region of parameters is obtained from the formula  
\begin{equation} 
1-1.96\cdot \frac{\sigma}{N_{c\bar{c}}} \leq  
\frac{\varepsilon_S}{\varepsilon_B} \cdot  r \leq 1+1.96 \cdot  
\frac{\sigma}{N_{c\bar{c}}}  \; ,
\end{equation} 
where $\sigma$ is defined as  
\begin{equation} 
\sigma = (\varepsilon^{2}_{stat}+\varepsilon^{2}_{syst}) 
\end{equation} 
and includes the error on the event counting from both a statistical and  
systematics sources. The factor 1.96 takes into account the required  
confidence level. 

In Fig.~\ref{fig:cont50ev}, for each plot,  
the two lines bound the region of coupling parameters where no significant  
excess/defect of associated charm-production events is found. In other words,  
an observation of a number of charm pair events in  
agreement with SM predictions excludes the regions outside the band.  
For each plot shown in Fig.~\ref{fig:cont50ev},  we report 
in Table~\ref{tab:limit} 
the $\Delta x$ and $\Delta y$ values, respectively the 
bandwidth at $y = -30$ and $x = -30$. 
\begin{table}[tbp] 
\small 
\begin{center} 
\begin{tabular}{||c|c|c|c|c||} 
\hline 
Scenario & \multicolumn{2}{l|}{Present experiments} & 
\multicolumn{2}{l||}{\nufact\ } \\
\hline\hline 
$\varepsilon$ & $\Delta x$  & $\Delta y $ & $\Delta x $ & $\Delta y$ \\
\hline 
0.01 & 12.0  & 5.5  & 2.5  & 1.5 \\
\hline 
0.10 & 12.5  & 6.0  & 3.5  & 2.0 \\
\hline 
0.25 & 14.0  & 6.5  & 6.5  & 3.0 \\
\hline  
0.50 & 18.0  & 8.0  & 13.0  & 5.5 \\
\hline 
\end{tabular} 
\caption{Band widths at $x = -30 $ and $y = -30 $ for all the sensitivity 
plots shown in Fig.~\ref{fig:cont50ev}. $\varepsilon$ indicates the systematic 
uncertainty.} 
\label{tab:limit} 
\end{center} 
\end{table}
We assume that $10^7$ CC interactions are collected at the \nufact. 
Figure~\ref{fig:cont50ev} shows the comparison between the sensitivity
achievable with  present experiments\footnote{Such a measurement
could also be exploited by the NuTeV experiment, which recently
measured NC charm production in $\nu_\mu$--Fe
scattering~\cite{nutev}.} (say CHORUS) and the one obtainable at a
\nufact. At a \nufact\  the systematic uncertainties
would play an important role if they are larger than about 10\%, while the
statistics is the leading contribution for present experiments. 

\subsection{Bounds on 4-fermion operators from a $\mathbf\nu$-Factory}
\label{sec:newph}
\def\beq{\begin{equation}}
\def\eeq{\end{equation}}
\def\bea{\begin{eqnarray}}
\def\eea{\end{eqnarray}}

\def \lsim{\mathrel{\vcenter
     {\hbox{$<$}\nointerlineskip\hbox{$\sim$}}}}
\def \gsim{\mathrel{\vcenter
     {\hbox{$>$}\nointerlineskip\hbox{$\sim$}}}}

\def\ffops{4-fermion operators~}

%\subsection{what}
Many types of Beyond-the-Standard-Model (BSM) physics 
appear at energy scales above that
of the neutrino-scattering process, so their tree-level effects
at a \nufact\  can
be parametrized by 4-fermion operators.
These can contribute in the $\mu^{\pm}$ decay, or in the scattering in the
detector.  All new physics induced by scalars or vectors can be
written, via Fierz transformations, as $(V \pm A)(V-A)$ vertices,
which we normalize as:
\beq
\eta_{PL}^{ \bar{\ell} e \bar{\nu}\nu} ( \bar{\ell} \gamma^{\alpha}P e)
(\bar{\nu} \gamma_{\alpha}P_L \nu), ~~~~
\eta_{PL}^{ \bar{u}_i d_j \bar{\ell}\nu} ( \bar{u}_i \gamma^{\alpha}
P d_j)
(\bar{\ell} \gamma_{\alpha}P_L \nu) \; ,
\label{s}
\eeq
where $P = P_{L} = (1 - \gamma_5)/2$ or
 $P = P_{R} = (1 + \gamma_5)/2$, and $\ell$
is a charged lepton.  Neutrinos
without index can be of any flavour. We assume that
the strongest neutrino interactions are
SM weak, so that the neutrino-flavour basis makes
sense perturbatively (see ref.~\cite{g}). 

Bounds on a variety of extensions of
the SM from neutrino
scattering were reviewed in \cite{LLM};
a brief update can be found in \cite{conrad}.
 New physics in neutrino 
scattering that cannot be parametrized
by \ffops has recently been discussed in \cite{bkbook}.
A catalogue of old constraints on \ffops can be found 
in Refs.~\cite{LLM} (many models) and \cite{dbc} (leptoquarks and
bileptons)---more up-to-date bounds are reviewed in
\cite{L} ($Z^{\prime}$s) and many other recent papers.
Constraints on $Z^{\prime}$s from
a neutrino factory were studied
in the previous section, and bounds on $R$-parity-violating couplings
from neutrino scattering have been discussed in
\cite{dgmm}.

%\subsection{why}
A \nufact\ can improve bounds on \ffops
involving neutrinos by many orders
of magnitude.  However, since neutrinos
are weakly interacting, the resulting bounds
are generically weaker than
bounds on \ffops involving charged leptons
rather than neutrinos. For instance,
new physics that can contribute both to
lepton-flavour violation in
neutrino scattering and to
$\mu \rightarrow e \gamma$
will be more strongly constrained
(or more readily detected) in
$\mu \rightarrow e \gamma$ than
in neutrino scattering, as can be seen from 
the bounds on lepton-flavour violation discussed 
in the Report of the Stopped-Muon Working Group~\cite{gianrep}. 
It could nonetheless be interesting to look
for BSM physics  in neutrino 
scattering at the near detector
of a neutrino factory. 

First of all, any observed
lepton-flavour violation  must be due to something other
than oscillations. Neutrinos are in fact produced 
only a few hundred metres away from the detector, 
and have no time to oscillate, given the current determinations of
the mixing parameters.
 New physics that can
be parametrized by \ffops can then be distinguished from oscillations.

Secondly, there is BSM physics that can induce $\nu \nu \ell \ell$
4-fermion operators, without inducing $\ell \ell \ell \ell$  
operators\footnote{For example by the exchange
of a singlet that couples to
$(\nu_i \ell_j - \nu_j \ell_i)$.}
($\ell$ is a charged lepton.)
For quarks,  BSM physics (\eg\ LQ, $Z'$)
that induces $\nu \bar{\nu} q q$
also induces $\ell \bar{\ell} q q$,
so we do not discuss operators involving quarks.
There could nonetheless be  interesting limits on
flavour-changing vertices involving
charm quarks, from neutrino scattering off nucleons,
because rare $D$-meson
decays are poorly measured.

As explained above, we only
consider the case of flavour violation induced in the muon decay.
Since we do not carry out a complete simulation of the event rates, we
shall assume that the kinematics of the
muon decay induced by the first
vertex in Eq.~(\ref{s}) is the {\it same}
as for the SM  $(V-A)(V-A)$ current. We shall then parameterize 
the muon-decay rate induced by the vertex (\ref{s}) as:
\beq
\Gamma (\mu \rightarrow e \bar{\nu}_i \nu_j )
= \left( \frac{\eta_{PL}^{ \bar{\mu}e \bar{\nu}_i \nu_j}}
 {2 \sqrt{2} G_F} \right)^2 
\Gamma_{SM} (\mu \rightarrow e \bar{\nu}_e \nu_\mu)
\equiv [\epsilon_{PL}^{ \bar{\mu}e \bar{\nu}_i \nu_j}]^2
\Gamma_{SM} (\mu \rightarrow e \bar{\nu}_e \nu_\mu)
\eeq
and will assume the neutrino beam shapes induced by these BSM decays
to be approximated by the SM one.

The new physics signal that we explore is given by a final-state
$\tau$ or by a wrong-sign muon (WSM; \eg\ a $\mu^+$ produced in the
detector by a neutrino beam produced in $\mu^-$ decay).
These signals do not suffer from any irreducible physics background. 
Some $\tau$'s can be produced in the decays of charmed mesons produced by
SM CC interactions, but in principle these events will also contain an
electron or muon from the CC vertex. The $\tau$ decays from charm
produced in NC neutrino interactions, can be vetoed by
identifying the second charm in the event. Similar considerations
apply to events with WSMs. As a consequence, a meaningful estimate of
the background rates is not possible without a concrete detector
study. To quantify the discovery potential, we therefore limit
ourselves to presenting a sensitivity reach for the detection of $N$
anomalous events, assuming 100\% reconstruction efficiency for the BSM
signal. We assume $10^{6} ~ \nu e$ and $10^{8} ~ \nu N$ scattering
interactions per year  
in the detector due to  SM CC
interactions. Then 
BSM physics could be detected in $\nu e$ scattering if:
\beq
 \frac{\eta}
 {2 \sqrt{2} G_F} \equiv 
  \epsilon > \sqrt{N} \times 10^{-3} 
\label{nue}
\eeq
and could be detected in $\nu N$ scattering if:
\beq
 \frac{\eta}{ 2 \sqrt{2} G_F} 
 \equiv \epsilon > \sqrt{N} \times 10^{-4}
\label{nuN}
\eeq
Notice that while the reach is lower with $\nu e$ scattering, the
detector backgrounds could be much smaller, and any future detailed
study will have to consider the possibility offered by this
channel. The results we present here are obtained assuming the event
rates from  $\nu N$ scattering.

We list in Table~\ref{tab:ffopt} the lepton-flavour-violating 
\ffops on which a $\nu$ factory could
set better bounds than are available now. 
 In the first column,
we list  $\bar{\mu} e \bar{\nu} \nu$ vertices of different chirality.
 This is a useful
way of listing 4-fermion vertices because
the new physics that generates them could
depend on fermion chirality. In the second
column are the best available
limits yet on the operators, and the processes
from which the bounds come. In the last column
are the limits that a \nufact\  could set,
assuming a 10 event sensitivity ($N = 10$ in 
Eqs.~\ref{nue} and \ref{nuN}). Horizontal lines
in the last column separate operators
that are distinguishable at a \nufact\ 
because they induce different
final-state particles  in the detector;
 polarized muons would be required 
to distinguish between LL and RL couplings.
Similarly, horizontal lines in 
the second column separate
operators that are distinguishable
in present experiments (where the
flavour of the outgoing neutrinos is
often not detected).

\begin{table}
\renewcommand{\arraystretch}{1.5}
\caption{ \label{tab:ffopt} 
Flavour changing four fermion vertices involving neutrinos,
which a $\nu$ factory could set  a bound on.
Neutrinos with index $i$ can be of any flavour.
The processes listed in parenthesis 
presently constrain (at 90 \% cl) the coefficients of the four fermion
vertices  to be less than 
$\epsilon \times 2 \sqrt{2} G_F$. 
A neutrino factory could set bounds given in the 
last column.  These bounds assume
that all \ffops other than the
contrained one are absent. Note that
there are much stronger bounds on
$\bar{\ell}_i \ell_j \bar{\ell}_k \ell_l$
vertices than the quoted bounds on
$\bar{\ell}_i \ell_j \bar{\nu}_k \nu_l$.}
\begin{center}
\begin{tabular}{||  c | c | c ||} \hline \hline
  Vertex& Current limits& \nufact\ limit\\ 
\hline \hline
   $(\bar{\mu} \gamma^{\alpha} P_L {e} )
   (\bar{\nu}_{\tau} \gamma_{\alpha} P_L \nu_i )$ 
  & $|\epsilon| < 0.06$ ($\nu$ oscillations) & $3 \times 10^{-4}$ \\ 
   \cline{1-2} 
   $(\bar{\mu} \gamma^{\alpha} P_R {e} )
   (\bar{\nu}_{\tau} \gamma_{\alpha} P_L \nu_i )$  
   & $|\epsilon| < 0.03$ ($G_{\mu} (g^S_{RR})$) &  $3 \times  10^{-4}$
 \\ \hline
   $(\bar{\mu} \gamma^{\alpha} P_L {e} )
   (\bar{\nu}_{\mu} \gamma_{\alpha} P_L \nu_i )$ 
   & 
       $|\epsilon| < 0.1$ ($\mu \rightarrow e {\nu}_e \bar{\nu}_\mu$)
       &  $3 \times 10^{-4}$ \\ 
   \cline{1-2} 
   $(\bar{\mu} \gamma^{\alpha} P_R {e} )
   (\bar{\nu}_{\mu} \gamma_{\alpha} P_L \nu_i )$  
   & $|\epsilon| < 0.03$ ($G_{\mu} (g^S_{RR})$) &  $3 \times 10^{-4}$
    \\  \cline{3-3}
   $(\bar{\mu} \gamma^{\alpha} P_R {e} )
   (\bar{\nu}_i \gamma_{\alpha} P_L \nu_\tau )$  
   & $|\epsilon| < 0.03$ ($G_{\mu} (g^S_{RR})$) &  $3 \times  10^{-4}$  
    \\ \cline{1-2}
   $(\bar{\mu} \gamma^{\alpha} P_L {e} )
   (\bar{\nu}_i \gamma_{\alpha} P_L \nu_{\tau} ) $ 
   & $|\epsilon| < 0.2$ ($G_{\mu}$) &   $3 \times  10^{-4}$ 
   \\    \hline
  \hline
\end{tabular}
\end{center}
\end{table}
\renewcommand{\arraystretch}{1}

\section{CONCLUSIONS}
This work documents our assessment of the physics
potential of detectors placed at the front-end of a high-current
muon storage ring. In most of the cases presented, we tried to
evaluate in quantitative terms the ultimate accuracies that can be
reached, given the available statistics and given the theoretical
knowledge available today. 

In the case of determinations of the partonic densities of the
nucleon, we proved that the \nufact\ could significantly improve the
already good knowledge we have today. In the unpolarized case, the
knowledge of the valence distributions would improve by more than one
order of magnitude, in the kinematical region $x\gsim 0.1$, which is
best accessible with 50~GeV muon beams. The individual components of
the sea ($\bar{u}$, $\bar{d}$, ${s}$ and $\bar{s}$), as well as the
gluon, would be measured with relative accuracies in the range of
1--10\%, for $0.1\lsim x \lsim 0.6$. The high statistics available over
a large range of $Q^2$ would furthermore allow the accurate
determination of higher-twist corrections, strongly reducing the
theoretical systematics that affect the extraction of $\as$ from sum
rules and global fits.  

In the case of polarized densities, we stressed the uniqueness of the
\nufact\ as a means of disentangling quark and antiquark distributions,
and their first moments in
particular. These can be determined at the level of few per cent for
up and down, and 10\% for the strange, sufficient to distinguish
between theoretical scenarios, and thus allowing a full understanding
of the proton spin structure. A potential ability
to pin down the shapes of individual flavour components with accuracies at the
level of few per cent is  limited by the mixing with the 
polarized gluon. To identify this possible
weakness of the \nufact\ polarized-target programme, it was crucial to
perform our analysis at the NLO; we showed in fact that any study
based on the LO formalism would have resulted in far too optimistic
conclusions. This holds true both in the
case of determinations based on global fits and on direct
extractions using flavour tagging in the final state. 
Our conclusion here is that a full exploitation of the \nufact\
potential for polarized measurements of the shapes of
individual partonic densities requires an a-priori knowledge of
the polarized gluon density. It is hoped that the new information
expected to arise from the forthcoming set of polarized DIS
experiments at CERN, DESY and RHIC will suffice. 

The situation is also very bright for measurements of C--even
distributions. Here, the first moments of singlet, triplet and octet
axial charges can be measured with
accuracies which are up to one order of magnitude better than the
current uncertainties. In particular, the improvement in the
determination of the singlet axial charge would allow a definitive
confirmation or refutation of the anomaly scenario compared to the
`instanton' or `skyrmion' scenarios, at least if the theoretical
uncertainty originating from the small--$x$ extrapolation can be kept under
control. The measurement of the octet axial charge with a few percent
uncertainty will allow a determination of the strange contribution to
the proton spin better than 10\%, and allow stringent tests of models
of $SU(3)$ violation when compared to the direct determination from
hyperon decays.

The measurement of two fundamental constants of nature, $\as(M_Z)$ and
$\sin^2\theta_W$, will be possible using a variety of techniques. At
best the
accuracy of these measurements will match or slightly improve
the accuracy available
today, although the
measurements at the \nufact\ are subject to different systematics and
therefore provide an important consistency check of current data.
In the case of $\as(M_Z)$, the dependence of the results on the
modeling of higher-twist corrections both in the structure function
fits and in the GLS sum rule is significantly reduced relative to
current measurements, as mentioned above. 
In the case of $\sin^2\theta_W$, its determination via $\nu e$
scattering at the \nufact\ has an uncertainty of approximately
$2\times 10^{-4}$, dominated by the statistics and the luminosity
measurement. This error is comparable to what already known
today from EW measurements in $Z^0$ decays. Compared to these,
however, this determination would
improve current low-energy extractions, and be 
subject to totally different systematic uncertainties. It would also be
sensitive to different classes of new-physics contributions.
The extrapolation
to $Q=M_Z$ is affected, at the same level of uncertainty, by the
theoretical assumptions used in the evaluation of the hadronic-loop
corrections to $\gamma$-$Z$ mixing. The determination via DIS, on the
other hand, is limited by the uncertainties on the heavy-flavour
parton densities. As shown earlier, these should be significantly
reduced using the \nufact\ data themselves.

In several other areas, the data from the \nufact\ will allow 
quantitative studies to be made of phenomena that, so far have only
been explored at a mostly qualitative level. This is the case of the
exclusive production of charmed mesons and baryons (leading to very
large samples, suitable for precise extractions of branching ratios
and decay constants), of the study of spin-transfer
phenomena, and of the study of nuclear effects in DIS. While nuclear
effects could be bypassed at the \nufact\ by using  hydrogen
targets directly, the flavour separation of partonic densities will require
using also targets containing neutrons. This calls for an accurate
understanding of nuclear effects. The ability to run with both $H$ and
heavier targets will in turn provide rich data sets useful for
quantitative studies of nuclear models.
The study of $\Lambda$ polarization both in the target and in the
fragmentation regions, will help clarifying the intriguing problem of
spin transfer. We reviewed several of the existing models, and
indicated how semi-inclusive neutrino DIS will allow the
identification of the
right ones, as well as providing input for the measurement of
polarized fragmentation functions.

Finally, we presented some cases of exploration for physics beyond the
SM using the \nufact\ data. Although the neutrino beam energies
considered in our work are well below any reasonable threshold for new
physics, the large statistics makes it possible to search for
manifestations of virtual effects. The exchange of new gauge bosons
decoupled from the first generation of quarks and leptons can be seen
via enhancements of the inclusive charm production rate, with a
sensitivity well beyond the present limits. Rare
lepton-flavour-violating decays of muons in the ring could be tagged
in the DIS final states
through the detection of wrong-sign electrons and muons, or of prompt
taus. Once again, the sensitivity at the \nufact\ goes well beyond
existing limits.

The work presented here has two clear weaknesses, which point to
two directions for further work. On one
side, a realistic evaluation of the experimental feasibility of the
proposed measurements should be performed. Concrete detector designs
should be proposed, and the detector performance should be evaluated
in the light of the statistical and theoretical accuracy reach set by
our study. As part of this work, an optimization of the
beam parameters (energy, length of the straight section, distance of
the detector from the ring) should be performed for each individual
physics task. On the other side, we made no any effort to present
a physics case justifying a set of goals for the performances. 
Considering that these results will
not be available before at least 10--15 years from now, 
some judgement on the merit of these
measurements has to be given. 
For example, why
should we want to know 15 years from now 
the strange density of the proton with an
accuracy of 1\%? How would our knowledge of fundamental physics, or
our ability to predict new phenomena,
improve if we could reach this goal? From the measurements listed in
this document, which ones will be the most important at the time when
the \nufact\ will be operating?
We hope that future studies of this new and fascinating
facility will address this important aspect of the case
for front-end experiments.
\vskip1cm
\noindent
\section*{ACKNOWLEDGEMENTS}
We are grateful to G.~Altarelli, V.~Barone, S.~Catani, D.~Harris, K.~McFarland,
and C.~Weiss for discussions. We thank A.~Blondel for providing the code
describing the neutrino beam. The work of S.I.A. and
S.A.K. was supported by the RFBR grant N 00-02-17432.
 The work of M.L.M., R.D.B., S.F., T.G. and G.R. 
was supported in part by the EU TMR contract
FMRX-CT98-0194 (DG 12 - MIHT). S.I.A., S.F. and S.A.K. acknowledge the
hospitality of the CERN Theory Division during some stages of this work.

\end{document}